\documentclass[twocolumn,aps,prl,superscriptaddress]{revtex4-2}

\usepackage{amsmath}
\usepackage{graphicx}
\usepackage{MnSymbol}
\usepackage{parskip}
\usepackage{makecell}

\begin{document}

\title{Rubber Friction: Theory, Mechanisms, and Challenges}

\author{B.N.J. Persson}
\affiliation{Peter Gr\"unberg Institute (PGI-1), Forschungszentrum J\"ulich, 52425, J\"ulich, Germany}
\affiliation{State Key Laboratory of Solid Lubrication, Lanzhou Institute of Chemical Physics, Chinese Academy of Sciences, 730000 Lanzhou, China}
\affiliation{MultiscaleConsulting, Wolfshovener str. 2, 52428 J\"ulich, Germany}

\author{R. Xu}
\affiliation{Peter Gr\"unberg Institute (PGI-1), Forschungszentrum J\"ulich, 52425, J\"ulich, Germany}
\affiliation{State Key Laboratory of Solid Lubrication, Lanzhou Institute of Chemical Physics, Chinese Academy of Sciences, 730000 Lanzhou, China}
\affiliation{MultiscaleConsulting, Wolfshovener str. 2, 52428 J\"ulich, Germany}

\begin{abstract}
Rubber friction is of major practical importance in applications such as tires, rubber seals, and footwear. This review article focuses on the theory and experimental studies of rubber friction on substrates with random roughness. We examine both steady sliding and accelerated motion, with particular attention to the origins of the breakloose friction force and the influence of pre-slip, elasticity, and flash temperature on friction dynamics. We further discuss rolling friction for cylinders and spheres, as well as sliding friction for triangular sliders on dry and lubricated rubber surfaces. Theoretical predictions are compared with experimental results obtained using different materials, geometries, and environmental conditions, highlighting the importance of accounting for multiscale roughness. Open challenges, such as the role of adhesion enhancement, energy dissipation due to crack opening, and the physical origin of the short-distance roughness cut-off, are 
discussed.
\end{abstract}

\maketitle

\setcounter{page}{1}
\pagenumbering{arabic}




Corresponding author: B.N.J. Persson, 

email: b.persson@fz-juelich.de

\vskip 0.2cm

{\bf 1 Introduction} 

Rubber friction plays a critical role in many applications, including tires, rubber seals, footwear, and damping systems \cite{HK1,HK2,HK3,HK4,HK5}. 
In such systems, the frictional behavior is governed by a complex interplay of viscoelastic energy dissipation and adhesion, influenced by factors such as surface roughness, lubrication, temperature, and sliding velocity.

The area of real contact when two elastic solids with random roughness are squeezed together by a normal force $F_{\rm N}$ is typically proportional to the normal force \cite{A1,A2,A3,A4,A5,A6,A7,A8,A9}. This proportionality holds even when adhesion increases the contact area, provided that the adhesion does not result in a (macroscopic) pull-off force, a condition satisfied in most practical applications (e.g., negligible adhesion when lifting a bottle from a table \cite{Adh}). When the real contact area remains proportional to $F_{\rm N}$, the sliding friction force is usually also proportional to $F_{\rm N}$. In such cases, one defines static and kinetic friction coefficients ($\mu_{\rm s}$ and $\mu_{\rm k}$), where the force required to initiate sliding (the breakloose friction force) equals $\mu_{\rm s} F_{\rm N}$, and the friction force during steady sliding equals $\mu_{\rm k} F_{\rm N}$.

Understanding rubber friction has long been a focus of tribological research. Pioneering work by Schallamach \cite{1,2}, Greenwood and Tabor \cite{Tabor1}, and Grosch \cite{Grosch} laid the foundation for interpreting the adhesive contribution (from shearing the real area of contact) to friction, as well as the viscoelastic contribution arising from macroscopic deformation \cite{Tabor1}. Schallamach proposed a stick-slip mechanism at the interface and discovered detachment waves (Schallamach waves) that characterize the sliding of soft rubber on smooth surfaces. Greenwood and Tabor investigated the role of viscoelasticity in both sliding and rolling configurations at the macroscopic level, while Grosch systematically explored the dependence of friction and wear on sliding velocity and temperature.

Subsequent developments incorporated more detailed descriptions of rubber rheology and surface roughness into theoretical models. For example, Chernyak and Leonov \cite{Chernyak}, and later Volokitin and Persson \cite{theory3}, presented theories that accounted for the adhesive contribution with frequency-dependent viscoelastic properties of rubber. Finally, Persson \cite{A1}, using his multiscale contact mechanics theory, introduced a comprehensive framework that includes the viscoelastic contribution due to roughness over multiple length scales, as well as the adhesive contribution from the real area of contact. This approach enabled more quantitative predictions of both adhesive and hysteretic friction components, including the effects of surface topography and thermal feedback such as flash temperature \cite{VG,flash}. Comparisons with experimental observations under various conditions, including dry and lubricated contacts, have provided strong support for the predictive power of this framework \cite{9,10,water,Nam}.

This review summarizes theoretical and experimental progress in the study of rubber friction on rigid, randomly rough substrates. We discuss key physical quantities and mechanisms, and explore how different geometries and boundary conditions affect the frictional response.

The article is organized as follows. Secs. 2 and 3 review two fundamental quantities that govern rubber friction in Persson contact mechanics: the viscoelastic modulus of rubber and the surface roughness power spectrum.

In Sec. 4, we introduce the analytical framework used to calculate the real contact area between a nominally flat elastic block and a rigid rough substrate, based on Persson contact mechanics theory. The theory predicts how the contact area depends on the applied pressure, the effective elastic modulus, and the surface roughness, which enters through the root-mean-square (rms) slope. The extension to viscoelastic sliding conditions is also briefly discussed.

In Sec. 5, we present the core of the analytical theory for rubber friction on rough surfaces. We begin with the general concepts, followed by the analytical formulation of rubber friction (Sec. 5.1), and then comparisons with experimental data (Sec. 5.2). Additional considerations, such as dynamic effects and their influence on the friction coefficient, are discussed in Sec. 5.3.

Sec. 6 addresses rolling friction, which can be regarded as a limiting case of sliding friction where the contribution from interfacial shear is negligible and friction arises solely from bulk viscoelastic deformation.

In Sec. 7, we review a special case involving a triangular slider on rubber surfaces under both dry and lubricated conditions. This case effectively summarizes the theoretical framework: under dry conditions, the behavior follows the sliding friction theory described in Sec. 5, while under lubrication, where interfacial shear is suppressed and the behavior closely resembles that of rolling friction.

In Sec. 8, we examine the effects of lubrication on rubber friction. This includes the role of fluid-induced separation between the rubber and substrate, the competition between wetting and dewetting mechanisms, and how these influence the transition from boundary to hydrodynamic lubrication. We also analyze how interfacial adhesion and surface roughness affect the wetting dynamics, especially under transient sliding conditions. Furthermore, the contribution of adhesion in lubricated contacts is discussed, with a focus on how lubricant layers modulate the adhesive interaction and the resulting shear stress. These insights are supported by both theoretical analysis and comparisons with experimental findings.

Sec. 9 reviews the similarities and differences between friction on ice and snow. While the mechanisms of friction on ice have been widely studied due to their relevance to winter mobility, rubber-snow interactions remain less well understood. The temperature dependence of the friction behavior on ice as well as the water content dependence on snow are discussed. Experimental findings are compared with theoretical predictions, highlighting the role of surface premelting, capillary forces, and viscoelastic dissipation.

Finally, Sec. 10 provides a summary and conclusion, reflecting on the current understanding of rubber friction and highlighting open questions and future directions.

\begin{figure}[tbp]
\includegraphics[width=0.45\textwidth,angle=0]{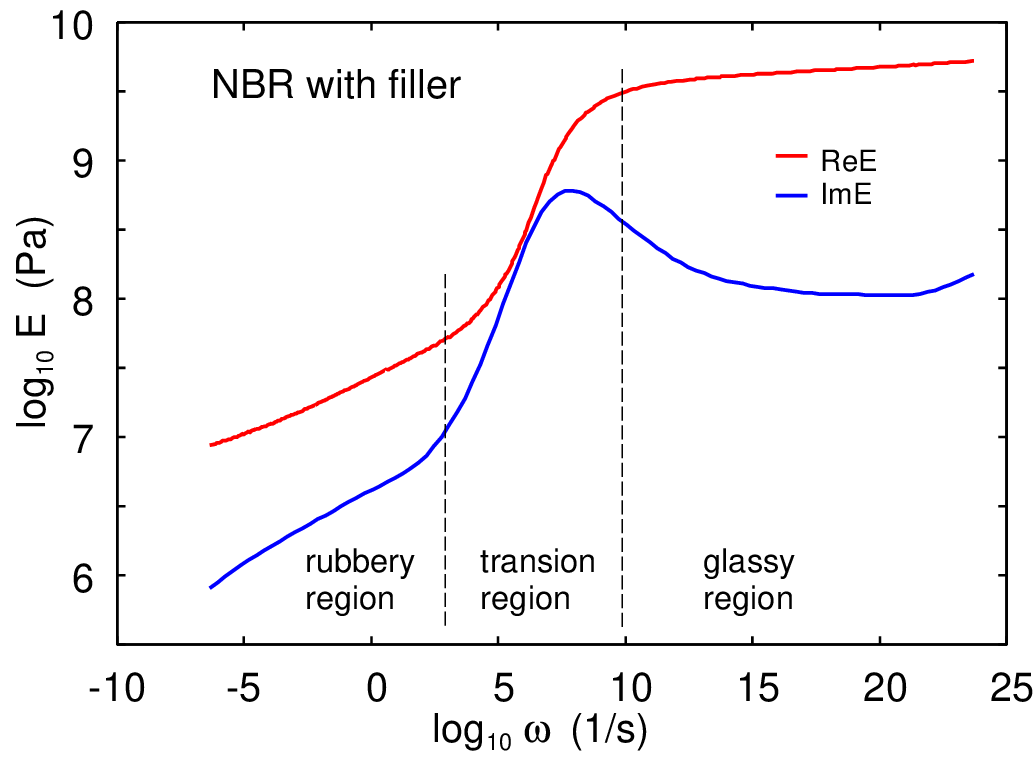}
\caption{
The real and the imaginary part of the (linear response) viscoelastic modulus as a function of frequency $\omega$ (log-log scale).
For a NBR rubber compound with filler particles at $T=20^\circ {\rm C}$.}
\label{NBR.1logOmega.2logE.eps}
\end{figure}

\begin{figure}[tbp]
\includegraphics[width=0.45\textwidth,angle=0]{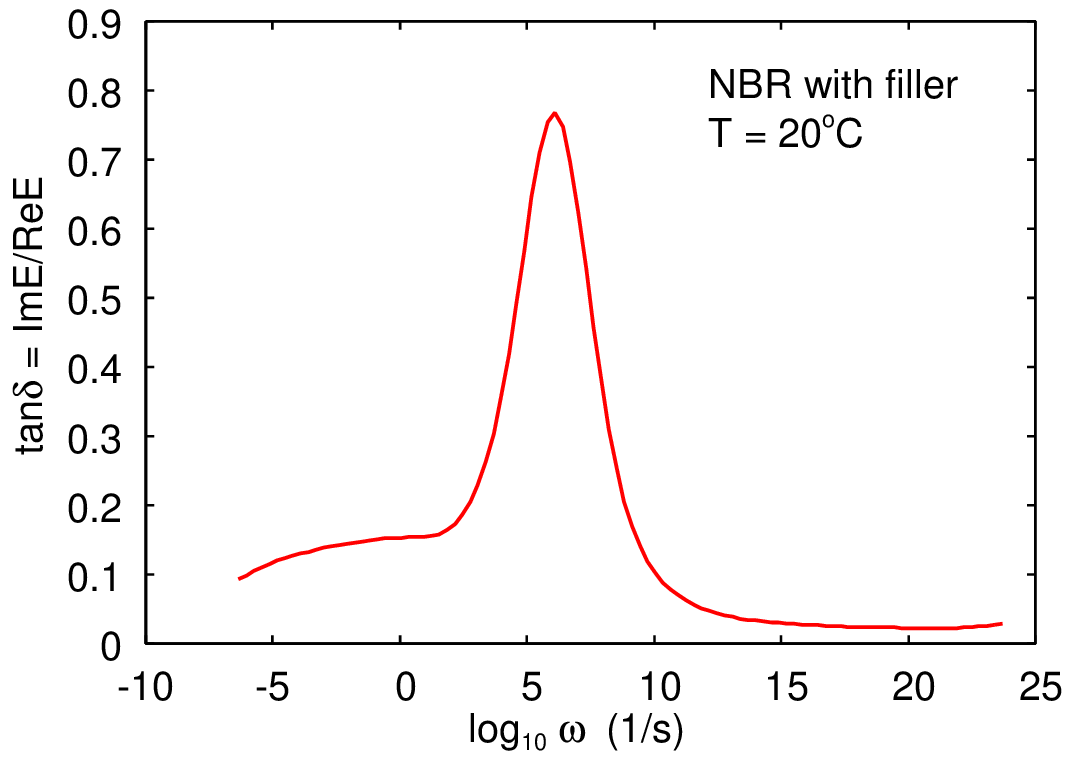}
\caption{
The loss tangent ${\rm tan}\delta = {\rm Im}E/{\rm Re}E$ as a function of the logarithm of the frequency $\omega$.
For a NBR rubber compound shown in Fig. \ref{NBR.1logOmega.2logE.eps}.}
\label{NBR.1logOmega.2logLosstangent.eps}
\end{figure}

\vskip 0.3 cm
{\bf 2 Viscoelastic modulus}

Rubber compounds consist of long molecular chains that are crosslinked, resulting in a viscoelastic solid.
When they undergo time-dependent deformation, mechanical energy is converted into thermal motion.
Here, we will be sloppy and use the term “energy dissipation” to describe the conversion of mechanical energy into heat.
In rubber, energy dissipation arises from internal friction between polymer chains as molecular segments slide past one another.

It is convenient to use complex notation for stress and strain:
$$\sigma (t) = \sigma (\omega) e^{-i\omega t}, \ \ \ \ \epsilon(t) = \epsilon (\omega) e^{-i\omega t} \eqno(1)$$
The physical stress and strain correspond to the real parts of these expressions. For a linear viscoelastic material, the viscoelastic modulus $E(\omega)$ is defined by
$$\sigma (\omega) = E(\omega) \epsilon (\omega)\eqno(2)$$
The modulus $E(\omega)$ is complex:
$$E(\omega) = E_{\rm R} (\omega) - i E_{\rm I} (\omega) = |E(\omega)| e^{-i\alpha} \eqno(3)$$
where $E_{\rm R}$, $E_{\rm I}$, and $\alpha$ are real quantities. Using (1) and (3), we see that if $\epsilon (t) = \epsilon_0 \cos (\omega t)$, then
$\sigma (t) = |E(\omega)| \epsilon_0 \cos (\omega t + \alpha)$. This indicates that the stress response lags behind the strain by a phase angle $\alpha$, a result of energy dissipation.

At low frequencies (or high temperatures), rubber behaves as a soft elastic solid with a typical modulus $E_0$ on the order of a few ${\rm MPa}$. At high frequencies (or low temperatures), it again behaves elastically, but with a much higher modulus, $E_1 \approx 1000 E_0$, typically on the order of a few ${\rm GPa}$. The low-frequency regime is called the rubbery region, while the high-frequency regime is referred to as the glassy region \cite{glassy0, glassy1}.

A simple model often used to describe the viscoelastic behavior of rubber is:
$${1\over E} = {1 \over E_1} +  \left ( {1 \over E_0}-{1 \over E_1} \right ) {1 \over 1 - i\omega \tau }\eqno(4)$$
Here, the model is characterized by a single relaxation time $\tau$. In reality, rubber exhibits a broad distribution of relaxation times, and the transition from the rubbery to the glassy region spans several decades of frequency. This behavior is illustrated in Figs. \ref{NBR.1logOmega.2logE.eps} (master curve) and \ref{NBR.1logOmega.2logLosstangent.eps} (loss tangent) for acrylonitrile butadiene rubber (NBR).

Rubber deformation involves stress augmented thermally activated processes, and $E(\omega)$ is strongly temperature-dependent. An important characteristic of rubber is its glass transition temperature $T_{\rm g}$. We define $T_{\rm g}$ as the temperature at which ${\rm tan} \delta = {\rm Im}E (\omega,T)/{\rm Re}E(\omega,T)$ reaches a maximum for a frequency $\omega = 0.01 \ {\rm s}^{-1}$. This definition gives values of $T_{\rm g}$ that are close to those obtained using standard viscosity or calorimetric measurements.

When deformed at temperatures well below $T_{\rm g}$, rubber is in the glassy state, where molecular motion is limited.  
At temperatures well above $T_{\rm g}$, it is in the rubbery state with high chain mobility, and would be a (high-viscosity) liquid if not for the crosslinks.

The viscoelastic master curve and the temperature dependence of $E(\omega)$ can be determined using a Dynamic Mechanical Analyzer (DMA), which subjects small 
rubber specimens to elongation or shear at various frequencies and temperatures. 

\begin{figure}[tbp]
\includegraphics[width=0.45\textwidth,angle=0.0]{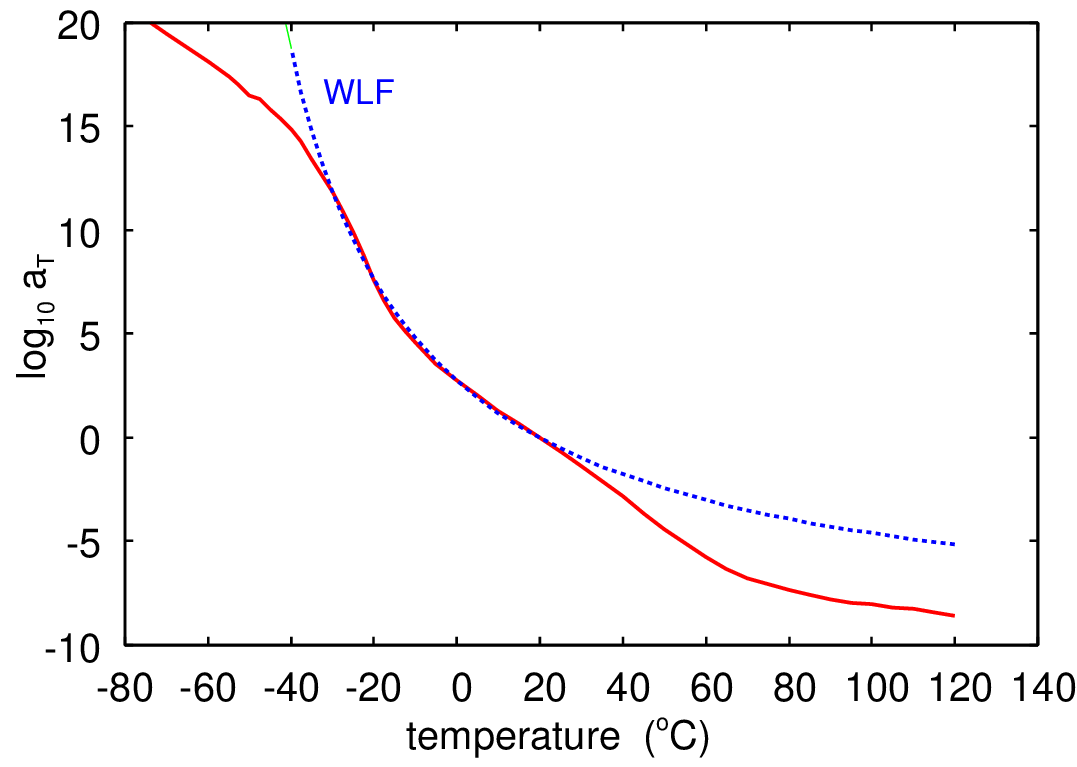}
\caption{\label{1Temp.2logaT.eps}
Horizontal shift factor $a_T$ as a function of temperature for the SBR compound used in Sec. 7. 
The reference temperature is $T_0 = 20^\circ {\rm C}$, with $a_{T_0} = 1$.
}
\end{figure}

By shifting the frequency segments measured at different temperatures $T$ (typically from
$-80^\circ {\rm C}$ to $120^\circ {\rm C}$) and in some frequency range $\omega$
(typically from $0.1 \ {\rm Hz}$ to $100 \ {\rm Hz}$), one can construct a master curve $E(\omega,T) = E(\omega a_T,T_0)$ that spans a wide frequency range.
Here, $a_T$ is the shift factor, with $a_T = 1$ at $T = T_0$. For ``simple'' rubber compounds, if the reference temperature
$T_0 = T_{\rm g}$, $a_T$ is approximately given by the Williams-Landel-Ferry (WLF) equation:
$${\rm log}_{10} a_T = - \frac{C (T - T_{\rm g})}{T + T_1 - T_{\rm g}} \eqno(5)$$
with typical values $C \approx 17.44$ and $T_1 \approx 51.6 \ {\rm K}$.
However, for most rubber compounds, this relation is valid only within a limited temperature range. 
This is illustrated in Fig. \ref{1Temp.2logaT.eps}, which shows the shift factor $a_T$ for a carbon-filled styrene-butadiene rubber (SBR) compound. 
It should also be noted that the frequency-temperature shifting procedure used to construct the master curve 
is strictly valid only at very small strains, where stress and strain are linearly related. In the linear response region, ${\rm Re}E$ and ${\rm Im}E$ are not independent functions: ${\rm Re}E$ can be obtained from ${\rm Im}E$ using a Kramers-Kronig relation.  
Hence, if shifting produces a smooth mastercurve for, e.g., ${\rm Im}E$, the same shift function $a_T$ will result in a smooth mastercurve for ${\rm Re}E$. This is not true at larger strains, where there is no simple relation between the real and imaginary parts of the effective modulus.

\begin{figure}
\includegraphics[width=0.45\textwidth,angle=0.0]{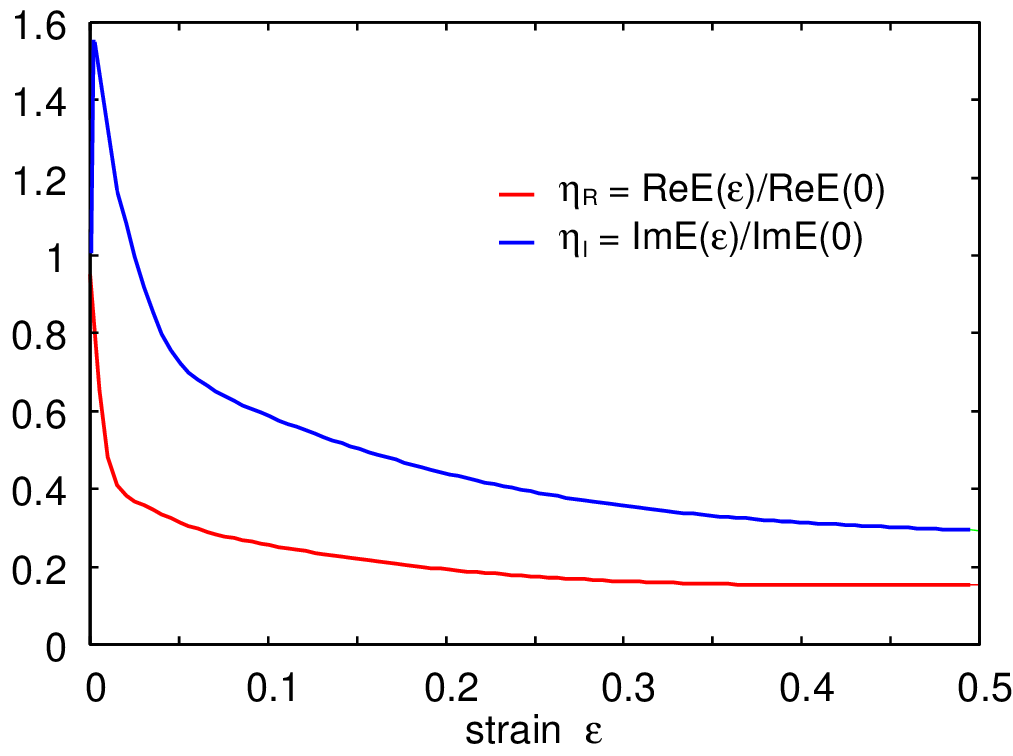}
\caption{\label{1strain.2reduction.eps}
Viscoelastic modulus at strain $\epsilon$, normalized by the modulus at $\epsilon = 10^{-4}$, 
measured at a frequency of $1 \ {\rm Hz}$ and at $T = 20^\circ {\rm C}$, for the same SBR compound shown in Fig. \ref{1Temp.2logaT.eps}.
}
\end{figure}

Most rubbers used in tires, seals, and other applications contain filler particles, typically carbon or silica nanoparticles. These fillers form networks within the rubber matrix, making the material stiffer at small deformations, typically when the strain is below $0.001$, where the filler network remains intact. As the strain increases the filler network breaks down and the real part of the modulus decreases, and at strain levels relevant to most applications (typically $0.1-0.5$), the modulus can be reduced by a factor of $\sim 5$.

To account for this strain-dependent softening, we define an effective modulus $E_{\rm eff}$, which depends on the strain $\epsilon$ and the temperature $T$:
$${\rm Re} E_{\rm eff} = E_{\rm R}(\omega,T) \, \eta_{\rm R} (\epsilon), 
\ \ \ \ {\rm Im} E_{\rm eff} = E_{\rm I}(\omega,T) \, \eta_{\rm I} (\epsilon) \eqno(6)$$
where $E_{\rm R}$ and $E_{\rm I}$ are the small-strain (linear response) modulus.  
The strain softening functions $\eta_{\rm R}(\epsilon)$ and $\eta_{\rm I}(\epsilon)$ can be measured using DMA.  
They also depend on $\omega$ and $T$, but this dependence is weaker than that of the small-strain viscoelastic modulus.  
We note that the modulus obtained in this way account for nonlinearity via the secant method, which we consider the most appropriate approach for incorporating nonlinear effects within a linear response framework.

In Fig. \ref{1strain.2reduction.eps}, we show $\eta_{\rm R}$ and $\eta_{\rm I}$  for a SBR compound with carbon-filler. The function $\eta_{\rm R}(\epsilon)$ decreases monotonically with increasing strain. In contrast, $\eta_{\rm I}(\epsilon)$ first increases and reaches a maximum around $\epsilon \approx 0.01$, then decreases continuously over the strain range studied.
For very large strain, where the polymer chains are fully stretched, $\eta_{\rm R} (\epsilon)$ increases rapidly, but this large strain region is not important in most applications.

\begin{figure}[tbp]
\includegraphics[width=0.45\textwidth,angle=0.0]{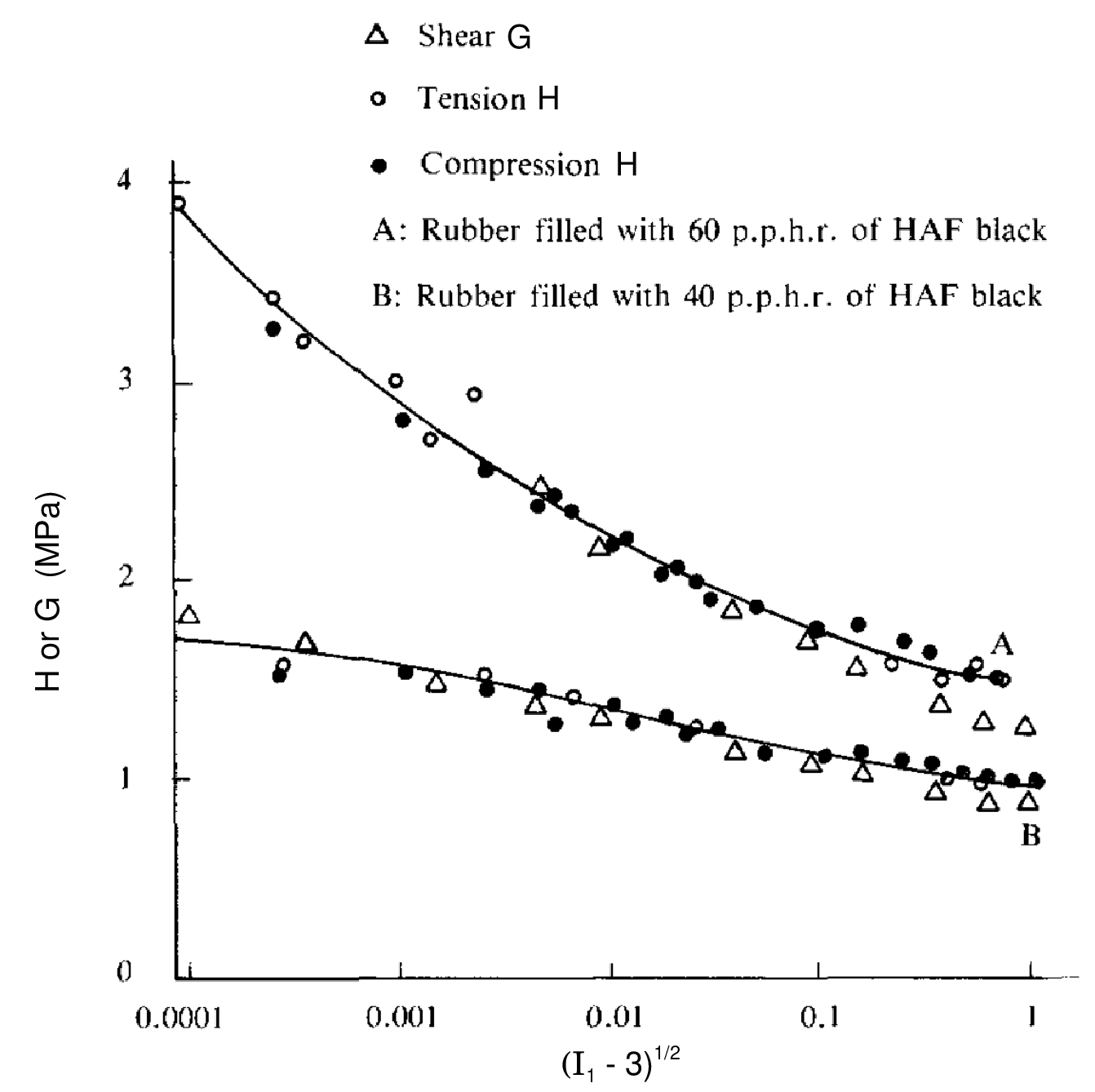}
\caption{\label{NonlinearRubberU.eps}
Dependence of the material parameters $H$ and $G$ on the strain invariant $I_1$ for natural rubber filled with 60 and 40 p.p.h.r. of HAF (N330) carbon black. Adapted from Ref. \cite{Greg1,Greg2}.
}
\end{figure}

In many applications, the strain field is complex, involving elongation, compression, and shear components. Experimental studies have shown that rubber nonlinearity is insensitive to
the mode of deformation, at least in the static limit. Gregory \cite{Greg1,Greg2} found that for a natural rubber (NR) compound with carbon black filler, the elastic energy stored in the deformed rubber can be accurately considered as a function of the strain invariant
$$I_1 = \lambda_1^2 + \lambda_2^2 + \lambda_3^2,$$
where $\lambda_1 = L_x/L_0$ is the extension ratio in the $x$-direction, and similarly for $\lambda_2$ and $\lambda_3$.

Fig. \ref{NonlinearRubberU.eps} shows the effective modulus in elongation and compression,
$H = \sigma / (\lambda - \lambda^{-2})$ (with $H = E/3$ in the linear response regime), 
and in shear, $G = \sigma / \gamma$, as a function of $(I_1 - 3)^{1/2}$. 
Clearly, the effective modulus is nearly independent of the mode of deformation up to $(I_1 - 3)^{1/2} \approx 1$.

Physically, if strain softening results from the breakdown of the filler network as the rubber deforms, this effect should occur regardless of whether the sample is extended, compressed, or sheared. Hence, one expects strain softening to occur at small strain for all deformation modes, as observed in Fig. \ref{NonlinearRubberU.eps}.

Assuming incompressible rubber, so that $\lambda_1 \lambda_2 \lambda_3 = 1$, we have for elongation or compression in the $x$-direction $\lambda_1 = L/L_0$ and $\lambda_2 = \lambda_3$, leading to
$$I_1 = \lambda_1^2 + {2 \over \lambda_1}.$$
Thus, $(I_1 - 3)^{1/2} = 1$ corresponds to $L/L_0 \approx 0.54$ (compression) or $L/L_0 \approx 1.68$ (elongation), 
which is similar to or larger than the strains expected in most engineering applications.

\begin{figure}[tbp]
\includegraphics[width=0.3\textwidth,angle=0]{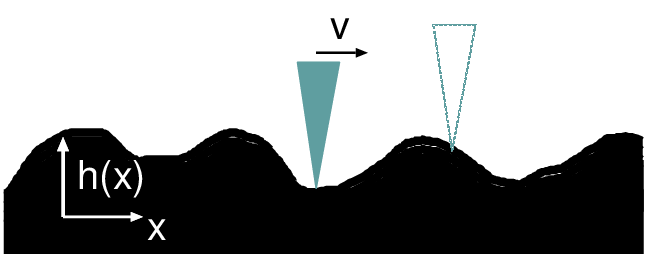}
\caption{
The surface height profile $h(x)$ along the $x$ axis, as measured using a stylus instrument.
}
\label{heightprofile.eps}
\end{figure}

\begin{figure}[tbp]
\includegraphics[width=0.45\textwidth,angle=0]{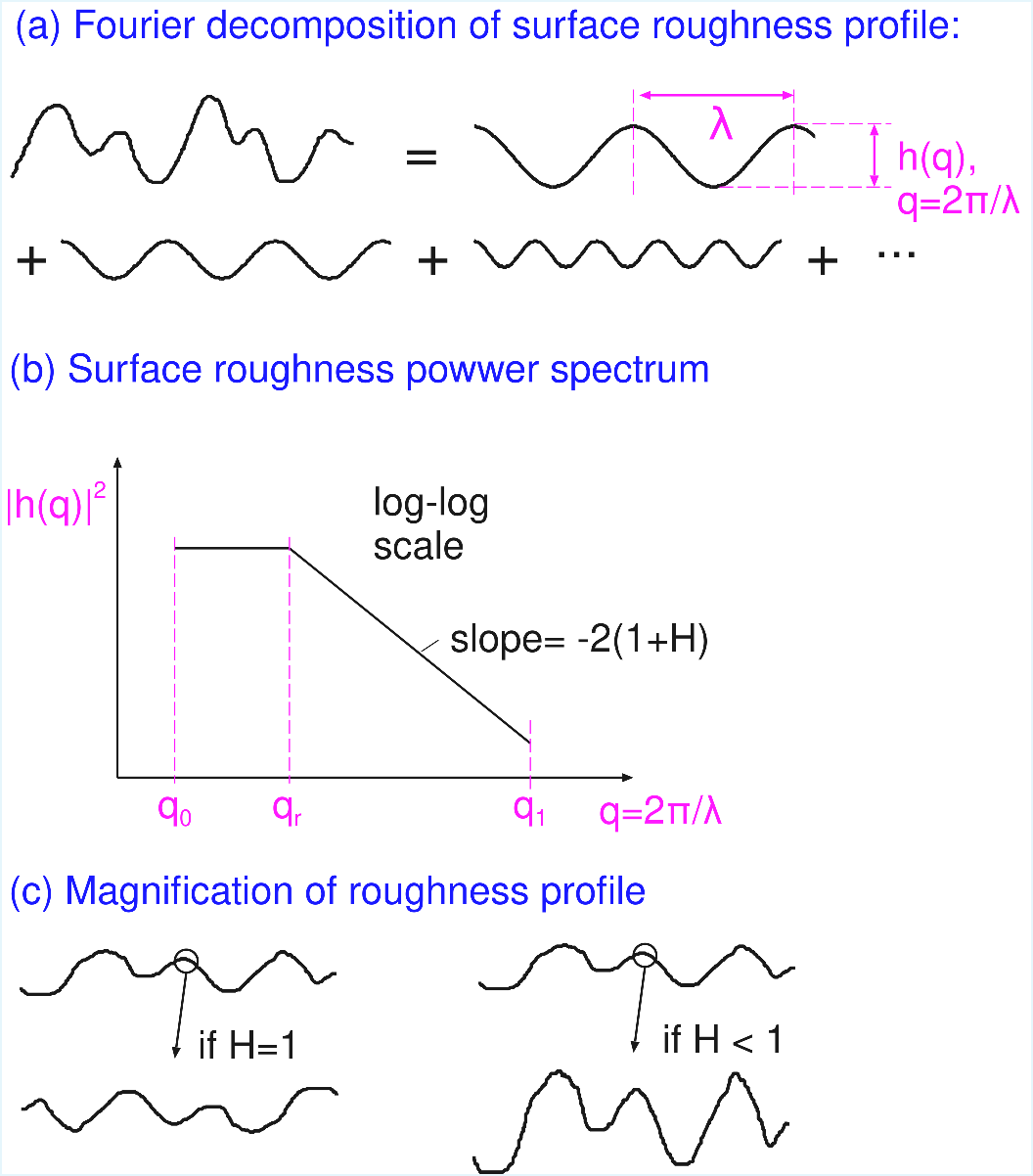}
\caption{
(a) A surface roughness profile $h(x)$ can be expressed as a sum of periodic functions with different wavelengths and amplitudes.  
(b) The two-dimensional surface roughness power spectrum is proportional to the square of the amplitude of the Fourier transform of the height $h(x,y)$, measured over a square area. Most engineering surfaces have a power spectrum that appears as a tilted line on a log-log plot, with the slope defined by the Hurst exponent $H$. A roll-off occurs for wavenumbers $q < q_{\rm r}$. The fractal dimension of the surface is $D_{\rm f} = 3 - H$.  
(c) Surfaces with a fractal dimension $D_{\rm f} = 2$ or Hurst exponent $H = 1$ are statistically self-similar, meaning that magnified segments resemble the original in a statistical sense. For $D_{\rm f} > 2$ or $H < 1$, the ratio of roughness amplitude to wavelength increases with magnification. Most engineering surfaces have $0.7 < H < 1$.
}
\label{FourierNewBo.cut1.eps}
\end{figure}

\vskip 0.3 cm
{\bf 3 Surface roughness power spectra}

There are many ways to study surface roughness \cite{challangeTOP}, but the most accurate methods are engineering stylus and Atomic Force Microscope (AFM) instruments. In such measurements, a sharp tip is moved along a straight line (see Fig. \ref{heightprofile.eps}), while being pressed lightly against the substrate. 
The tip follows the surface contour, moving up and down, and records the topography $z = h(x)$.

If the surface roughness has isotropic statistical properties, then all statistical properties of the surface can be captured in a 
sufficiently long line scan of length $L$. J. B. Fourier showed that a (nearly) arbitrary function $h(x)$ can be expressed as a sum or integral of ${\rm cos}(2 \pi x/\lambda)$ and ${\rm sin}(2 \pi x/\lambda)$ waves with different amplitudes and wavelengths $\lambda$ (see Fig. \ref{FourierNewBo.cut1.eps}a). It is more convenient to use complex notation:
$$h(x) = \int dq \ h(q_x) e^{-i q_x x} \eqno(7)$$
where the wavevector $q_x = 2\pi / \lambda$. Since $h(x)$ is real, we must have $h^*(q_x) = h(-q_x)$. The one-dimensional (1D) power spectrum is defined as \cite{Prev,Pastewka}
$$C_{\rm 1D} (q_x) = \frac{2\pi}{L} |h(q_x)|^2 \eqno(8)$$

The two-dimensional (2D) power spectrum $C_{\rm 2D}$, which appears in analytical theories of rubber friction, can be derived from $C_{\rm 1D}$ through an integral transform when the surface roughness is isotropic. However, it is more conveniently obtained from measurements of the surface height $h(x,y)$ over a square area $A_0 = L^2$. The 2D power spectrum is given by
$$C_{\rm 2D}({\bf q}) = \frac{(2\pi)^2}{A_0} |h({\bf q})|^2 \eqno(9)$$
where $h({\bf q})$ is the Fourier transform of the surface height $h({\bf x})$, and ${\bf q} = (q_x, q_y)$ is the wavevector. The range of $q_x$ is $\pi / a < q_x < \pi / a$, and similarly for $q_y$, where $a$ is the lateral spacing between adjacent height data points along the $x$ axis.

In the isotropic case, $C_{\rm 2D}({\bf q})$ depends only on the magnitude $q = |{\bf q}|$ of the wavevector. For simplicity, we will refer to $C_{\rm 2D}({\bf q})$ as $C({\bf q})$ in the following.

Surfaces often exhibit power-law spectra of the form $C(q) \sim q^{-\beta}$, which appear as straight lines on a log-log plot [see Fig. \ref{FourierNewBo.cut1.eps}(b)], but with a roll-off at long wavelengths (i.e., small wavenumbers). This roll-off arises because most engineering surfaces are designed to be smooth on the scale of the object. 
For example, for asphalt road surfaces, the roll-off is determined by the largest stone particles in the mixture. In this case the 
roll-off wavelength is on the order of a few millimeters, corresponding to a roll-off wavenumber of $q_{\rm r} = 2\pi / \lambda_{\rm r} \approx 10^3 \ {\rm m}^{-1}$.

Self-affine fractal surfaces are characterized by a power-law spectrum $C(q) \sim q^{-\beta}$, where $\beta = 2(1 + H)$ and $H$ is the Hurst exponent. The Hurst exponent is related to the fractal dimension by $D_{\rm f} = 3 - H$. When $D_{\rm f} = 2$ or $H = 1$, the surface is self-similar, meaning that a magnified segment statistically ``looks the same'' as the original surface [see Fig. \ref{FourierNewBo.cut1.eps}(c)]. 

For $D_{\rm f} > 2$ or $H < 1$, the ratio between roughness amplitude and wavelength increases with magnification. Most engineering surfaces have $0.7 < H < 1$, and such surfaces appear rougher at higher magnifications [see Fig. \ref{FourierNewBo.cut1.eps}(c)].

\begin{figure}[tbp]
\includegraphics[width=0.45\textwidth,angle=0]{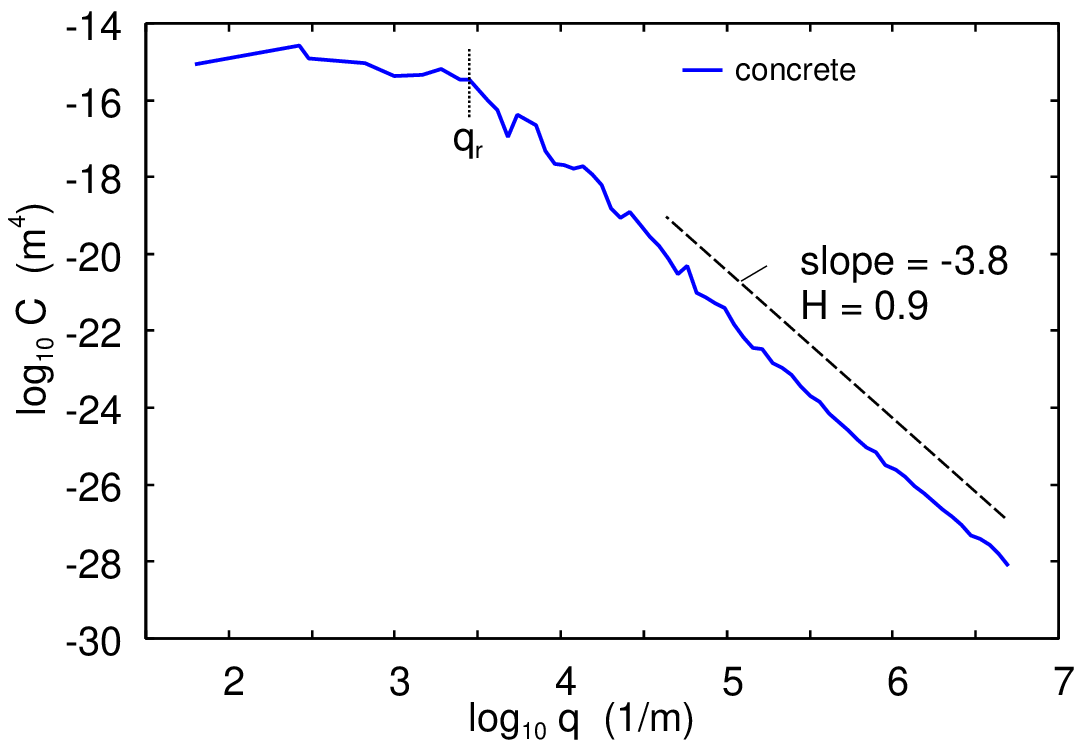}
\caption{
The 2D surface roughness power spectrum of a concrete surface, plotted as a function of the wavenumber on a log-log scale.
}
\label{1logq.2logC.concrete.steelmold.eps}
\end{figure}

As an example, Fig. \ref{1logq.2logC.concrete.steelmold.eps} shows the 2D surface roughness power spectrum of a concrete surface. The sloped region is well approximated by a straight line with a slope of $-3.8$, corresponding to a Hurst exponent $H = 0.9$. The roll-off wavenumber is $q_{\rm r} \approx 3000 \ {\rm m}^{-1}$, corresponding to a wavelength of $\lambda_{\rm r} = 2\pi / q_{\rm r} \approx 2 \ {\rm mm}$.

\begin{figure}[tbp]
\includegraphics[width=0.45\textwidth,angle=0]{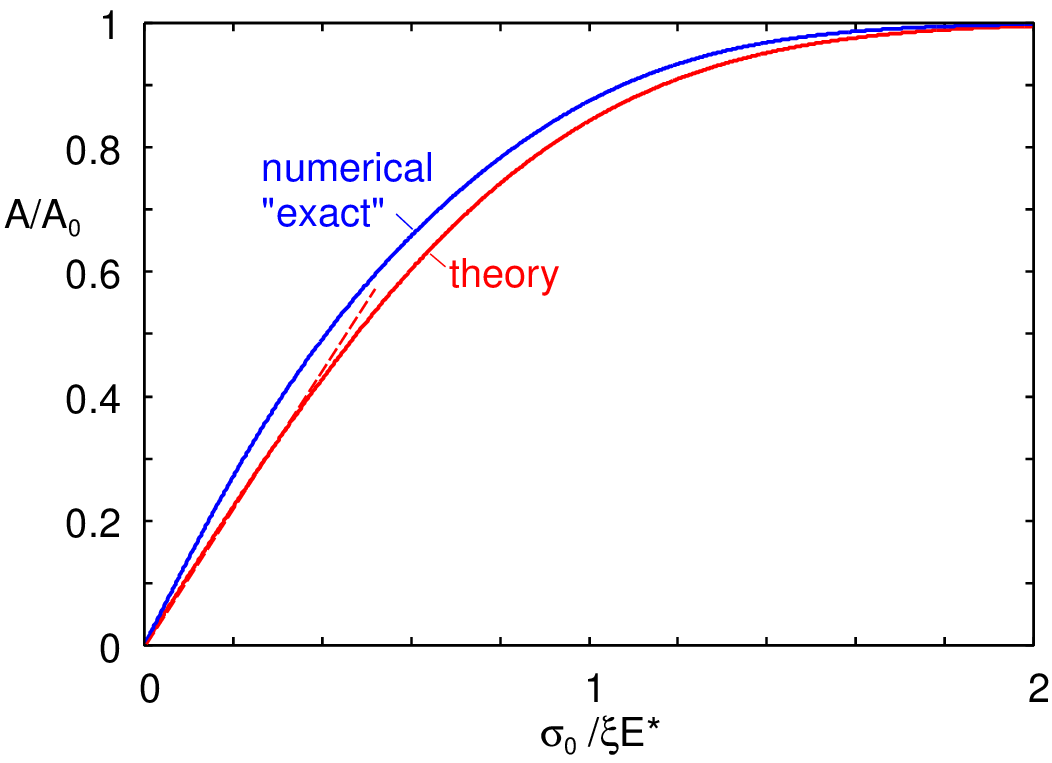}
\caption{
The normalized area of real contact $A/A_0$ as a function of $\sigma_0/\xi E^*$, where $\sigma_0$ is the nominal contact pressure,
$\xi$ the rms slope, and $E^* = E/(1-\nu^2)$ the effective elastic modulus. The blue line represents the result of
an ``exact'' numerical study \cite{A2,A3}, and the red line shows the theoretical prediction from (10) \cite{A1}.
}
\label{1pressure.2area.eps}
\end{figure}

\vskip 0.3cm
{\bf 4 Elements of contact mechanics}

Consider a rectangular block squeezed against a substrate with nominal pressure $\sigma_0$.
For simplicity, we assume that the substrate is rigid and randomly
rough, and that the block is elastic with a flat surface. 
The normalized area of real contact predicted by Persson
contact mechanics theory \cite{A1} is given by
$${A\over A_0} = {\rm erf} \left ({\surd 2 \sigma_0 \over \xi E^*}\right ), \eqno(10)$$
where ${\rm erf}(x)$ is the error function, 
$A_0$ is the nominal contact area (i.e., the area of the bottom surface of the block), and 
$E^* = E/(1-\nu^2)$ is the effective elastic modulus ($E$ is the Young’s modulus and $\nu$ the Poisson ratio).
The rms slope $\xi$ can be obtained 
directly from the measured height profile $z = h(x,y)$ or calculated from
the surface roughness power spectrum using
$$\xi^2 = \int d^2q \ q^2 C_{\rm 2D}({\bf q}).$$
The red line in Fig. \ref{1pressure.2area.eps} corresponds to (10).

Using the approximation ${\rm erf}(x) \approx 2x/\surd \pi$ for $x \ll 1$, we obtain, for small arguments of the error function,
$${A\over A_0} \approx {\kappa \sigma_0 \over \xi E^*}, \eqno(11)$$
where $\kappa = (8/\pi)^{1/2}$.
The linear relation in (11) between $A$ and $\sigma_0$ holds approximately
for $A/A_0 < 0.3$ (see the dashed line in Fig. \ref{1pressure.2area.eps}).

The analytical expression in (10) depends on the surface roughness power spectrum
only through the rms slope $\xi$. Numerical studies \cite{A2,A3,Camp,Put} have shown that different power spectra with the same
rms slope give nearly the same (universal) relation between $A/A_0$ and $\sigma_0/\xi E^*$, as predicted by the theory.
The blue line in Fig. \ref{1pressure.2area.eps} shows the result of one ``exact'' numerical study, with the slope parameter $\kappa \approx 2$.

In the case of sliding contact (velocity $v$) 
for viscoelastic solids, $A/A_0$ can be calculated
using equations that will be presented later. Here, we note that
if the roughness occurs on a single length scale (in the sliding direction)
$\lambda$, then (10) remains approximately valid if $E^*$ is 
replaced by $|E(qv)/(1-\nu^2)|$, where $q = 2 \pi/\lambda$, 
and $E(\omega)$ is the complex viscoelastic modulus.
This relation follows from the observation that the viscoelastic
solid is deformed at the frequency $\omega = qv$ when sliding on
a rigid corrugation with wavelength $\lambda$.

\begin{figure}[tbp]
\includegraphics[width=0.45\textwidth,angle=0]{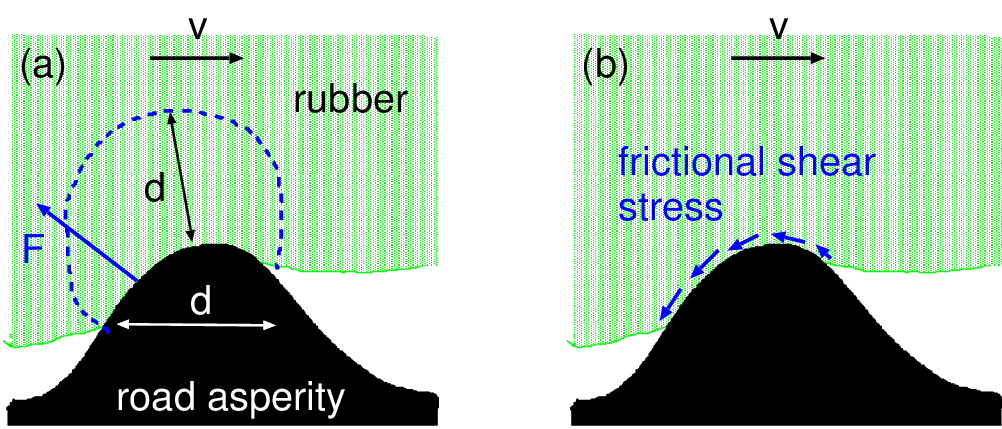}
\caption{
Several mechanisms contribute to rubber friction. (a) During sliding, surface asperities on the road cause time-dependent deformations in the rubber. These deformations extend into the rubber to a depth comparable to the size of the asperity contact regions. The resulting energy dissipation produces a contact force with a tangential component that opposes the direction of motion. (b) A frictional shear stress acts in the area of real contact. This stress can originate from different mechanisms (see Fig. \ref{ShearStress.eps}).
}
\label{TwoFriction.eps}
\end{figure}

\vskip 0.3 cm
{\bf 5 Rubber friction on rough surfaces}

We consider a rubber block sliding on a hard rough substrate, such as a tire tread block on a road surface. There are two main contributions to the friction force $F_{\rm f}$, as illustrated in Fig. \ref{TwoFriction.eps}.

One contribution arises from the viscoelasticity of the rubber. During sliding, the asperities on the road surface induce time-dependent deformations in the rubber, which result in energy dissipation (conversion of mechanical energy into heat).
This results in a contact force with a tangential component that opposes the sliding direction [see Fig. \ref{TwoFriction.eps}(a)].
The asperity induced deformations occur mainly within volume elements that extend into the rubber to a distance comparable to the 
lateral size of the asperity contact regions. 

In addition, there is a contribution from the frictional shear stress acting in the area of real contact, often referred to as the adhesive contribution.

Defining the friction coefficient in the usual way as $F_{\rm f} = \mu F_{\rm N}$, where $F_{\rm N}$ is the normal force, we write
$$\mu = \mu_{\rm visc} + \mu_{\rm cont} \eqno(12)$$

Experiments have shown that for sliding on randomly rough surfaces, both the viscoelastic contribution $\mu_{\rm visc}$ and the contact-area contribution $\mu_{\rm cont}$ are independent of $F_{\rm N}$, unless the normal force is large enough to approach full contact, or the sliding speed is high enough for frictional heating to become significant, or the surfaces are so smooth that adhesion manifest itself on the macroscopic scale as a pull-off force.

\begin{figure}[tbp]
\includegraphics[width=0.35\textwidth,angle=0]{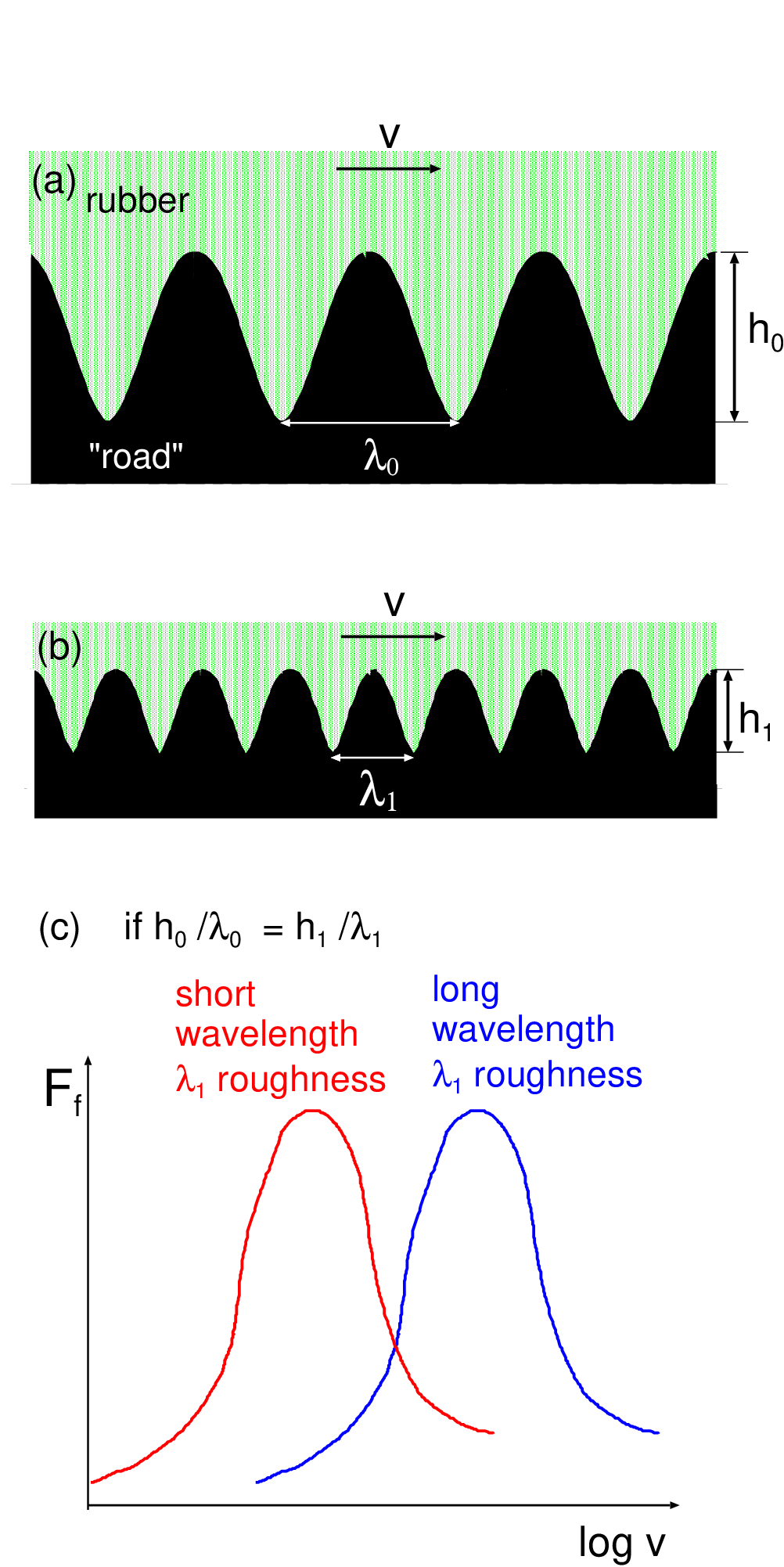}
\caption{
If rubber slides in complete contact with a sinus corrugated surface, the friction force $F_{\rm f}(v)$ as a function of the logarithm of the sliding speed appears the same in both cases, except for a shift along the velocity axis. On a ${\rm log}v$-scale this shift is proportional to $\log(\lambda_1/\lambda_0)$ and results from the different excitation frequencies $\omega = v/\lambda$ generated by the two corrugated profiles.
}
\label{1x.2h.eps}
\end{figure}

\vskip 0.1 cm
{\bf Viscoelastic contribution}

Assume first that adhesion can be neglected. The contribution from rubber viscoelasticity depends on the surface roughness across multiple length scales. This is illustrated in Fig. \ref{1x.2h.eps}, which shows the case of a rubber block sliding in contact with periodically corrugated substrates having (a) long-wavelength and (b) short-wavelength roughness.

If the ratio $h/\lambda$ between the amplitude $h$ and the wavelength $\lambda$ of the roughness is the same, and full contact occurs, then the friction force $F_{\rm f}(v)$ acting on the block as a function of the logarithm of the sliding speed will be the same in both cases, except for a shift along the frequency axis [see Fig. \ref{1x.2h.eps}(c)]. This shift results from the fact that the deformation frequency induced in the rubber is $\omega \approx v/\lambda$. To achieve the same viscoelastic energy dissipation in both cases, the velocities must be chosen such that the deformation frequency is equal, giving $v_1/v_0 = \lambda_1/\lambda_0$.

In reality, when sliding on a randomly rough surface, the rubber rarely makes full contact with the roughness profile at all length scales. The apparent contact area decreases as shorter-wavelength roughness components are taken into account. For surfaces with a Hurst exponent $H < 1$, the ratio $h/\lambda$ increases with decreasing wavelength, which would lead to larger contributions from the short-wavelength components if full contact occurred. 

In most practical applications, $H$ is close to 1 (typically $H \approx 0.8$), and in this case the viscoelastic contribution from each decade of 
roughness length scale is approximately equal. This raises a key question: down to what length scale should roughness 
be included when calculating the viscoelastic contribution to the friction force?

Analysis of experimental data indicates that, for tire tread rubber sliding on road surfaces, most of the viscoelastic contribution arises from interactions with asperities down to a cutoff length $\lambda_{\rm min}$ on the order of micrometers \cite{1,9,add}. However, the origin and precise value of this short-wavelength cutoff are not yet fully understood.

The viscoelastic contribution to the friction from asperities at a single length scale can be estimated as follows. When a rubber block slides in contact with an asperity, as illustrated in Fig. \ref{TwoFriction.eps}, the asperity induces a deformation in the rubber with a typical frequency $\omega \approx v/d$, where $d$ is the characteristic size of the contact region.

The deformation field extends into the rubber to a depth of order $d$, so the volume of the deformed region scales as $d^3$. Over a sliding distance $d$, the time-dependent stress can be approximated as $\sigma(t) = \sigma_1 \cos(\omega t)$ over a time interval $T = \pi d/v$ (half of one oscillation period). Letting $\dot{\epsilon} = d\epsilon/dt$ denote the strain rate, the energy dissipated during this process is
$$
\Delta E \approx d^3 \int_0^T dt \ \sigma(t) \dot{\epsilon}(t)
$$
Using complex notation for harmonic stress and strain, we have
$$
\Delta E \approx d^3 \sigma_1^2 {1\over 4} \int_0^T dt \ \left(e^{i\omega t} + e^{-i\omega t} \right)
(i\omega) \left( \frac{e^{i\omega t}}{E^*(\omega)} - \frac{e^{-i\omega t}}{E(\omega)} \right)
$$
where we have used $E(-\omega) = E^*(\omega)$. Performing the integral yields
$$
\Delta E \approx \frac{\pi}{2} d^3 \sigma_1^2 \, {\rm Im} \frac{1}{E(\omega)} 
$$

This energy equals the sliding distance $d$ multiplied by the friction force $F_{\rm f}$, so that
$$
F_{\rm f} \approx \frac{\pi}{2} d^2 \sigma_1^2 \, {\rm Im}  \frac{1}{E(\omega)}
$$

If there are $N$ macroasperity contact regions the friction force
$$
F_{\rm f} \approx N\frac{\pi}{2} d^2 \sigma_1^2 \, {\rm Im}  \frac{1}{E(\omega)}
$$
Using the relation $A \approx N d^2$ for the area of real contact, we can express the total normal force as $F_{\rm N} = A \sigma_1 =N d^2 \sigma_1$. The friction coefficient is then
$$
\mu = \frac{F_{\rm f}}{F_{\rm N}} \approx \frac{\pi}{2} \sigma_1 \, {\rm Im} \frac{1}{E(\omega)} 
$$

The normal force can also be written as $F_{\rm N} = \sigma_0 A_0$, where $A_0$ is the nominal contact area and $\sigma_0$ the nominal contact pressure,
which implies $\sigma_1 = \sigma_0 A_0 / A$. 
Using the known result (see Sec. 4) $A / A_0 \approx 2 \sigma_0 / (\xi |E(\omega)|)$, we obtain $\sigma_1 \approx \xi |E(\omega)| / 2$, so that
$$
\mu \approx \frac{\xi \pi}{4} |E(\omega)| \, {\rm Im}  \frac{1}{E(\omega)} 
= \frac{\xi \pi}{4} \frac{{\rm Im} E(\omega)}{|E(\omega)|} \eqno(13)
$$
where $\xi$ is the cumulative rms slope for the considered roughness.

In the transition region between the rubbery and glassy regimes, the ratio ${\rm Im} E(\omega) / |E(\omega)| \approx 1$. Therefore, the maximum viscoelastic contribution to the friction coefficient is expected to be of order $\xi$, which is typically of order 1.

\begin{figure}[tbp]
\includegraphics[width=0.49\textwidth,angle=0]{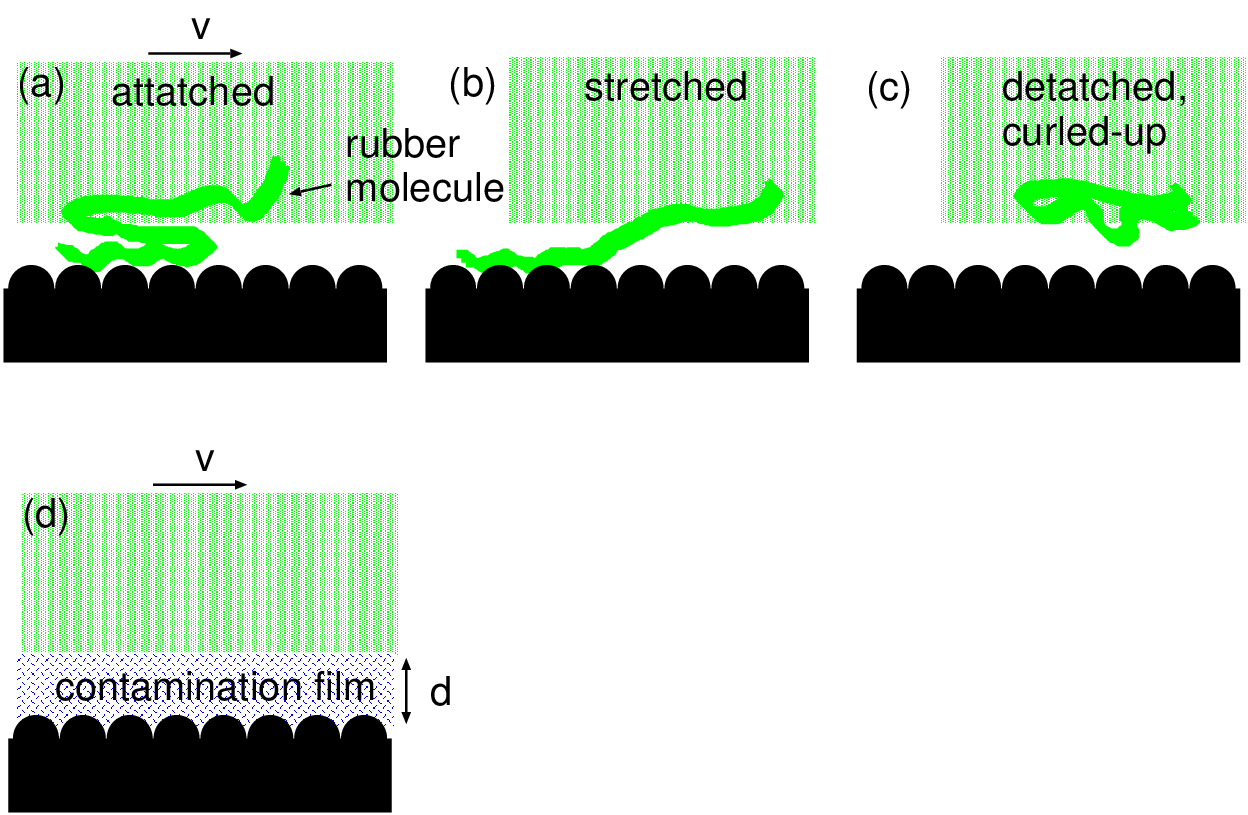}
\caption{
The frictional shear stress in the area of real contact [see Fig. \ref{TwoFriction.eps}(b)] can arise from (a)-(c) binding-stretching-detachment cycles of rubber molecules at the sliding interface, or (d) from shearing of thin contamination films located between the substrate and the rubber.
}
\label{ShearStress.eps}
\end{figure}

\begin{figure}[tbp]
\includegraphics[width=0.45\textwidth,angle=0]{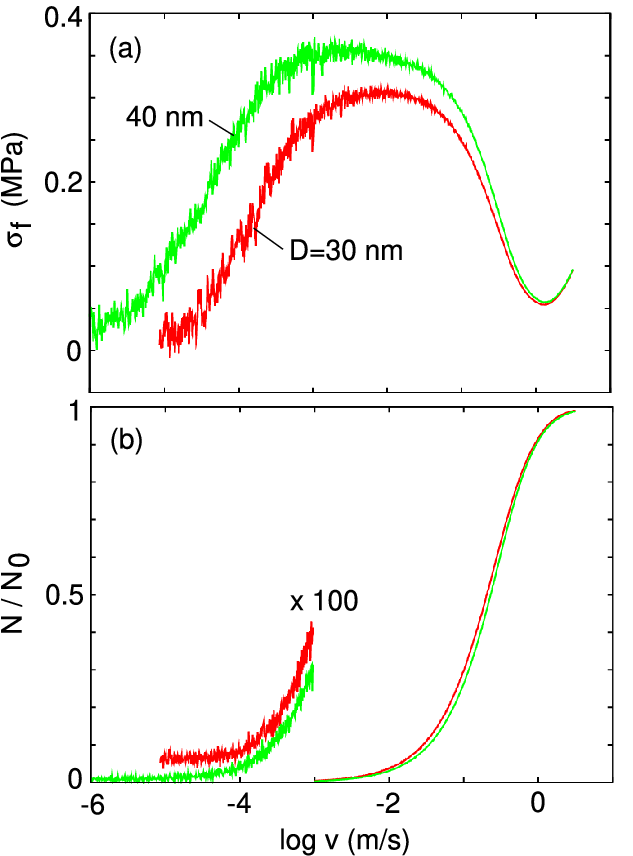}
\caption{
Calculated frictional shear stress (a) and the fraction $N/N_0$ of the rubber surface in the sliding state (b), shown as a function of the logarithm of the sliding speed. $D$ is the characteristic size of the interfacial units that undergo stick-slip-stick motion. Results are for styrene-butadiene (SB) rubber. Adapted from Ref. \cite{theory3}.
}
\label{vary.D=30.40nm.eps}
\end{figure}

\begin{figure}[tbp]
\includegraphics[width=0.45\textwidth,angle=0]{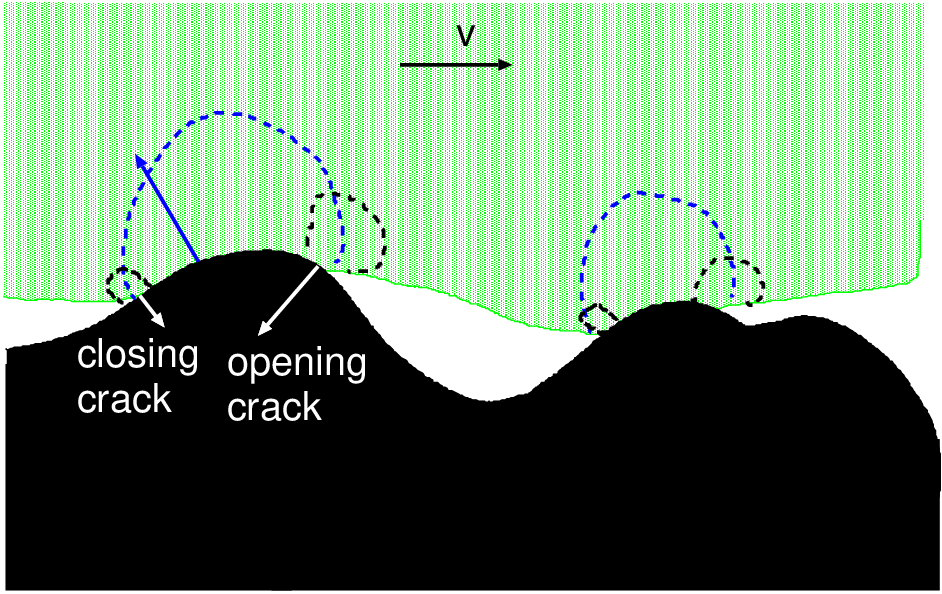}
\caption{
The adhesive interaction between rubber and substrate leads to the formation of opening and closing cracks at the edges of asperity contact regions during sliding. Due to adhesion hysteresis, more energy is dissipated at the opening crack than is recovered at the closing crack, resulting in a net contribution to the sliding friction force.
}
\label{Opening.eps}
\end{figure}

\begin{figure}[tbp]
\includegraphics[width=0.45\textwidth,angle=0]{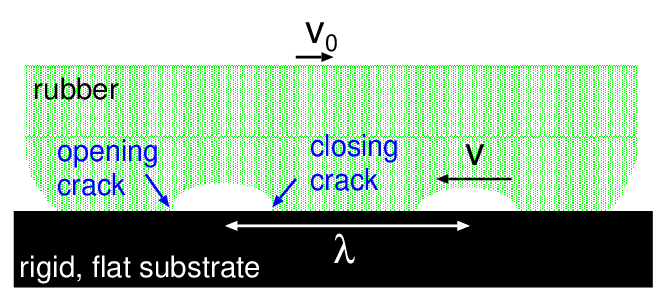}
\caption{
When a block of soft rubber slides on a flat, smooth substrate (with negligible roughness), Schallamach waves are observed above a critical sliding speed. The rubber adheres to the substrate, and no direct slip occurs at the interface. Instead, localized detachment zones form at the leading edge and propagate opposite to the sliding direction.
}
\label{SchallamachWAVES.eps}
\end{figure}

\vskip 0.1 cm
{\bf Area of real contact contribution}

We first consider dry and clean surfaces. In sliding contact, rubber molecules and substrate atoms interact as illustrated in Fig. \ref{ShearStress.eps}(a)-(c). In many cases, the interfacial interactions are weak, such as van der Waals forces. During stationary contact, the rubber chains at the interface adjust to the substrate potential in order to minimize the free energy. This bond formation may involve overcoming energy barriers and therefore does not occur instantaneously, but requires a finite relaxation time.

During sliding at low velocities, thermal fluctuations assist in formation and breaking of the rubber-substrate bonds, 
resulting in a friction force that approaches zero as the sliding velocity tends to zero. 
At high velocities, there is not enough time for the rubber molecules to adjust to the substrate potential. As a result, for high sliding speeds 
the rubber interface effectively  
``floats'' above the substrate in an incommensurate-like state relative to the corrugated potential. In this regime, the frictional shear stress is also small.

Therefore, the frictional shear stress as a function of sliding speed must exhibit a maximum at some intermediate velocity $v^*$. This friction mechanism was first studied in a simplified model by Schallamach \cite{1}, and later extended by Chernyak and Leonov \cite{Chernyak}, and in a more realistic model by Persson and Volokitin \cite{theory3}. Because of its conceptual importance, we present in Appendix~A a derivation of the Schallamach result using an approach that differs slightly from the one originally used by Schallamach.

The actual elementary stick-slip units are not likely to be individual molecular segments, but rather small patches of molecular segments, which we refer too
as stress domains, with a characteristic size $D$. In Fig. \ref{vary.D=30.40nm.eps}, we show the results of calculations 
(using the model of Persson and Volokitin) for $D = 30$ and $40 \ {\rm nm}$. Fig. \ref{vary.D=30.40nm.eps}(a) shows the frictional shear stress, and Fig. \ref{vary.D=30.40nm.eps}(b) shows the fraction $N/N_0$ of the rubber surface in the sliding state, both as functions of the logarithm of the sliding speed \cite{theory3}. 

In the simple Schallamach model (see Appendix~A), the frictional shear stress is given by $\sigma_{\rm f} = \sigma^* f(v/v^*)$, where $\sigma^*$ is the maximum shear stress (with $D = 2R$):
$$\sigma^* \approx \left ( {2E k_{\rm B} T \over R^3} \right )^{1/2}\eqno(14)$$
and the reference velocity:
$$v^* =  \left ({ 2k_{\rm B} T \over E R \tau_0^2}\right )^{1/2}\eqno(15)$$ 

Here, $k_{\rm B}$ is the Boltzmann constant, $T$ the absolute temperature, $E$ the elastic modulus, and $\tau_0$ the average time a 
stress domain spends in the detached state. 

The function $f(v/v^*)$ depends on $\Delta E = \epsilon_1 - \epsilon_0$, where $\epsilon_1$ is the energetic barrier to break the 
rubber-substrate bond in a stress domain, and $\epsilon_0$ the energetic barrier for forming the bond. For $\Delta E / k_{\rm B} T \approx 3$,
$f(v/v^*)$ is approximately a Gaussian function of ${\rm log}_{10} v$, with the full width at half maximum of about 2 velocity decades.

The values for $\sigma^*$ and $v^*$ are consistent with experimental observations (see Appendix~A). Experiments show that the full width at half maximum of $\sigma_{\rm f}(v)$ is about 4 velocity decades, 
which is larger than predicted by the Schallamach model, 
but similar to the prediction by the Persson-Volokitin theory (see Fig.~\ref{vary.D=30.40nm.eps}).

Another contribution to the friction force from the area of real contact, which can be significant in particular for smooth surfaces and soft rubber compounds, is illustrated in Fig. \ref{Opening.eps}. During sliding, adhesive interactions between the rubber and the substrate give rise to opening and closing cracks at the edges of asperity contact regions. Due to adhesion hysteresis, more energy is dissipated at the opening crack than is recovered at the closing crack. This asymmetry results in a contribution to the sliding friction force.

The energy dissipation at the crack tip may spatially overlap with bulk viscoelastic dissipation (see Fig. \ref{Opening.eps}), in which case the total energy loss may not equal the sum of the two mechanisms calculated independently \cite{Carb,AddCarb1,AddCarb2,AddCarb3,AddCarb4}.

In classical theories of interfacial crack propagation, it is typically assumed that the 
stress field extends far from the crack tip into the contact region and where it remains in a relaxed state. 
However, this assumption does not hold in situations where an indenter, 
such as a rigid sphere, is rapidly brought into and lifted out of contact with a viscoelastic solid. 
A similar condition applies to most sliding friction scenarios, where the rubber at a given surface 
point is initially nearly undeformed but undergoes rapid deformation as it enters 
and exits contact with substrate asperities, particularly at high sliding speeds. 
In such cases, the stress remains confined to the immediate vicinity of the contact region and 
does not penetrate deeply into the bulk. 
The influence of time-dependent deformation on interfacial crack propagation has been discussed 
qualitatively in Ref. \cite{Tribo} and quantitatively in Ref. \cite{Muser} and \cite{NewCarb}.

Opening and closing crack propagation govern the friction behavior when soft rubber slides on a substrate, where both the rubber and the substrate surface are very smooth. This is illustrated in Fig. \ref{SchallamachWAVES.eps}, where a smooth rubber block slides on a flat substrate with negligible roughness. In this case, the rubber adheres to the substrate, and no direct slip occurs at the interface. Instead, localized detachment zones form at the leading edge and propagate opposite to the sliding direction. These are known as Schallamach waves.

Schallamach wave motion can be understood by analogy with the motion of a rug on a carpet: one can either pull the entire carpet or create a local deformation and move it forward. Due to adhesion hysteresis, more energy is required to break the bonds at the opening crack than is recovered at the closing crack. This results in a friction force that can be estimated as follows.

Assume a rectangular rubber block of length $L_x$ in the sliding direction and width $L_y$ in the transverse direction. Let the Schallamach waves move with speed $v$, and assume that their spacing is $\lambda$. If $v_0$ is the sliding speed of the block, then the frictional work over a time interval $\Delta t$ is $F_{\rm f} v_0 \Delta t$. This must equal the energy dissipated by peeling during that time.

There are $N = L_x / \lambda$ waves, each of length $L_y$, and the energy dissipated is $N L_y \gamma_{\rm eff} v \Delta t$, where $\gamma_{\rm eff} = \gamma_{\rm open} - \gamma_{\rm close} \approx \gamma_{\rm open}$ is the effective adhesion energy. Equating this expression for
the energy dissipation with $F_{\rm f} v_0 \Delta t$ gives
$$
F_{\rm f} = A_0 \gamma_{\rm eff} \frac{v}{v_0 \lambda} \eqno(16)
$$
where $A_0 = L_x L_y$ is the surface area. Roberts and Thomas \cite{RobT} have shown that (16) is in good agreement with experimental results.

In many practical cases, surfaces are not perfectly clean, and thin, liquid-like contamination films may exist within the area of real contact. In such cases, there is an additional contribution to the friction force from shearing of the contamination film. These confined films generally exhibit velocity-dependent viscosity, so the resulting shear stress is not linearly proportional to the strain rate.

\vskip 0.3 cm
{\bf 5.1 Analytical theory of rubber friction on rough surfaces}

As mentioned above, there are two main contributions to the rubber friction force $F_{\rm f}$ on rough surfaces: one arising from the area of real contact, and the other from viscoelastic deformations induced by asperities.

We express the total friction force as
$$
F_{\rm f} = F_{\rm visc} + F_{\rm cont}
$$

If the normal force is given by $F_{\rm N} = A_0 \sigma_0$, where $A_0$ is the nominal contact area and $\sigma_0$ the nominal contact pressure, the friction coefficient becomes
$$
\mu = \mu_{\rm visc} + \frac{\sigma_{\rm f} A}{\sigma_0 A_0} 
$$
where $A$ is the area of real contact and $\sigma_{\rm f}$ is the frictional shear stress acting within the contact area.

The Persson contact mechanics theory predicts the viscoelastic contribution $\mu_{\rm visc}$ and the area of real contact $A$. However, the dependence of the frictional shear stress $\sigma_{\rm f}$ on sliding velocity and temperature must be determined using other theoretical approaches or extracted from experimental data.

\vskip 0.1cm
{\bf Viscoelastic contribution}

We first consider sliding at constant speed and neglect frictional heating, assuming the temperature is uniform and equal to $T = T_0$. In Persson theory, the viscoelastic friction force acting on a rubber block pressed with nominal stress $\sigma_0$ against a hard, randomly rough surface is given by \cite{A1,Scagg}
$$
\mu_{\rm visc} \approx \frac{1}{2} \int_{q_0}^{q_1} dq \, q^3 C(q) S(q) P(q)
$$
$$
\times \int_0^{2\pi} d\phi \, \cos\phi \, 
{\rm Im}\frac{E(qv(t) \cos \phi, T_0)}{(1 - \nu^2)\sigma_0} \eqno(17)
$$

Here, $P(q)=A(q)/A_0$ is the normalized contact area observed at magnification $\zeta = q/q_0$:
$$
P(q) = \frac{1}{\surd \pi} \int_0^{\surd G(q)} dx \, e^{-x^2 / 4}
= {\rm erf} \left( \frac{1}{2 \surd G(q)} \right) \eqno(18)
$$

The function $G(q)$ is defined as
$$
G(q) = \frac{1}{8} \int_{q_0}^q dq \, q^3 C(q)
\int_0^{2\pi} d\phi \, 
\left| \frac{E(qv \cos \phi, T_0)}{(1 - \nu^2) \sigma_0} \right|^2 \eqno(19)
$$

The correction factor $S(q)$ accounts for the incomplete contact and is given by
$$
S(q) = \gamma + (1 - \gamma) P^2(q) \eqno(20)
$$
with $\gamma \approx 1/2$. Note that $S\to 1$ as $P \to 1$, which recovers the exact result for full contact. In fact, (17) becomes exact in the limit of complete contact.

The general theory also includes temperature effects, both the flash temperature and the gradual heating of 
the sliding rubber block with increasing
sliding speed \cite{VG}. The theory can also be applied to layered materials, e.g., for a rubber film on another solid. 
For a layered material consisting of two materials ${\bf 0}$ and ${\bf 1}$, with {\bf 0} forming a film of 
thickness $d$ on top of a semi-infinite {\bf 1},
one needs to replace $E^*(\omega)$ in (17)-(19) with $E^*_0(\omega)/K(\omega)$, where
$$K = \frac{1 + 4mqde^{-2qd} - mne^{-4qd}}{1 - (m + n + 4mq^2d^2)e^{-2qd} + mne^{-4qd}}$$
and
$$m = \frac{\mu_0/\mu_1 - 1}{\mu_0/\mu_1 + 3 - 4\nu_0}, \ \ \ \ \ \ n = 1 - \frac{4(1 - \nu_0)}{1 + (\mu_0/\mu_1)(3 - 4\nu_1)}$$

The shear modulus $\mu_0 = E_0/2(1 + \nu_0)$, and similarly for $\mu_1$.
Note that $K$ is dimensionless and depends only on $qd$, $\nu_0$, $\nu_1$, and $E_0/E_1$.
Note also that $K = 1$ for $d = \infty$, and $K = E_0^*/E_1^*$ for $d = 0$.

\begin{figure}[tbp]
\includegraphics[width=0.45\textwidth,angle=0]{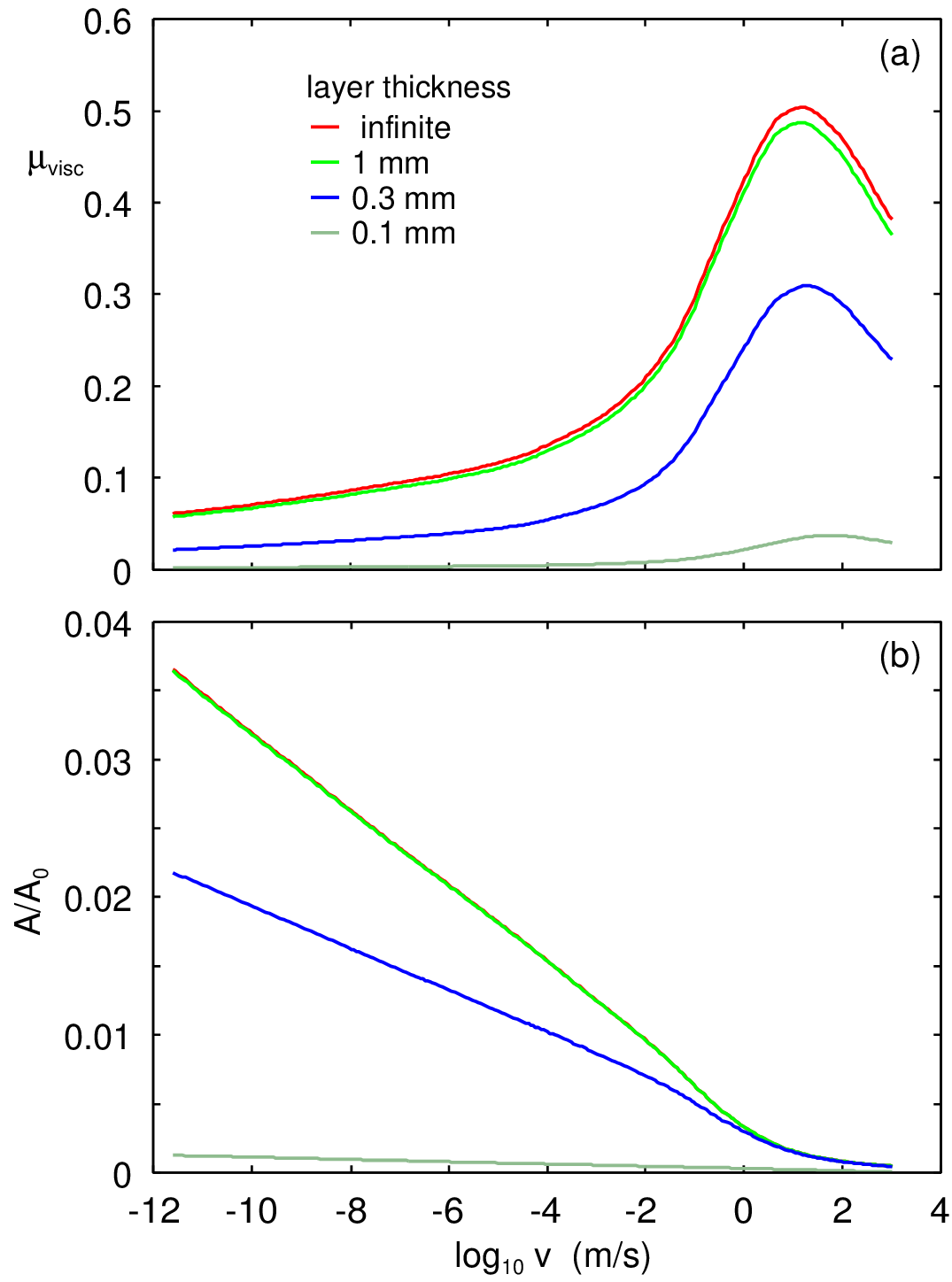}
\caption{
The viscoelastic contribution $\mu_{\rm visc}$ to the sliding friction coefficient,
and the relative contact area $A/A_0$, as functions of the sliding speed
for a flat rigid solid covered with a tread rubber film of thickness $d$ sliding on a concrete surface. 
The concrete surface
is relatively smooth, with an rms roughness of $0.05 \ {\rm mm}$, and the calculation includes all roughness components
down to a cut-off wavenumber of $\approx 10^6 \ {\rm m^{-1}}$, where the rms slope is $\approx 1$. 
Frictional heating is not included in the calculation. The nominal contact pressure is $\sigma_0 = 0.2 \ {\rm MPa}$.
}
\label{1logv.2mu.layered.eps}
\end{figure}

As an example, consider a thin rubber film (thickness $d$) on a rigid substrate sliding on a concrete surface. 
Fig.~\ref{1logv.2mu.layered.eps} shows the viscoelastic contribution $\mu_{\rm visc}$ 
to the sliding friction coefficient, and the relative contact 
area $A/A_0$ as a function of the sliding speed for the film thicknesses $d = 0.1$, $0.3$, and $1 \ {\rm mm}$, and for
$d = \infty$. The concrete surface is relatively smooth, with an rms roughness $h_{\rm rms} \approx 0.05 \ {\rm mm}$, 
which is considerably smaller than all the film thicknesses used. However, $h_{\rm rms}$ does not determine the
depth of the deformation field into the rubber; this is determined by the width of the asperity contact regions.
The surface used has a small-wavenumber cut-off $q_0 \approx 615 \ {\rm m^{-1}}$, corresponding to a wavelength
$\lambda_0 = 2\pi/q_0 \approx 10 \ {\rm mm}$, and the width of the largest asperity contact regions will be some fraction
of this. This explains why the friction drops already for $d < 1 \ {\rm mm}$.

\vskip 0.1cm
{\bf Adhesive contribution}

We now consider the contribution to the friction force arising from the area of real contact $A$. In sliding contact, rubber molecules interact with substrate atoms as illustrated in Fig. \ref{TwoFriction.eps}(a)-(c).

For this case, theory \cite{theory3} (see also Appendix A) 
predicts that the frictional shear stress exhibits a Gaussian-like dependence on the logarithm of the sliding speed. The curve has a full width at half maximum of $\sim 4$ (or more) frequency decades, and, at room temperature, is centered around a characteristic sliding speed typically on the order of $v^* \approx 1 \ {\rm cm/s}$.

In Ref. \cite{Nam}, we found that using the following empirical shear stress law $\sigma_{\rm f}(v, T)$, which is similar to the theoretical prediction \cite{theory3}, resulted in good agreement between theoretical results and experimental measurements:
$$
\sigma_{\rm f}(v) \approx \sigma_{\rm f0} \, \exp\left( -c \left[ \log_{10} \left( \frac{v}{v^*} \right) \right]^2 \right) \eqno(21)
$$
where for passenger car tread rubber,
typically $c \approx 0.17$, $\sigma_{\rm f0} \approx 4$-$8 \ {\rm MPa}$, and the reference sliding speed is $v^* \approx 1 \ {\rm cm/s}$ at $T = 20^\circ {\rm C}$.

The full width at half maximum of $\sigma_{\rm f}(v)$ as a function of $\log_{10} v$ is given by $\Gamma=2({\rm ln}2/c)^{1/2} \approx 4$.

The master curve described by (21) is valid at the reference temperature $T_{\rm ref} = 20^\circ {\rm C}$. The frictional shear stress at other temperatures can be obtained by replacing $v$ with $v a'_T$, where $a'_T$ is a shift factor derived during the construction of the friction 
master curve. This shift factor is often close to the bulk viscoelastic shift factor $a_T$, but in some cases we have found that an Arrhenius-type expression provides a more accurate description:

$$
\ln a'_T \approx \frac{\epsilon}{k_{\rm B}} \left( \frac{1}{T} - \frac{1}{T_{\rm ref}} \right) \eqno(22)
$$
where $\epsilon \approx 1 \ {\rm eV}$ is an activation energy, and $k_{\rm B}$ is the Boltzmann constant.
This temperature dependence is slightly weaker than that predicted by the bulk viscoelastic shift factor and may reflect a higher mobility of the polymer chains at the rubber surface compared with the bulk.

The contribution of the area of real contact to rubber friction depends sensitively on the presence of contamination particles and fluids. On a wet road surface, at sufficiently high sliding or rolling speeds, the surfaces in the apparent contact regions can become separated by a thin fluid film. In such cases, viscoelastic deformation of the rubber becomes the dominant contribution to friction.

\vskip 0.1 cm
{\bf Role of frictional heating}

Temperature has a strong influence on rubber sliding friction, especially when the sliding speed varies with time (see Sec. 5.3). 
The temperature distribution within a rubber block sliding on a rough surface is highly inhomogeneous.

This distribution can be decomposed into two components: the {\it background} temperature $T_0({\bf x}, t)$, which varies slowly in space and time, and the {\it flash} temperature $\Delta T({\bf x}, t)$, which is nonzero only in the vicinity of asperity contact regions, where frictional energy dissipation occurs.

The background temperature $T_0({\bf x}, t)$ depends on the {\it sliding history} and on external conditions. For example, in the case of a tire, $T_0({\bf x}, t)$ is dependent on the ambient air and road temperatures, whether the road is dry or wet, the rolling resistance of the tire, and past driving conditions such as braking, cornering, and speed.

In contrast, the flash temperature $\Delta T$ is largely independent of external conditions and sliding history. It arises from the energy dissipation in the asperity contact regions.

At low sliding speeds $v$ (typically below $1$-$10 \ {\rm mm/s}$), thermal diffusion significantly reduces the flash temperature. 
However, at higher sliding speeds, the temperature increase $\Delta T$ is large, and will shifts both the viscoelastic and adhesive friction contributions toward higher sliding speeds.

As with the friction force, both the frictional shear stress acting in the area of real contact (adhesive contribution) and the viscoelastic deformations contribute to frictional heating. We denote the corresponding friction coefficients as $\mu_{\rm cont}$ and $\mu_{\rm visc}$, respectively. The adhesive contribution results in energy dissipation within a nanometer-thick surface layer.

\begin{figure}[tbp]
\includegraphics[width=0.35\textwidth,angle=0]{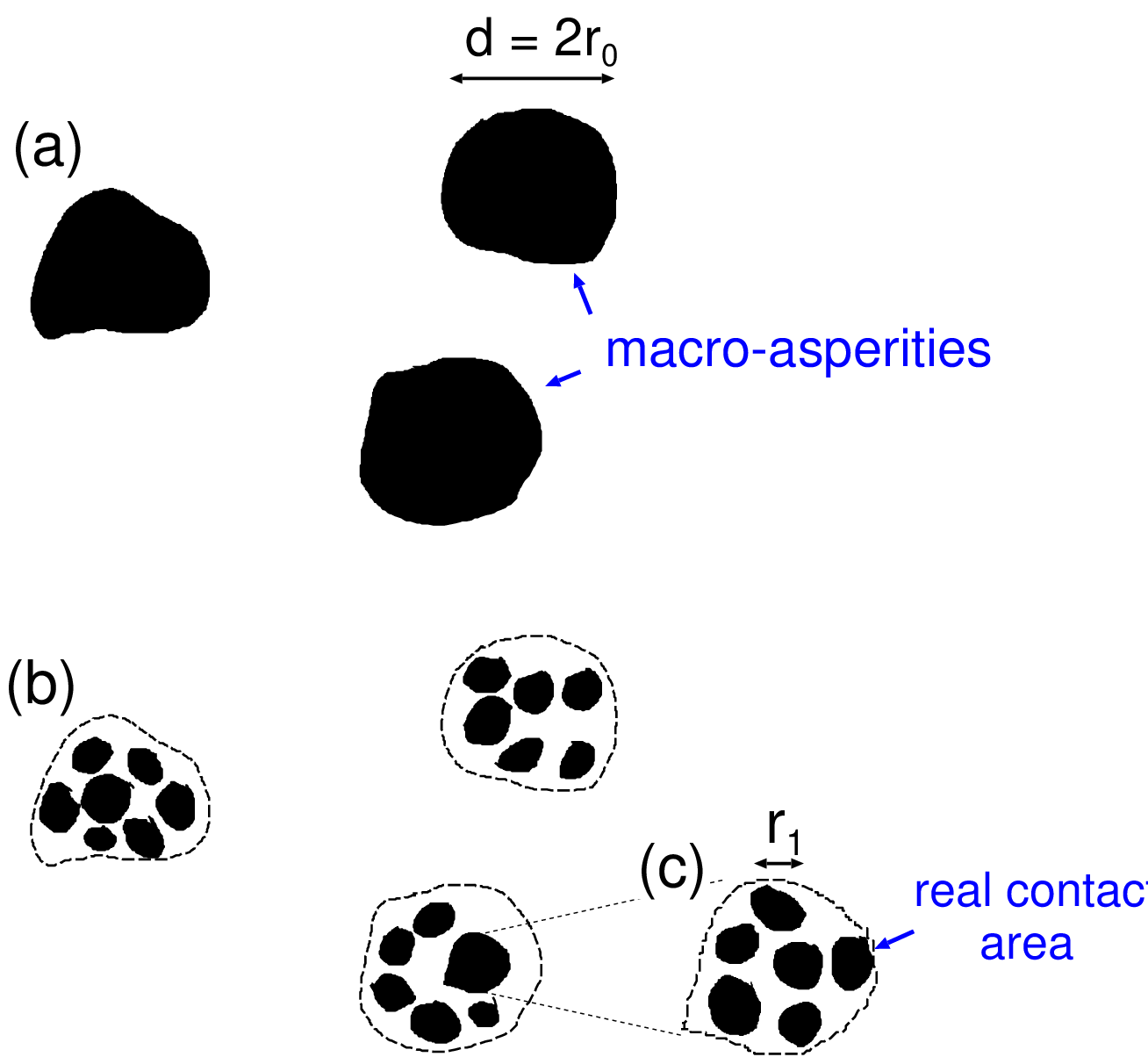}
\caption{
Multiscale nature of contact between two solids as a function of magnification.  
(a) At low magnification, contact appears to occur uniformly across macroasperities.  
(b) At intermediate magnification, smaller non-contact regions are revealed within the nominal contact patches.  
(c) At high magnification, even the apparent contact areas are found to consist of discrete microcontacts separated by non-contact gaps, illustrating the true real contact area.
}
\label{MacroAndReal.eps}
\end{figure}


To estimate the flash temperature, one must determine the size of the contact regions, which depends on the magnification. At low magnification, only the contact with the largest road asperities is visible, as illustrated in Fig.~\ref{MacroAndReal.eps}(a). These are referred to as the macroasperity contact regions. As the magnification increases, smaller asperities located on top of the larger ones become visible, and contact is observed in these regions as well [Fig.~\ref{MacroAndReal.eps}(b)]. The true area of real contact is revealed only at the highest magnification, as shown in Fig.~\ref{MacroAndReal.eps}(c). 

In Ref.~\cite{10}, a theory was developed that predicts the size of contact regions as a function of magnification. However, unless the sliding speed is very high, a simpler approach can be used. In this approach, the heat generated in the contact regions within the macroasperity contacts is assumed to spread laterally, and the resulting energy dissipation is expressed as a function of depth into the rubber. Using this heat distribution as a source term, the temperature increase in the rubber at the asperity contacts can be calculated by solving the heat diffusion equation, which includes heat exchange with the road surface. This method was introduced in Ref.~\cite{VG}, and here we only provide a qualitative overview. 

In the following analysis, we assume that the macroasperities have a diameter of $d \approx 1 \ {\rm mm}$, which is typical for rubber in contact with rough road surfaces.

Consider first the adhesive contribution. The frictional energy dissipated per unit area and unit time in the asperity contact 
regions is given by $q = \mu_{\rm cont} \sigma v$, where $\sigma = F_{\rm N}/A$ is the normal contact stress. 
The asperity contact time is $\tau = d/v$, and the energy dissipated per unit volume is 
$Q = q \tau /h = q d / h v = \mu_{\rm cont} \sigma d / h$, 
where $h \approx 1 \ {\rm nm}$ is the thickness of the region at the surface
where the adhesive frictional energy is deposited. 

At very high sliding speeds, thermal diffusion is negligible, and the temperature increase 
in the thin surface layer of thickness $h$ is given by
$$\Delta T = {Q \over \rho C_V} = {q d \over h v \rho C_V} = {q D  d \over \kappa_{\rm th} v h} \eqno(23) $$
where $D = \kappa_{\rm th} / \rho C_V$ is the thermal diffusivity, and 
$\rho$, $C_V$, and $\kappa_{\rm th}$ are the rubber mass density, heat capacity, and thermal conductivity, respectively.
For rubber with carbon fillers, typical values are $\rho \approx 1500 \ {\rm kg/m^3}$, $C_V \approx 1500 \ {\rm J/(kg\,K)}$, and
$\kappa \approx 0.3 \ {\rm W/(m\,K)}$, giving a thermal diffusivity of $D \approx 10^{-7} \ {\rm m^2/s}$.

For this equation to be valid, the heat must spread only over a distance of order $h$ during the contact time $\tau = d/v$.
If the phonon mean free path is of the order of $h$ or smaller, then the thickness of the affected layer can be estimated from the heat diffusion equation and is given by $\approx D/v$. Introducing the dimensionless parameter $J_h = v h / D$, the condition becomes $J_h > 1$, and (23) can be written as 
$$\Delta T = {q d \over \kappa_{\rm th} J_h} \eqno(24)$$

In the opposite limit of very small sliding speeds, the flash temperature corresponds to that of a stationary heat source.
In this case, for a circular contact region with diameter $d$, 
the maximum temperature increase, which occurs at the surface in the center of the contact region, is given by \cite{flash1,flash2,flash3}:
$$\Delta T = {q d \over \kappa_{\rm th}} \eqno(25)$$
This result is valid only when the sliding speed is sufficiently low such that the heat diffusion length, approximately $D/v$, is larger than the lateral size of the contact region, $d$. Introducing the dimensionless parameter $J_d = v d / D$, the condition becomes $J_d < 1$.

An interpolation formula that is approximately valid across all sliding speeds was presented by Greenwood in Ref.~\cite{Greenwood}, and is extended here to include a $J_h^2$ term as follows:
$$\Delta T \approx { q d \over \kappa_{\rm th}} \left ({1 \over 1 + 0.39 J_d +J_h^2 } \right )^{1/2} \eqno(26)$$
Note that $J_d = J_h^2$ when $v = Dd/h^2$, which, for the adhesive contribution where $h \approx 1 \ {\rm nm}$, corresponds to an extremely high velocity in most cases.
For example, if $d = 1 \ {\rm mm}$, then $v \approx 10^8 \ {\rm m/s}$. Hence, for the adhesive contribution, the $J_h^2$ term can always be neglected. 

In contrast, for the viscoelastic contribution, energy is dissipated in a volume element extending a distance $h \approx d$ into the rubber, so that $v = Dd/h^2 \approx D/d \approx 0.1 \ {\rm mm/s}$ for $d = 1 \ {\rm mm}$. 
Thus, for the viscoelastic case, the $J_h^2$ term cannot be neglected. However, in this case a more detailed analysis (as presented in Ref.~\cite{VG}) is required, since viscoelastic energy dissipation decays rapidly with depth into the rubber.

Since $J_d = 1$ corresponds to the sliding speed $v = D/d \approx 10^{-4} \ {\rm m/s}$ (using $d = 1 \ {\rm mm}$),
and since for this velocity $\Delta T \approx q d / \kappa_{\rm th} = \mu \sigma v d / \kappa_{\rm th} \approx 0.1$-$1 \ {\rm K}$ for $\sigma \approx 1$-$10 \ {\rm MPa}$ (see below),
we conclude that frictional heating is negligible at sliding speeds below $v \approx 10^{-4} \ {\rm m/s}$. 

For higher sliding speeds, the maximum temperature increase due to adhesive energy dissipation is approximately given by:
$$\Delta T \approx {q d \over \kappa_{\rm th}} (0.39 J_d)^{-1/2} \approx 1.6 {\mu_{\rm cont} \sigma \over \kappa_{\rm th}} (vDd)^{1/2} \eqno(27)$$

\begin{figure}[tbp]
\includegraphics[width=0.45\textwidth,angle=0]{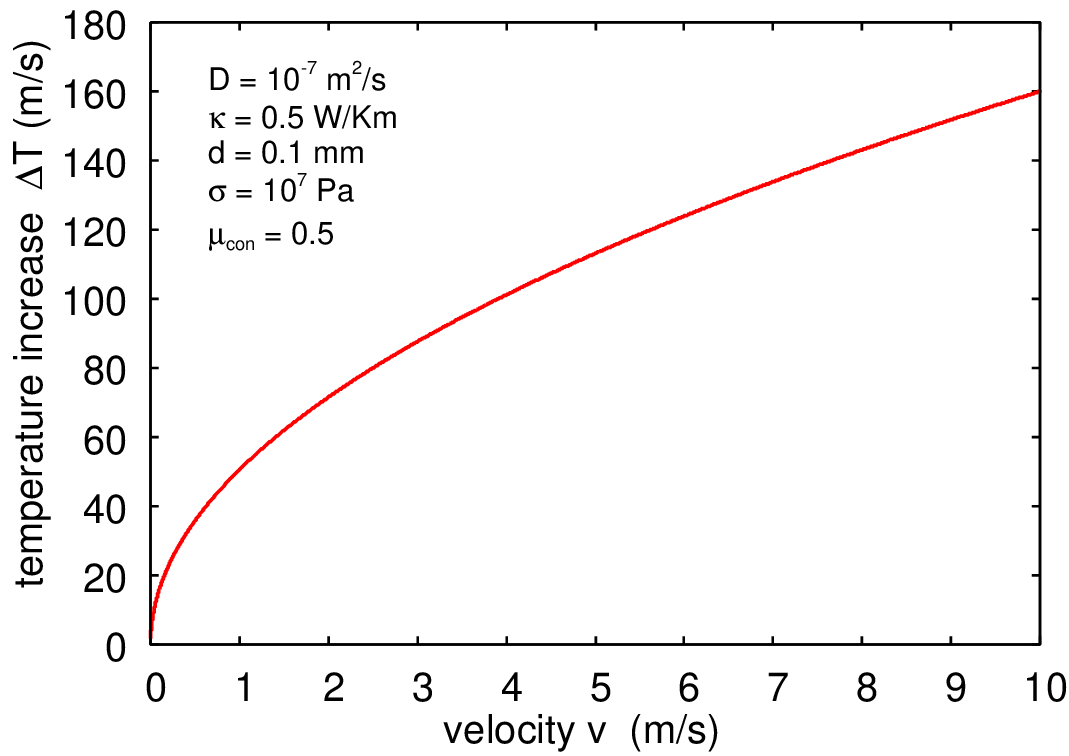}
\caption{
Flash temperature as a function of sliding speed, estimated using (27) with typical parameter values:
$D=10^{-7} \ {\rm m^2/s}$, $\kappa_{\rm th} = 0.5 \ {\rm W/(K \cdot m)}$, $d=0.1 \ {\rm mm}$, $\sigma = 10^7 \ {\rm Pa}$, and
$\mu_{\rm cont} = 0.5$.
}
\label{1v.2Temp.adhesive.eps}
\end{figure}

To estimate the temperature increase, we use the contact mechanics result [see (11)]  
$\sigma \approx F_{\rm N}/A \approx \xi |E(\omega)|/2$, where $\omega \approx v/d$. For typical sliding speeds  
(including strain softening) $|E(\omega)| \approx 2\times 10^7 \ {\rm Pa}$ and using
the typical rms-slope of $\xi \approx 1$ gives $\sigma \approx 10 \ {\rm MPa}$.
Using $\sigma \approx 10 \ {\rm MPa}$ in Fig.~\ref{1v.2Temp.adhesive.eps} shows the flash temperature predicted from (27), assuming typical thermal parameters. 
The calculation neglects heat exchange with the substrate and the contribution from viscoelastic energy dissipation. 
It also assumes that the contact area and the friction coefficient remain constant, although in reality both depend on the sliding speed and temperature. 
The predicted temperature rise can be significant, e.g., $\approx 40^\circ{\rm C}$ at $v \approx 1 \ {\rm m/s}$, 
which would shift the viscoelastic mastercurve toward higher frequencies (and thus higher velocities). This is an important effect in rubber sliding friction
and is included in the full theory\cite{VG}.

\begin{figure}[tbp]
\includegraphics[width=0.45\textwidth,angle=0]{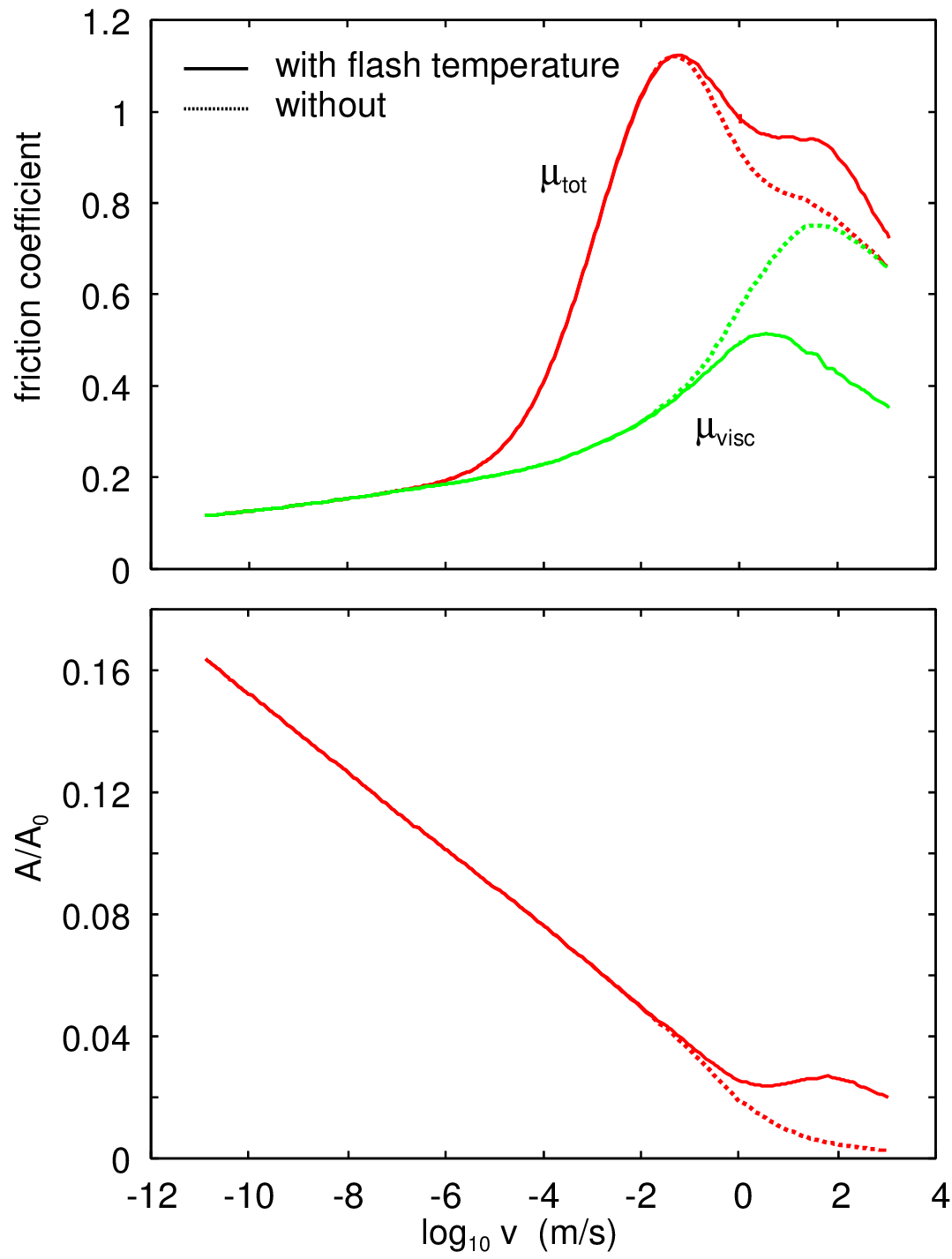}
\caption{
The influence of the flash temperature on (a) the sliding friction coefficient $\mu$ and
(b) the relative contact area $A/A_0$. The green and red lines in (a) represent the viscoelastic contribution
to the friction, $\mu_{\rm visc}$, and the total friction coefficient, $\mu = \mu_{\rm visc} + \mu_{\rm cont}$.
The solid lines correspond to calculations with the flash temperature included, and the dashed lines to those without it.
The results are for a tire tread compound on a concrete surface at a background temperature of $30^\circ {\rm C}$ and a
nominal contact pressure of $0.3 \ {\rm MPa}$.
}
\label{1logv.2mu.Area.compoundB.concrete.eps}
\end{figure}

Fig.~\ref{1logv.2mu.Area.compoundB.concrete.eps}  
shows the influence of the flash temperature on the sliding friction coefficient and the area of real contact.  
The green and red lines in (a) represent the calculated viscoelastic contribution  
to the friction, $\mu_{\rm visc}$, and the total friction coefficient, $\mu = \mu_{\rm visc} + \mu_{\rm cont}$.  
The solid lines correspond to calculations with the flash temperature included, and the dashed lines to those without it.  
Fig.~\ref{1logv.2mu.Area.compoundB.concrete.eps}(b) shows the relative contact area $A/A_0$, with and without the flash temperature.  
Note that the flash temperature makes the rubber elastically softer, which results in an increase in the contact area with  
increasing sliding speed for $v > 1 \ {\rm m/s}$.


\begin{figure}[tbp]
\includegraphics[width=0.45\textwidth,angle=0]{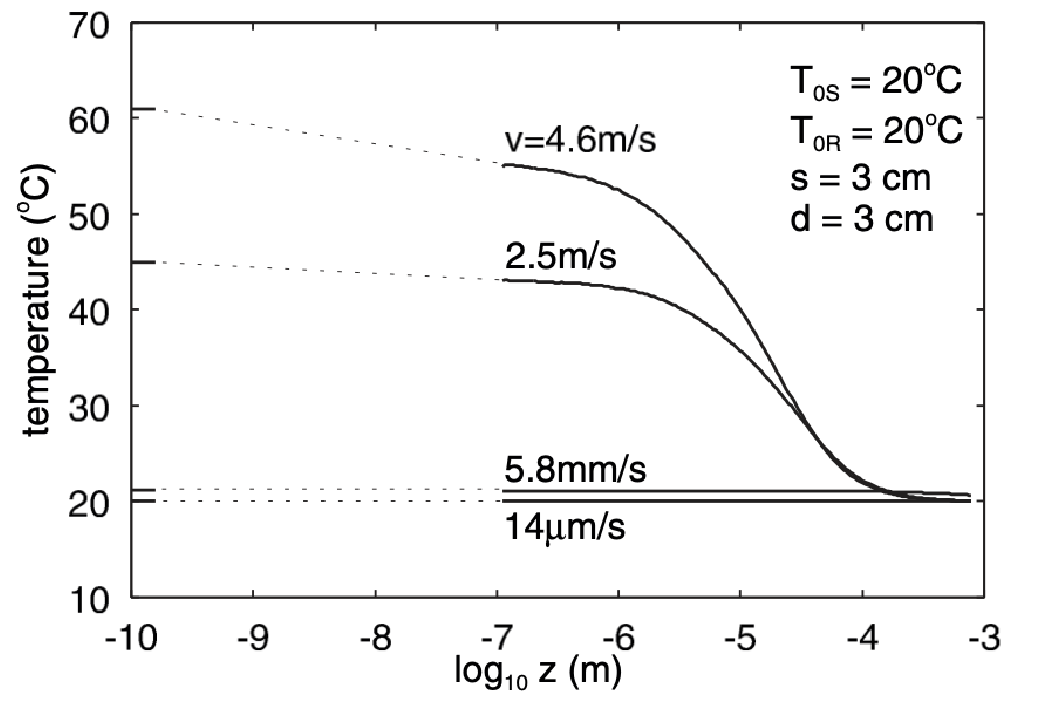}
\caption{
Temperature as a function of the distance into the rubber within a macroasperity contact region, for different sliding speeds. Results are for a rubber block sliding on an asphalt road surface. Adapted from Ref. \cite{VG}.
}
\label{FrictionalHeatingInto.eps}
\end{figure}

The equations for the sliding friction force and frictional heating, incorporating both viscoelastic and adhesive energy dissipation, were presented in Ref.~\cite{VG}. The theory in Ref.~\cite{VG} predicts the non-uniform temperature distribution and the corresponding friction force during accelerated motion. As an example, Fig.~\ref{FrictionalHeatingInto.eps} shows the temperature profile as a function of depth into the rubber within a macroasperity contact region, for several sliding speeds. The results correspond to a total sliding distance of $3 \ {\rm cm}$ on an asphalt road surface.

In the simulation, the rubber block has a length $L = 5 \ {\rm cm}$ in the sliding direction, and the macroasperity contact region is located $3 \ {\rm cm}$ from the leading edge of the block. The initial temperatures of both the rubber and the road are $20^\circ {\rm C}$. Note that for the two higher sliding speeds ($v = 4.6$ and $2.5 \ {\rm m/s}$), the maximum temperature occurs at the rubber surface, while for the lower sliding speeds, the temperature remains nearly constant throughout the depth.

\vskip 0.1 cm
{\bf Open problems}

Analysis of experimental data indicates that, for tire tread rubber sliding on road surfaces, most of the viscoelastic contribution to friction arises from interactions with asperities down to a cut-off length $\lambda_{\rm min} = 2\pi/q_1$, which is typically on the order of micrometers \cite{1,9,add}. However, the origin and precise value of this short-wavelength cut-off remain poorly understood.

Understanding of this viscoelastic cut-off length is essential for any theory of rubber friction, whether analytical or numerical. This requirement is not limited to Persson’s contact mechanics theory \cite{A1}, but also applies to numerical methods such as the Finite Element Method (FEM) \cite{3,4} or Boundary Element (BE) methods\cite{Scagg,BE}.

The physical origin of $\lambda_{\rm min}$ depends on the nature of the surface roughness. For relatively smooth surfaces, viscoelastic contributions may arise from roughness features down to atomic dimensions. On such surfaces, adhesion also becomes significant and influences the viscoelastic component of friction \cite{6,7}. 

Real rubber materials often contain mobile molecules, such as waxes, that diffuse to the surface and form a thin film. This film thickness may act as an effective cut-off length, at least for short sliding distances before being sheared off \cite{8,9}. On very rough surfaces, the high stresses in the asperity contact regions can lead to the formation of a thin surface layer with properties different from those of the bulk rubber, effectively defining the cut-off length. For instance, formation of micrometer-scale cracks\cite{10}, or breaking of molecular chains \cite{France}, 
may occur in a thin layer at the rubber surface during the run-in process.

In our analysis of rubber friction on road surfaces, we have found that including surface roughness down to length scales where the cumulative rms slope $\xi$ is of order 1 yields results in good agreement with experimental data. This typically corresponds to a cut-off wavenumber of $q_1 = 2\pi / \lambda_{\rm min} \approx 10^6 \ {\rm m}^{-1}$.
For a further discussion of the large-wavenumber cut-off, see Ref.~\cite{Ciav3}.

It is important to note that all analytical theories of rubber friction are based on the small-slope approximation. Therefore, including roughness beyond the $\xi \approx 1$ threshold lacks theoretical justification. Moreover, there is currently no surface topography measurement technique that can reliably capture height profiles at resolutions where $\xi > 1$.

Even if the exact value of $\lambda_{\rm min}$ is not known, the total friction coefficient can still be expressed as in (12), where the contact area contribution is written as $\mu_{\rm cont} = \sigma_{\rm eff} A / \sigma_0 A_0$. Here, $A$ is the real contact area, calculated by including all surface roughness down to the assumed cut-off length. 
In this case the shear stress is an effective shear stress which could have contributions both from viscoelasticity and from the adhesive interaction in the area of real contact.

Thus, in some studies, $A$ has been chosen as the nominal contact area $A_0$, in which case $\mu_{\rm visc}$ arises solely from macroscopic deformations of the contact region. This approach was employed in Ref. \cite{11}, where a rigid sphere was slid on human skin, a viscoelastic material. In that case, the nominal contact was Hertzian-like, and the macroscopic contact area had a distorted circular shape. Under such conditions, all asperity-induced viscoelastic energy dissipation is effectively captured within the definition of $\sigma_{\rm eff}$.

Another important issue concerns the role of adhesion. Here, we are not referring to the adhesive interaction between rubber molecules and the substrate within the real contact area, which gives rise to the frictional shear stress $\sigma_{\rm f}$. Instead, we refer to two distinct adhesion-related mechanisms: (i) the increase in real contact area due to adhesion, and (ii) the energy dissipation from opening cracks at the edges of asperity contacts, occurring at length scales where adhesion is relevant.

The contribution of adhesion to the enhancement of contact area was investigated by Plagge and Hentschke \cite{6}, 
while the contribution from opening cracks was studied by Le Gal et al. \cite{Kl} and by Carbone et al. \cite{Carb}. 
In  Ref. \cite{Carb} and \cite{7} Carbone et al. also showed that adhesion increases the contact area.
These mechanisms may significantly affect friction in very soft rubber compounds sliding on smooth surfaces. 
However, for rough substrate surfaces with roughness
on many length scales, such as road surfaces, experimental evidence suggests that this form of adhesion is often negligible.

For the relatively stiff rubber compounds used in practical applications such as tires and seals, there is consistent experimental evidence that crack-opening mechanisms contribute little to the overall sliding or rolling friction. This is primarily due to the multiscale roughness of the surface, which strongly suppresses adhesion \cite{Kill}. 
In Sec. 8it is shown that for rubber compounds which exhibiting strong adhesion on smooth surfaces under dry conditions, but no adhesion in water, 
display nearly identical sliding friction on rough surfaces under both dry and wet conditions \cite{water}. 
Similar observations have been made by Greenwood and Tabor \cite{Tabor1}.

\begin{figure}[!ht]
\includegraphics[width=0.47\textwidth,angle=0]{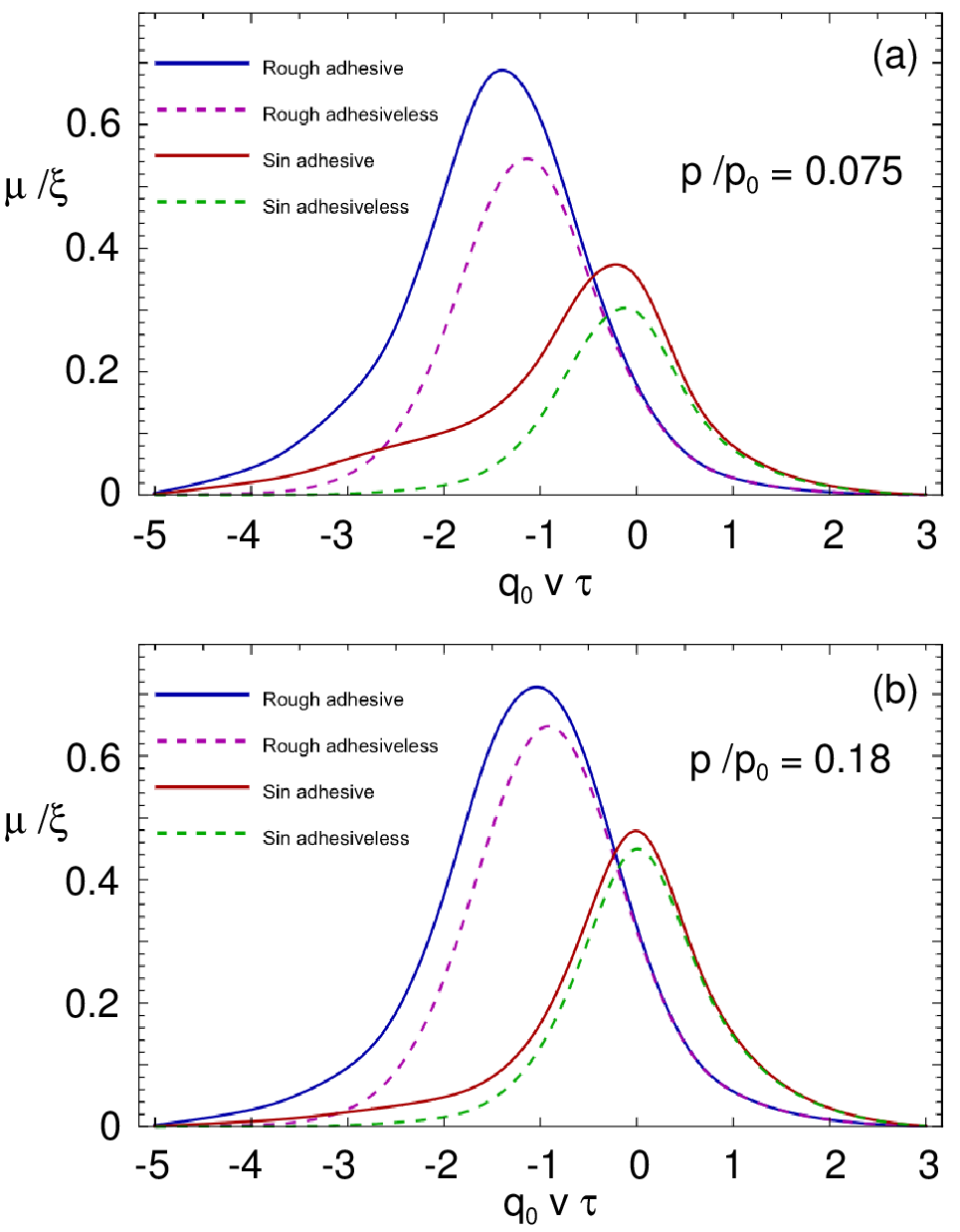}
\caption{\label{AdhesionCarboneHappy.eps}
The friction coefficient, divided by the surface rms slope $\xi$, as a function of the dimensionless sliding speed $q_0 v \tau$,  
for a surface with the height profile $h_0 \sin (q_0 x)$ (dashed lines), and for a surface with one-dimensional  
self-affine roughness, with the ratio between the large and small wavenumber cut-offs $q_1/q_0 = 60$ and the Hurst exponent $H = 0.8$ (solid lines).  
The viscoelastic modulus is of the form given in (4), with $E_1/E_0 = 10$ and relaxation time $\tau$.  
The applied pressure is $p = 0.075\,p_0$ in (a) and $0.18\,p_0$ in (b), where the reference pressure is $p_0 = E_0^* q_0 h_{\rm max}/2$,  
and $h_{\rm max}$ is the height of the highest asperity relative to the average surface plane.
Adapted from Ref. \cite{Menga}.
}
\end{figure}

The influence of adhesion on the sliding friction for a viscoelastic solid with the simple modulus given by (4),  
with $E_0 = 10 \ {\rm MPa}$ and $E_1/E_0 = 10$, was studied theoretically in Ref.~\cite{Menga}.  
The sliding friction was calculated for a surface with the height profile $h_0 \sin (q_0 x)$, and for a surface with one-dimensional (1D)  
self-affine roughness, with the ratio between the large and small wavenumber cut-offs $q_1/q_0 = 60$ and the Hurst exponent $H = 0.8$.  
The interfacial work of adhesion was $\gamma = 0.01 \ {\rm J/m^2}$, as is typical for rubber in contact with many solids (smooth surfaces).  

Fig.~\ref{AdhesionCarboneHappy.eps} shows the friction coefficient $\mu$ divided by the surface rms slope $\xi$  
as a function of the dimensionless sliding velocity $v/v_0$ for two different values of the dimensionless  
pressure $p/p_0$. The reference velocity is $v_0 = 1/(q_0 \tau)$, and the reference pressure is $p_0 = E_0^* q_0 h_{\rm max}/2$,  
where $h_{\rm max}$ is the height of the highest asperity relative to the average surface plane.  

Results are shown both with and without adhesion. As the pressure increases, the influence of adhesion decreases.  
It is not clear how these results would be modified for real surfaces with 2D roughness and with larger values of $q_1/q_0$.

\begin{figure}[!ht]
\includegraphics[width=0.3\textwidth,angle=0]{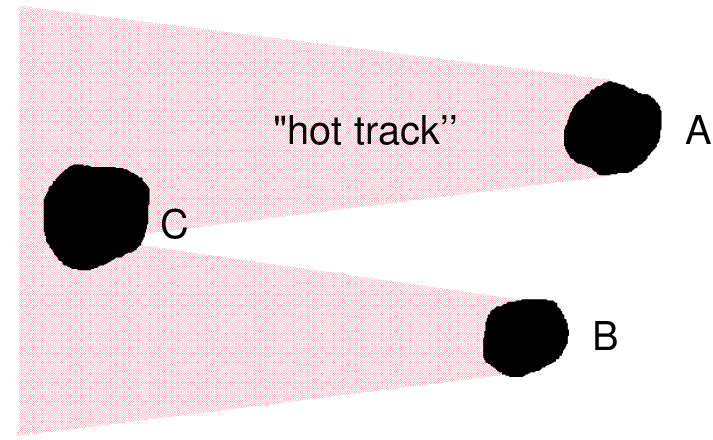}
\caption{\label{HotTrack.eps}
The asperity contact region C experiences a temperature increase due to the frictional heating generated by asperity contacts A and B located ahead of it in the sliding direction. The “hot tracks” originating from asperity contacts spread laterally with increasing distance from the contact due to thermal diffusion.
}
\end{figure}

Another important challenge is how to incorporate the multiscale nature of frictional heating. In the treatment presented in Ref.~\cite{VG}, the frictional energy dissipation occurring in the (small) contact regions within the macroasperity contacts was approximated as a laterally uniform heat source within the macroasperity contact regions. 
However, this simplification may not always be valid. The problem is inherently complex because, during sliding, a given contact region may be influenced by the thermal “hot tracks” generated by asperity contact regions located ahead of it in the sliding direction (see Fig.~\ref{HotTrack.eps}). In Ref.~\cite{VG}, this effect was accounted for at the level of macroasperity contacts, but a more comprehensive theory should also include such interactions between the smaller contact regions nested within the macroasperity zones.

\begin{figure}[!ht]
\includegraphics[width=0.45\textwidth,angle=0]{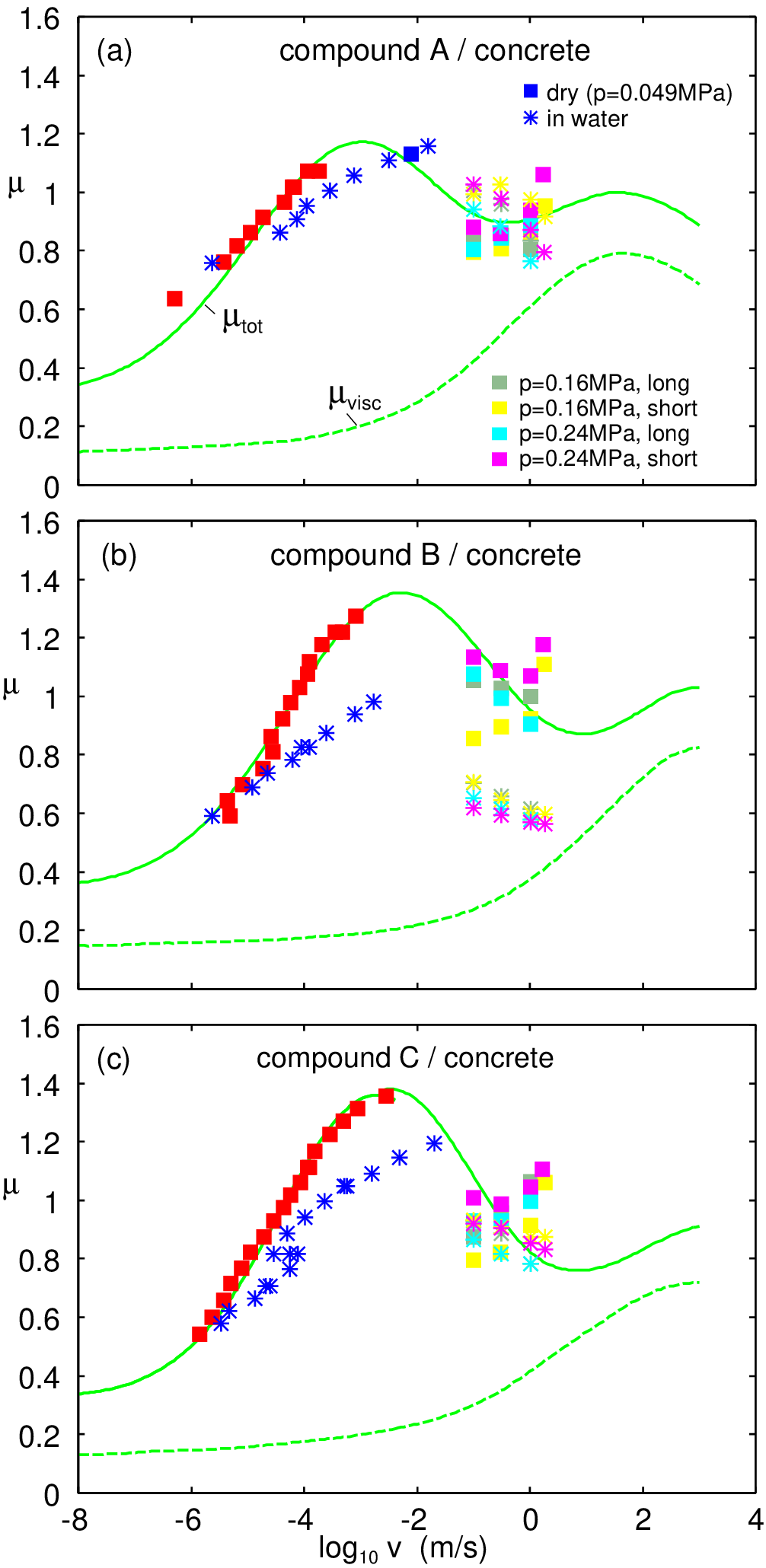}
\caption{\label{XXX.1logv.2mu.all.Leonardo.and.Tanta.1.eps}
Measured friction coefficients as a function of the logarithm of sliding speed for three different rubber compounds sliding on a concrete surface. The square symbols correspond to dry conditions, and the star symbols to wet conditions (in water). The dashed green lines represent the predicted viscoelastic contribution, while the solid lines show the sum of the viscoelastic contribution and the contribution from the area of real contact.
}
\end{figure}

\begin{figure}[!ht]
\includegraphics[width=0.45\textwidth,angle=0]{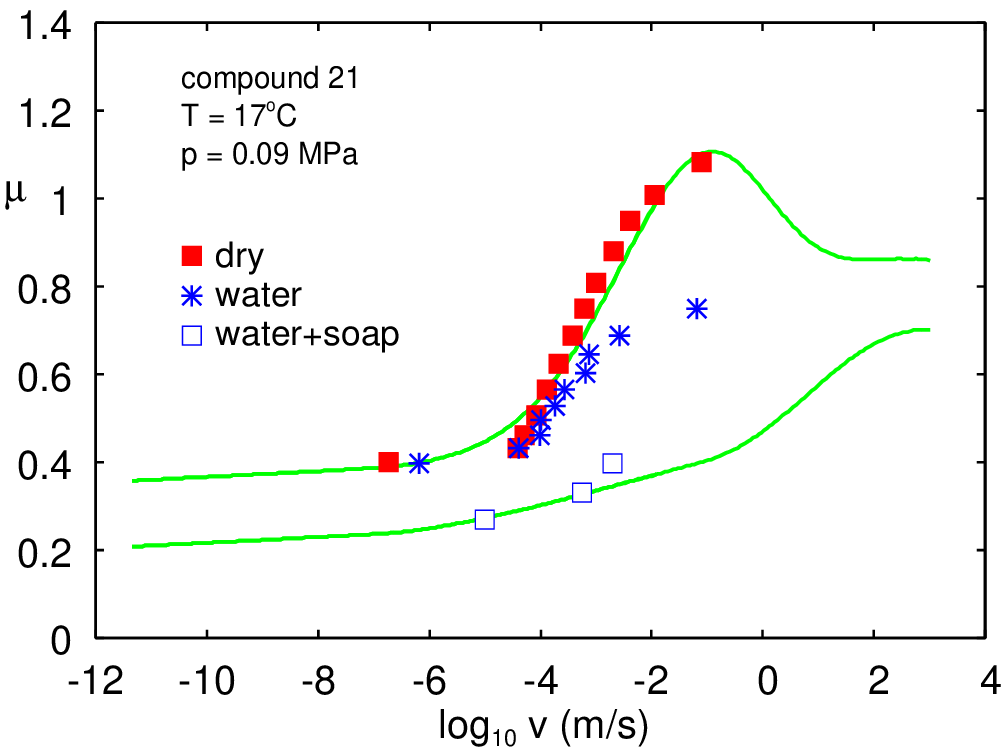}
\caption{\label{Cabout2.eps}
Measured friction coefficients as a function of the logarithm of sliding speed for a rubber compound sliding on an asphalt road surface composed of large stones with relatively flat and smooth upper surfaces. The filled square symbols represent dry conditions, the stars represent wet conditions (in water), and the open squares correspond to lubrication by soapy water. The lower green line shows the calculated viscoelastic contribution, and the upper green line represents the total friction coefficient.
}
\end{figure}

\vskip 0.1 cm
{\bf 5.2 Comparing the theory with experimental results}

Fig. \ref{XXX.1logv.2mu.all.Leonardo.and.Tanta.1.eps} 
shows the measured friction coefficient as a function of the logarithm of sliding speed for three different rubber compounds sliding on a concrete surface. The square symbols represent measurements under dry conditions, while the star symbols correspond to measurements in water. The dashed green lines indicate the theoretical viscoelastic contributions, and the solid lines show the sum of the viscoelastic contribution and the contribution from the area of real contact.

The results in Fig. \ref{XXX.1logv.2mu.all.Leonardo.and.Tanta.1.eps} demonstrate that, in most cases, at low sliding speeds, water is almost completely expelled from the asperity contact regions, resulting in a friction coefficient similar to that observed under dry conditions. However, at much higher sliding speeds, there is insufficient time for the water film to be fully squeezed out, leading to a significant reduction in friction. At very high speeds, hydroplaning may occur, and the friction becomes negligible.

While water is generally a poor lubricant on most surfaces, the addition of soap can produce effective lubrication, particularly when the contact pressure is not too high. On hydrophilic surfaces, lubrication may arise from osmotic pressure \cite{IRAEL} or hydration effects \cite{KLEIN}. To illustrate this, Fig. \ref{Cabout2.eps} shows the measured friction coefficient as a function of the logarithm of sliding speed for a rubber compound sliding on an asphalt road surface composed of large stones with relatively flat and smooth upper surfaces. The filled square symbols represent dry conditions, stars indicate in-water conditions, and open squares correspond to lubrication by soapy water. The lower green line shows the calculated viscoelastic contribution, and the upper green line represents the total friction coefficient. Notably, the friction coefficient in soapy water is in close agreement with the theoretical prediction based solely on the viscoelastic contribution.

\begin{figure}
\includegraphics[width=0.45\textwidth,angle=0.0]{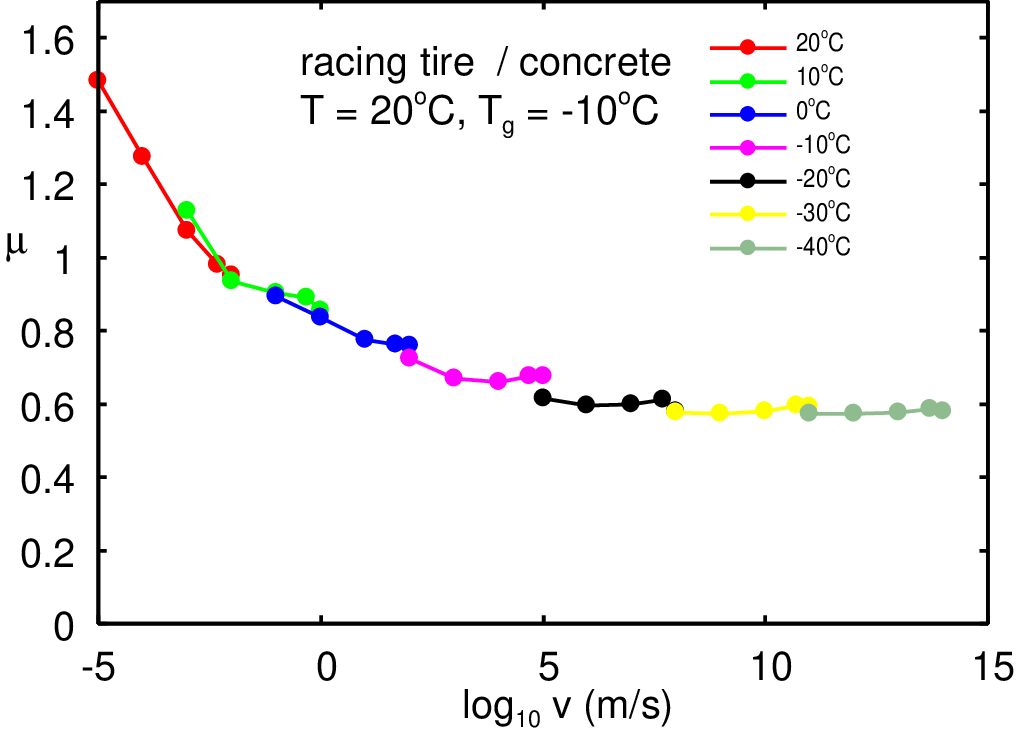}
\caption{\label{WW.RACING.TR12masterwithTheory.1.eps}
The sliding friction master curve for the racing compound. The velocity-dependent segments of the friction data were measured at temperatures between $-40^\circ {\rm C}$ and $20^\circ {\rm C}$, and were shifted along the velocity axis using the viscoelastic shift factor $a_T$, obtained from the construction of the viscoelastic master curve $E(\omega, T_0)$. The reference temperature is $T_0 = 20^\circ {\rm C}$.
}
\end{figure}

\begin{figure}
\includegraphics[width=0.45\textwidth,angle=0.0]{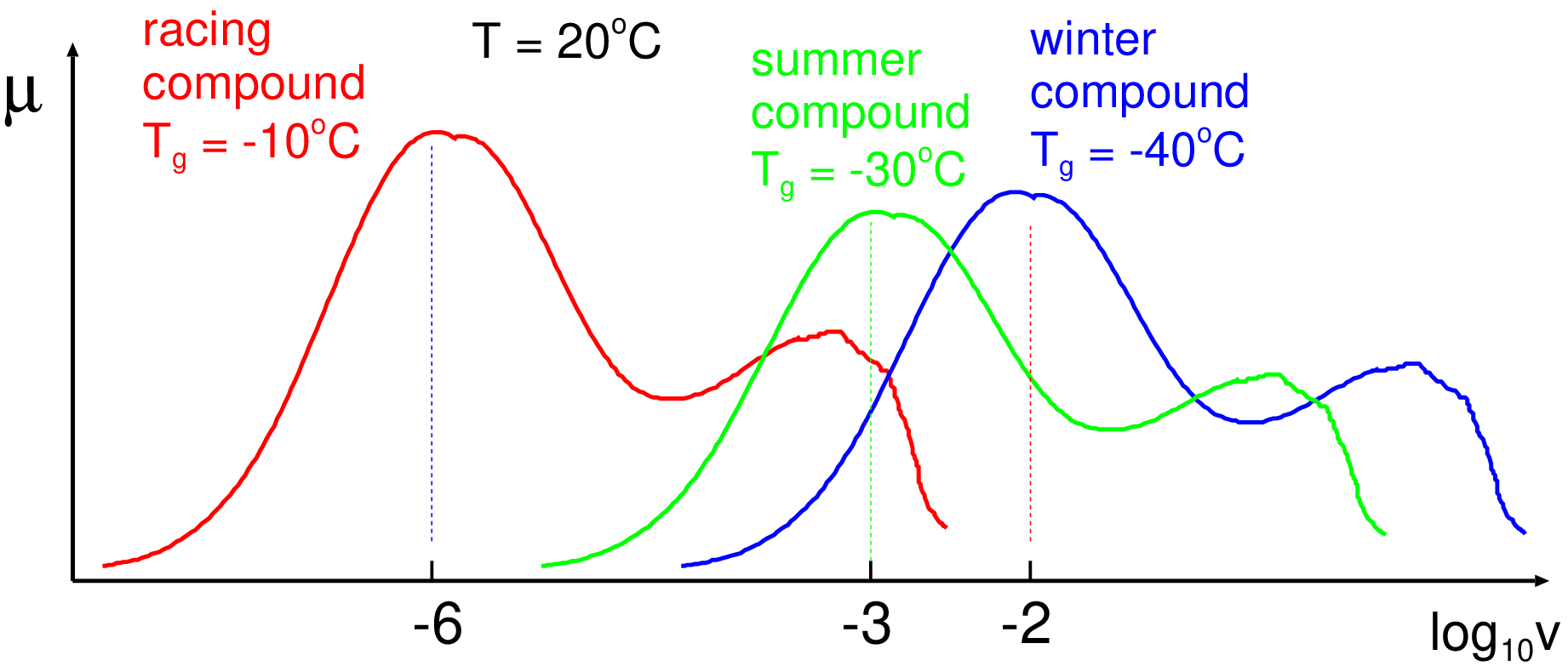}
\caption{\label{WW.ShowPrinciple.eps}
The friction coefficient at room temperature as a function of the logarithm of the sliding speed for a racing compound (red line),  
a passenger summer compound (green line), and a winter compound (blue line) (schematic).
}
\end{figure}

One important quantity that characterizes rubber is its glass transition temperature, $T_{\rm g}$, defined as the temperature at which ${\rm tan} \, \delta$ reaches its maximum for a 
given reference frequency (we use $\omega = 0.01 \ {\rm s}^{-1}$). 
Due to the WLF temperature-frequency relationship, $T_{\rm g}$ also provides information about the frequency at which ${\rm tan} \, \delta$ is maximal.

To get a large viscoelastic contribution to friction, the perturbation frequencies generated by the road asperities should lie within the transition region between the rubbery and glassy states, i.e., the region where ${\rm tan} \, \delta$ is large. Therefore, $T_{\rm g}$ is a critical parameter in the design of tread compounds for tires. Since different types of tires operate at different working temperatures, rubber compounds with different $T_{\rm g}$ values are used for winter, summer, and racing tires. As a result, the maximum of the friction coefficient as a function of sliding speed at a fixed temperature differs significantly among tire types. The friction master curve typically reaches its maximum at sliding speeds on the order of a few meters per second when the tire is used at its designed operating temperature.

The friction curves in Fig. \ref{XXX.1logv.2mu.all.Leonardo.and.Tanta.1.eps} 
and Fig. \ref{Cabout2.eps} correspond to summer tires for passenger cars at room temperature, where the maximum occurs at $\sim 1 \ {\rm cm/s}$. 
The typical operating temperature for summer tires is around $50^\circ {\rm C}$, and at such temperatures, the maximum shifts to a few meters per second. 
In contrast, the operating temperature for racing tires is $\sim 100^\circ {\rm C}$ or higher. To ensure that the maximum in the friction coefficient falls within a relevant velocity range ($\sim 1-10 \ {\rm m/s}$), racing tires are formulated with significantly higher $T_{\rm g}$ than passenger car tires. This is illustrated in Fig. \ref{WW.RACING.TR12masterwithTheory.1.eps} where, for a Formula 1 - like racing compound, the maximum friction occurs at extremely low speed at room temperature. 
Fig. \ref{WW.ShowPrinciple.eps} schematically illustrates the differences in friction behavior between summer and winter compounds for passenger car tires, and a racing compound.

Adhesion and viscoelasticity can have a strong influence on the area of real contact~\cite{Krick,Scheib}.  
The stiffening of the elastic properties of rubber when perturbed at high frequencies  
results in a drastic decrease in the contact area with increasing sliding speed  
in a region of intermediate velocities [see Fig.~\ref{1logv.2mu.Area.compoundB.concrete.eps}(b)].  
At very high sliding speeds, the contact area can increase due to frictional heating,  
which reduces the elastic modulus of the rubber.  

\begin{figure}
\includegraphics[width=0.48\textwidth,angle=0.0]{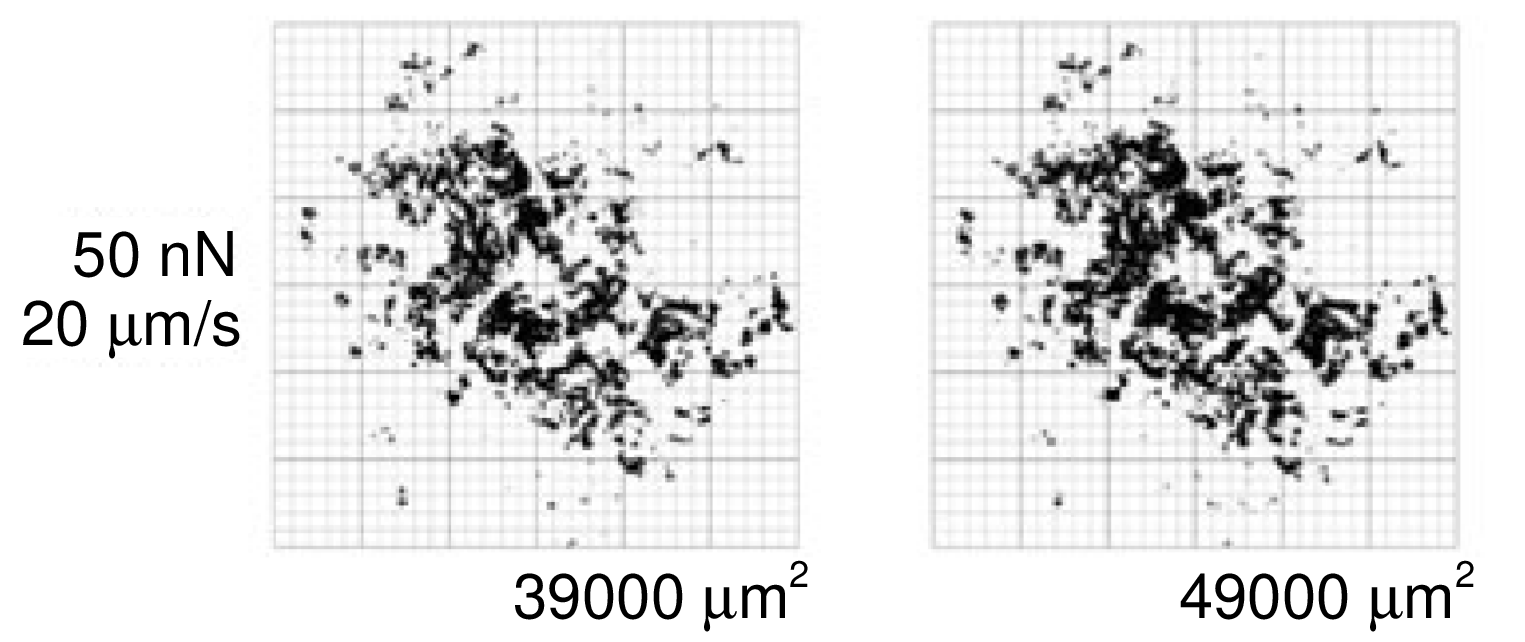}
\caption{\label{picKrick.eps}
Two contact footprints between a Buna-N nitrile rubber sphere (diameter $4.8 \ {\rm mm}$,  
with rms roughness $h_{\rm rms} = 5.2 \ {\rm \mu m}$) and a borosilicate float glass optical window.  
(a) shows the contact area before sliding, and (b) at a sliding speed of $20 \ {\rm \mu m/s}$.  
The contact area during sliding is larger by a factor of $49/39 \approx 1.26$.
}
\end{figure}

At low sliding speeds on smooth surfaces, experimental observations also indicate  
that adhesion can increase the contact area with increasing sliding speed~\cite{Krick}  
[see Fig.~\ref{picKrick.eps}]. This may be due to the increased energy required to propagate opening cracks  
at higher sliding speeds.

\begin{figure}
\includegraphics[width=0.48\textwidth,angle=0.0]{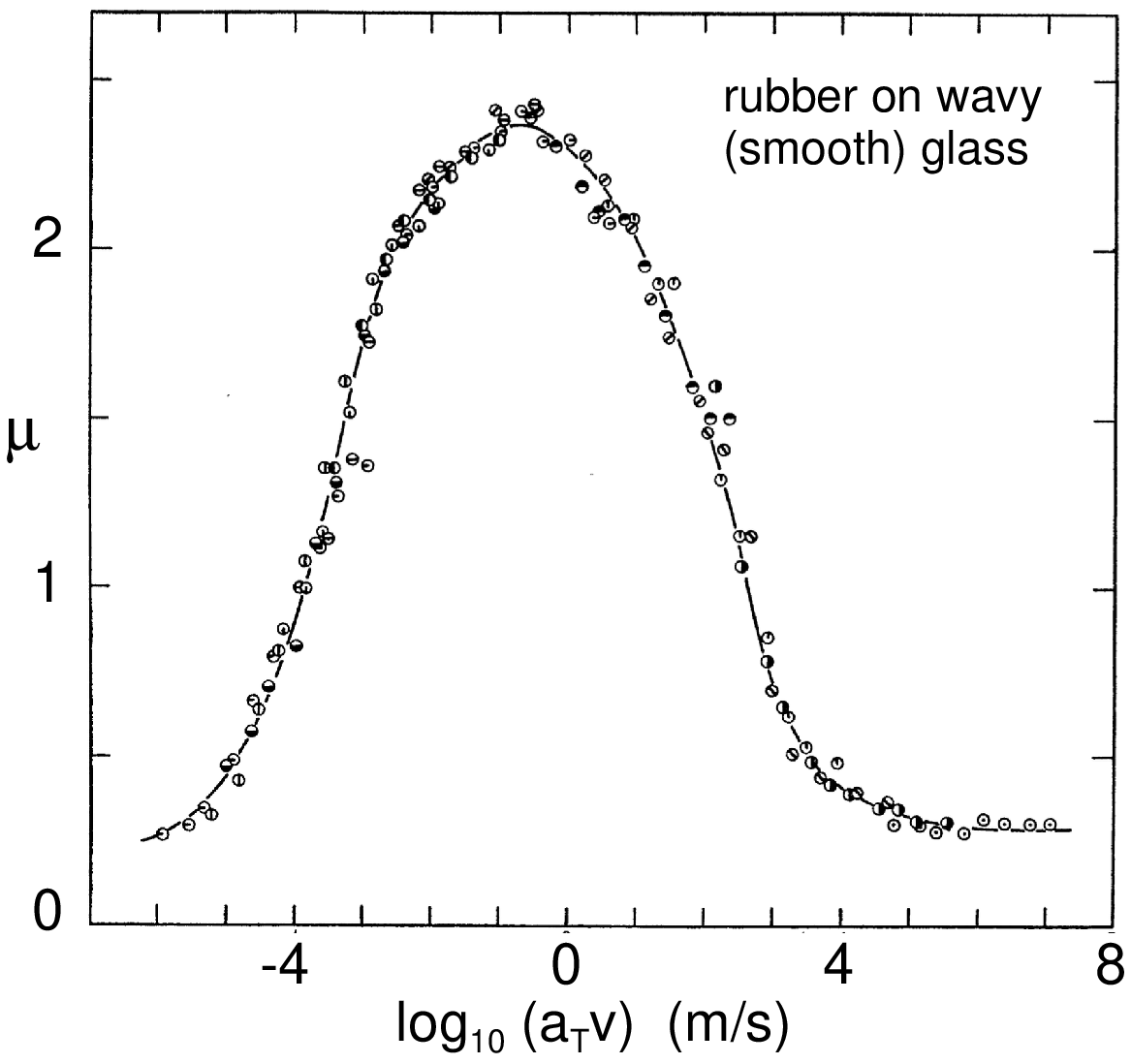}
\caption{\label{GROSCH.wavyGLASS.eps}
Master curve of the friction coefficient for acrylonitrile-butadiene rubber  
on smooth but macroscopically wavy glass. The reference temperature is $T_0 = 20^\circ {\rm C}$.  
The master curve was obtained by shifting velocity segments (velocity range approximately  
from $1 \ {\rm \mu m/s}$ to $1 \ {\rm cm/s}$), measured at temperatures between $-15^\circ$ and $85^\circ {\rm C}$,  
using the bulk viscoelastic shift factor $a_T$.  
When the friction coefficient decreases with increasing sliding speed, stick-slip occurs.  
For these velocities, the reported friction coefficient corresponds to the peak friction force during  
the stick-slip oscillations.  
Adapted from Ref.~\cite{Grosch}.
}
\end{figure}

\begin{figure}
\includegraphics[width=0.45\textwidth,angle=0.0]{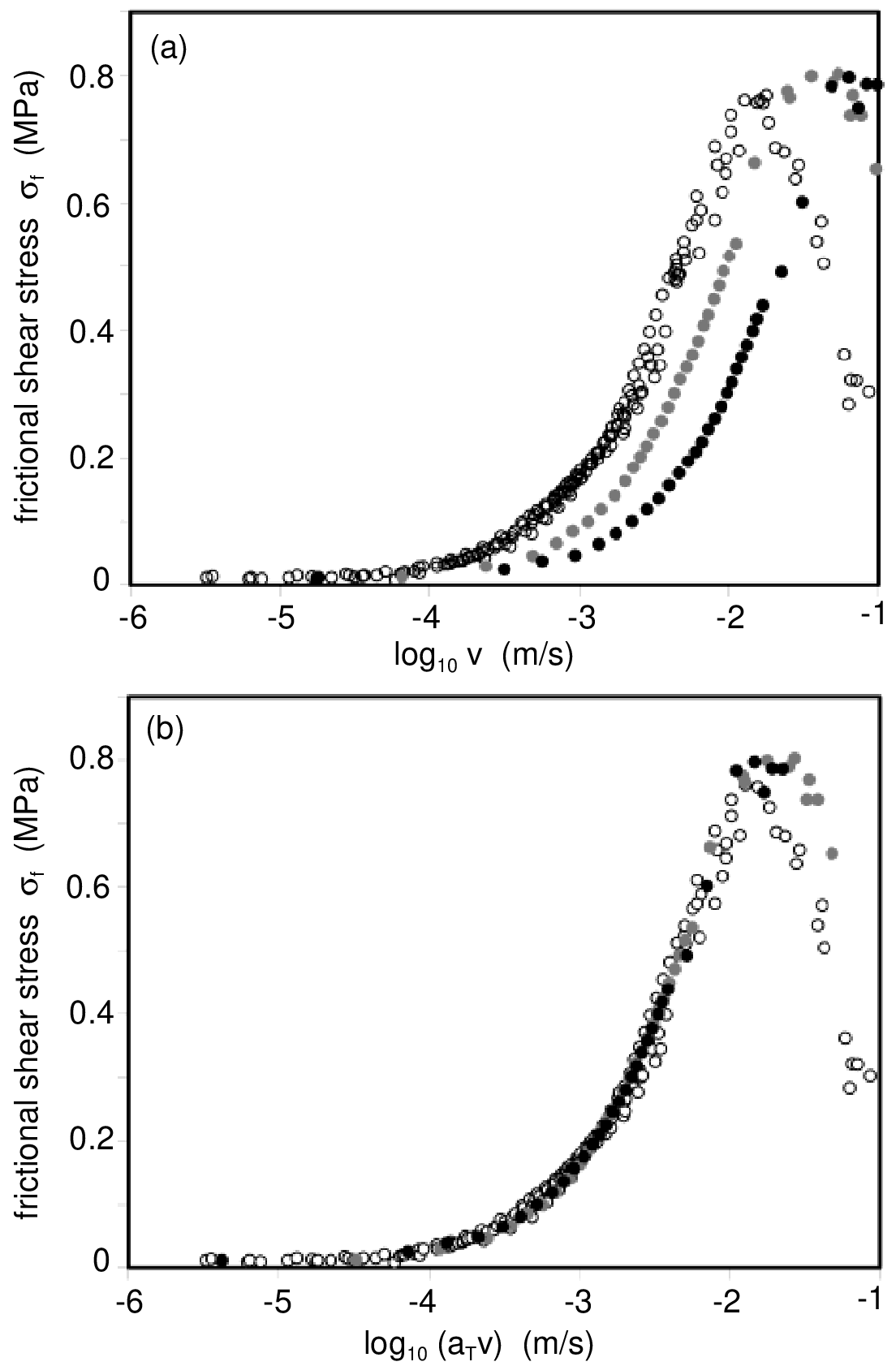}
\caption{\label{ChaudhuryShearStressPDMS.eps}
(a) Shear stress as a function of velocity and temperature for sliding between PDMS  
($M = 2.7 \ {\rm kg/mol}$) and a silicon wafer coated with a hexadecylsiloxane self-assembled monolayer (SAM).  
Open, gray, and black circles represent data measured at 298, 318, and 348~K, respectively.  
(b) The temperature-dependent shear stress data  
shifted to room temperature using an Arrhenius shift factor.  
The activation energy for sliding between PDMS and the SAM is estimated to be $25 \ {\rm kJ/mol}$.  
Open, gray, and black circles again represent data measured at 298, 318, and 348~K, respectively.
}
\end{figure}

\vskip 0.1 cm
{\bf Sliding friction on smooth surfaces}

Rubber friction for tire tread compounds on smooth surfaces is a complicated topic,  
because mobile components (waxes), which are added to the rubber compounds, diffuse to the rubber surface and are slowly sheared  
off during sliding, resulting in a friction coefficient that depends sensitively on the sliding distance.  
Experiments on ``simpler'' rubber compounds such as poly(dimethylsiloxane) (PDMS) elastomers give more reproducible results.  

Grosch~\cite{Grosch} studied the friction of acrylonitrile-butadiene rubber  
on smooth but macroscopically wavy glass.  
Fig.~\ref{GROSCH.wavyGLASS.eps}  
shows the master curve of the friction coefficient at the reference temperature $T_0 = 20^\circ {\rm C}$.  
The master curve was obtained by shifting velocity segments measured at temperatures between $-15^\circ$ and $85^\circ {\rm C}$,  
using the bulk viscoelastic shift factor $a_T$.  
The friction master curve has the general form expected from the Schallamach  
model~\cite{1} of the adhesive contribution to friction, but the width of the  
curve is larger than predicted by the Schallamach model and instead agrees better with the  
prediction of the Persson-Volokitin model~\cite{theory3}.  

Vorvolakos and Chaudhury~\cite{Chaud} performed experiments in which spherical cups of  
PDMS were slid on Si wafers coated with a hexadecylsiloxane monolayer, or on thin films of polystyrene.  
Elastomers with various molecular weights were used on both surfaces.  
Fig.~\ref{ChaudhuryShearStressPDMS.eps} shows the shear stress as a function of velocity and temperature  
between PDMS and a Si wafer with a self-assembled monolayer of hexadecylsiloxane.  
From the temperature-dependent studies, the activation energy of friction  
on both surfaces was estimated to be $25 \ {\rm kJ/mol}$, which  
compares well with values obtained from studies of melt dynamics.  

The friction first increases with sliding velocity, reaches a maximum, and then decreases, as expected  
from the Schallamach model. The maximum of the frictional shear stress  
$\sigma^*$ was found to be roughly proportional to the square root of the elastic modulus, as expected  
from the Schallamach model if the size of the stress domains is independent of the modulus  
[see ()]. The sliding velocity $v^*$ at which $\sigma_{\rm f}$ is maximum differs for the two substrates,  
which may reflect different relaxation times $\tau_0$,  
as expected since the free energy barrier for attachment depends on the substrate.  

\begin{figure}
\includegraphics[width=0.45\textwidth,angle=0.0]{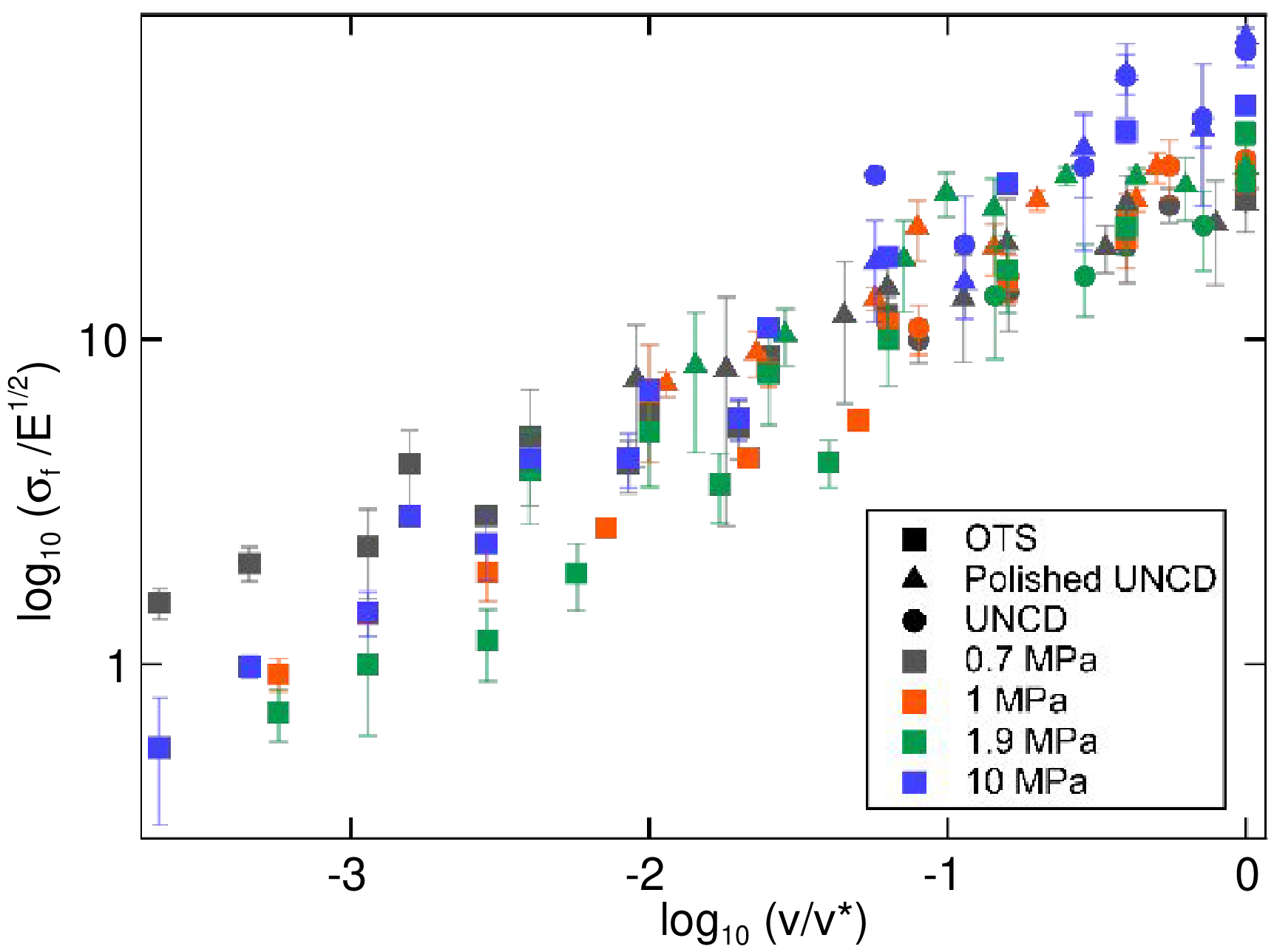}
\caption{\label{AliShearStress.eps}
The frictional shear stress divided by the elastic modulus raised to the $1/2$ power  
as a function of the normalized sliding speed $v/v^*$ (log-log scale). The data points  
are for PDMS rubber with different moduli sliding on three different surfaces (see inset).  
Adapted from Ref.~\cite{Ali3}.
}
\end{figure}

Maksuta et al.~\cite{Ali3}  
measured the frictional shear stress for PDMS rubbers  
with different elastic moduli sliding on three surfaces with similar interfacial chemistry  
but different roughness. Analysis of the experiments showed that  
friction on all the surfaces  
was dominated by adhesive contributions rather than viscoelastic dissipation,  
and that in all cases the rubber made complete contact with the substrate.  

Contrary to the expectation that roughness only  
influences the available contact area for adhesive friction,  
it was found that roughness caused a shift in the  
velocity at which the adhesive friction is maximum.  
The authors suggested that this resulted from roughness-induced  
vertical oscillations of the rubber surface during sliding.  
The frictional shear stress scaled with the elastic modulus as  
$E^{1/2}$ (see Fig.~\ref{AliShearStress.eps}), as  
expected from the Schallamach model for adhesive friction on smooth surfaces.

\begin{figure}
\includegraphics[width=0.45\textwidth,angle=0.0]{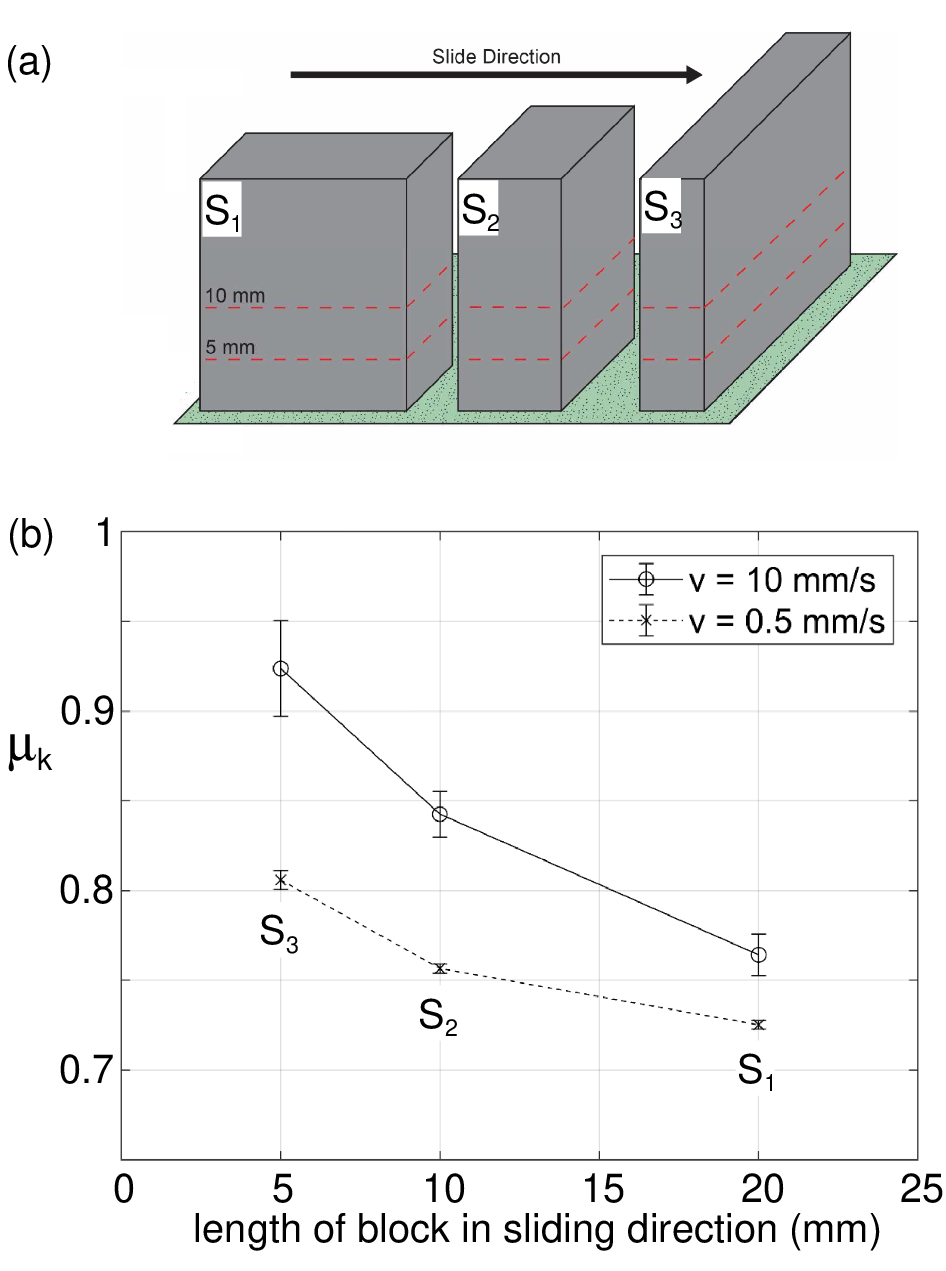}
\caption{\label{DependencySlidingDirection.eps}
(a) Three rectangular tread rubber blocks sliding on a rough substrate surface. The rubber blocks have the same bottom surface area. They were clamped at different distances from the bottom surface, but no significant dependence on the clamping position was observed.  
(b) The kinetic friction coefficient decreases with increasing length of the rubber block in the sliding direction. At the same time, the rubber wear rate decreased from $4.6$ to $1.9$ to $1.6 \ {\rm mg/Nm}$ for the blocks with the shortest to the longest length in the sliding direction. This strong dependence of wear on the friction coefficient (roughly $\sim \mu^{11}$) is consistent with the rubber wear theory presented in Ref.~\cite{10}. Adapted from Ref.~\cite{HK5}.
}
\end{figure}

\begin{figure}
\includegraphics[width=0.45\textwidth,angle=0.0]{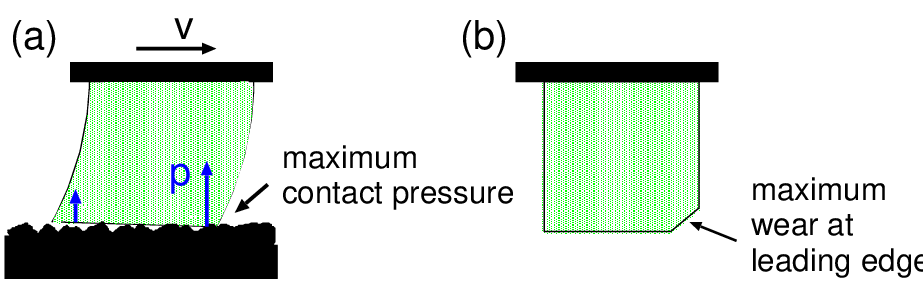}
\caption{\label{BLOBbend.eps}
During sliding, the rubber block deforms such that the maximum contact pressure occurs at the leading edge.  
Visual inspection after sliding shows that most of the wear has occurred at the leading edge \cite{HK5}.
}
\end{figure}

\vskip 0.1 cm
{\bf Dependency of the kinetic friction on the shape of the rubber block}

For contact between most solids with surface roughness, the area of real contact is proportional to the normal
force and independent of the size or shape of the nominal contact area. However, experiments have shown that for
rubber, the friction force may depend on the shape of the nominal contact area \cite{HK5}. 
In one study, Hale and Lewis slid rectangular blocks of different shapes [see Fig. \ref{DependencySlidingDirection.eps}(a)]  
on rough surfaces. Fig. \ref{DependencySlidingDirection.eps}(b) shows that the friction coefficient decreased with increasing  
length of the blocks in the sliding direction. One mechanism for this is frictional heating. Rubber friction is highly temperature dependent, and the longer a rubber surface remains in contact
in the sliding direction with road asperities, the warmer the asperities become. This, in turn, increases the temperature
of the rubber in the asperity contact regions. For rubber sliding on smooth surfaces, frictional heating
in the rubber asperity contact regions produces a similar effect and becomes more significant as the duration of contact 
with the smooth counter surface increases. 

At velocities around $10 \ {\rm mm/s}$, frictional heating is expected to occur and may explain why longer shapes (oriented in the sliding direction) 
result in lower friction coefficients. However, even at the lower sliding speed $v = 0.5 \ {\rm mm/s}$, where frictional heating is negligible, 
the friction force still decreases with increasing block length along the sliding direction [see Fig. \ref{DependencySlidingDirection.eps}(b)]. 
This cannot be due to frictional heating but may be an ``edge'' effect. Thus,
an investigation of the wear rate and its spatial distribution showed that the length of the leading edge also influences wear, with longer front-edge geometries 
oriented perpendicular to the sliding direction producing the highest wear mass. This increased wear for wider rubber blocks 
can be explained by shear deformation of the block, which leads to the highest contact pressure at
the leading edge where most of the wear was observed \cite{HK5} (see also Fig. \ref{BLOBbend.eps}). The substrate used in the friction study reported in
Ref. \cite{HK5} had very sharp roughness, consisting of a sand and acrylic paint mixture on a wood plate. This configuration resulted in very high wear rates, 
ranging from $1.6$ to $4.6 \ {\rm mg/Nm}$, compared to the much lower wear rates observed in Ref. \cite{10} for SB rubber sliding on a concrete surface,
which were between $0.02$ and $0.04 \ {\rm mg/Nm}$. These values were approximately 100 times smaller. In most cases, wear contributes negligibly to the
friction force. However, in the case of the large wear rates observed in Ref. \cite{HK5}, the wear contribution to friction may no longer be negligible.

\begin{figure}
\includegraphics[width=0.45\textwidth,angle=0.0]{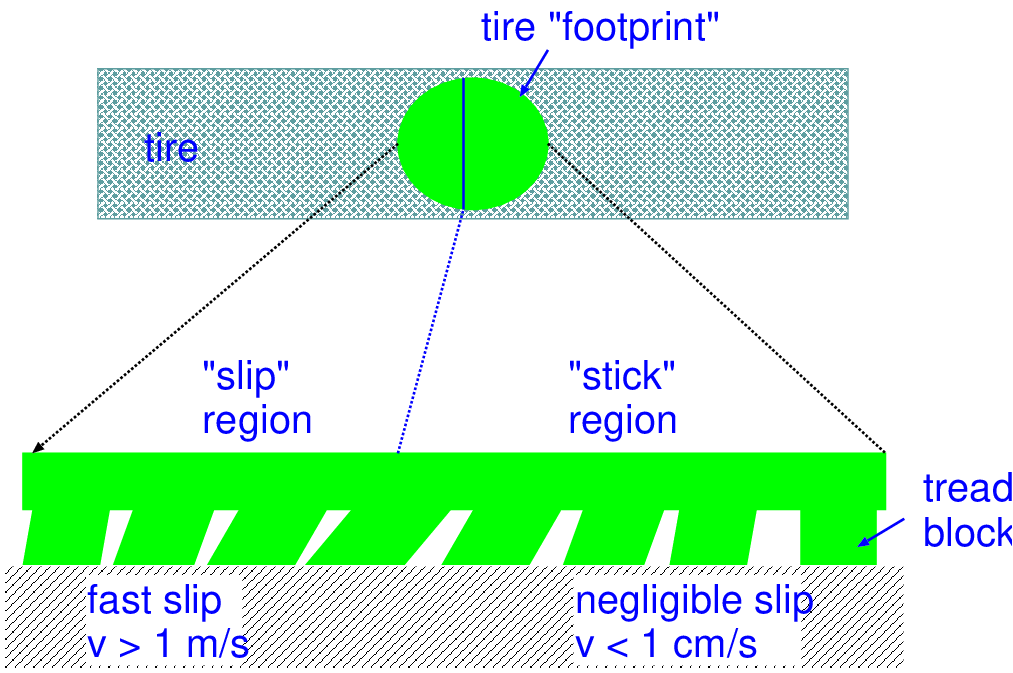}
\caption{\label{WW.macro1a.eps}
The deformation of the tire tread blocks as they pass through the tire 
footprint under finite tire slip, defined as $s = (v_{\rm c} - v_{\rm r})/v_{\rm c}$, where
$v_{\rm c}$ is the velocity of the car and $v_{\rm r}$ is the tire rolling velocity. The blocks initially deform elastically without slip,
but begin to slip near the exit of the tire-road footprint.
}
\end{figure}

\vskip 0.3cm
{\bf 5.3 Dynamical friction}

Up to now, we have assumed stationary conditions 
in which the sliding speed was constant. Here, we describe two dynamical effects that become crucial 
when the sliding speed changes, especially when slip initiates after some time of stationary contact ($v=0$). 
This situation arises in many practical applications. For example, when a tread block of a tire comes
into contact with the road surface, it initially does not slip but only deforms elastically. 
Slip begins when the shear stress becomes higher the static frictional shear stress. 
Hence, in general, for small tire slip 
(e.g., during ABS braking), the tread blocks will begin to slip near the exit of
the tire-road footprint (see Fig. \ref{WW.macro1a.eps}).

The breakloose friction force is defined as the maximum friction force before the onset of steady sliding, 
and is often referred to as the static friction force. It depends on the history of 
contact between two solids. For instance, if a block slides on a substrate, the resulting temperature 
increase in the block, particularly near asperity contact regions, may affect the friction force. 
If the block motion stops, the temperature distribution in the block and in the asperity contact 
regions of the substrate evolves over time, leading to a breakloose friction force that depends 
on the contact duration. This effect is particularly significant for rubber materials, where rapid 
flash temperature changes can cause substantial variations in friction during non-stationary sliding.

Other processes that can lower the interfacial free energy and increase the breakloose friction force 
include a slow increase in contact area, thermally activated (and hence contact-time-dependent) 
bond formation at the interface, or structural relaxation of the rubber 
chain molecules at the contact interface.

\begin{figure}
\includegraphics[width=0.3\textwidth,angle=0.0]{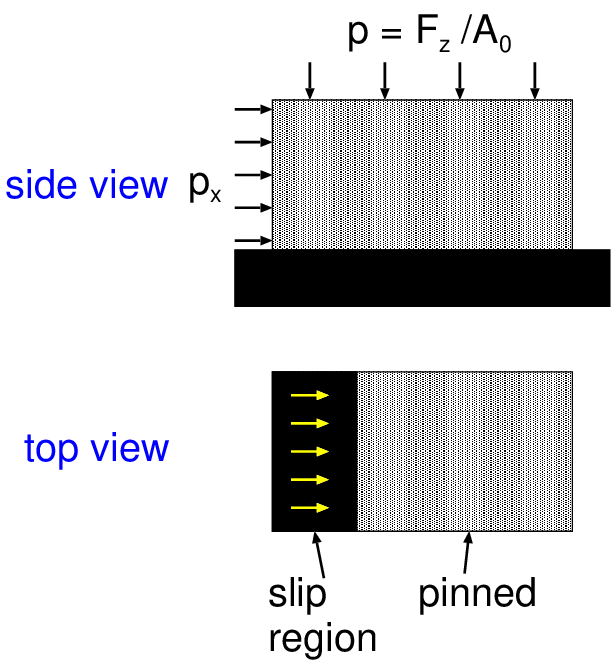}
\caption{\label{ZZZ.ThreeCases1.eps}
	Slip at the onset of sliding when the driving force is applied on one side of the rubber block.
        The region where slip occurs gradually expands as the system approaches the onset of global slip, at which point the entire contact area is sliding.
	The frictional shear stress in regions where the slip displacement $u > \lambda_{\rm asp}$ 
	is equal to the kinetic frictional shear stress $\sigma_{\rm k}$.
	Consequently, the breakloose friction force is smaller than $F_{\rm s} = A \sigma_{\rm s}$ and approaches
	$F_{\rm k}= A \sigma_{\rm k}$ as the elastic modulus decreases.
}
\end{figure}

\begin{figure}
\includegraphics[width=0.45\textwidth,angle=0.0]{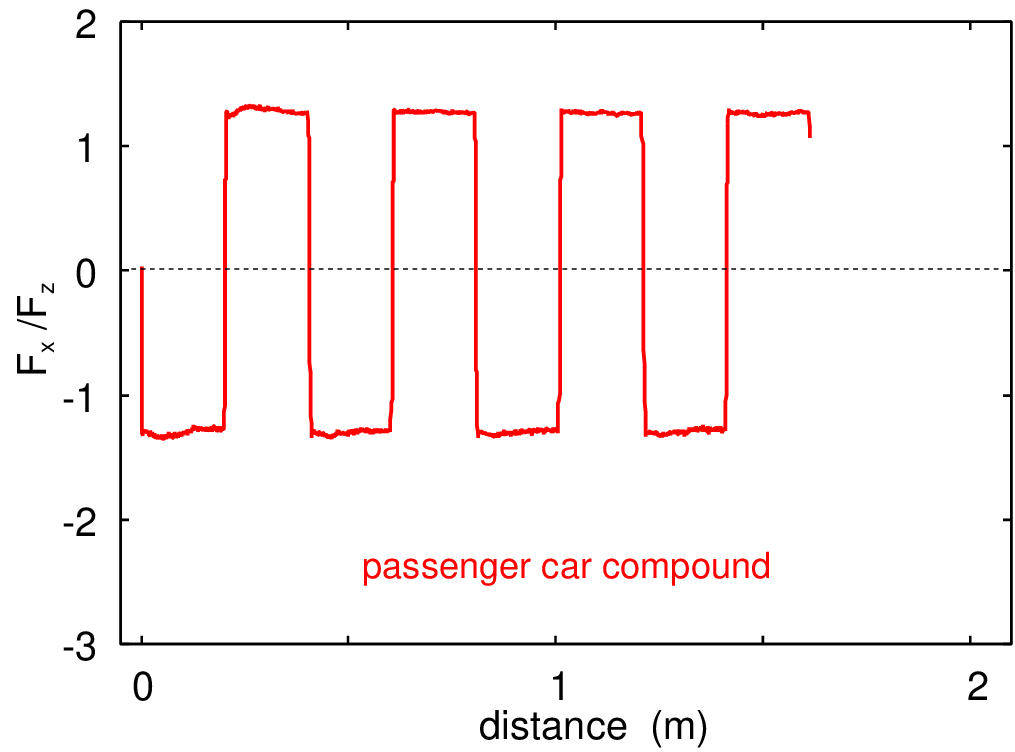}
\caption{\label{ZZZ.p.1x.2mu.last.eps}
	Ratio of tangential (friction) force $F_x$ to normal force $F_{\rm N}$ 
	as a function of sliding distance during oscillatory movement of rubber blocks.
        Data correspond to a passenger car tread compound sliding on a concrete surface at $T=20^\circ {\rm C}$. 
	The rubber blocks first move forward at $v= 3 \ {\rm mm/s}$ for $20 \ {\rm cm}$, and then backward 
	at the same velocity, returning to the starting point. 
	This motion is repeated four times.
	The rubber blocks are $0.5 \ {\rm cm}$ high with lateral dimensions of $4 \ {\rm cm} \times 4 \ {\rm cm}$.
}
\end{figure}

\vskip 0.1cm
{\bf Elasticity effects and pre-slip}


Strengthening of the contact during stationary contact does not always result in a breakloose friction force 
exceeding the kinetic friction force, especially for elastically soft materials. 
We now analyze this phenomenon with a particular focus on rubber friction.

Rubber is an elastically soft material, and as a result, the interfacial slip at the onset of sliding between a rubber block and a substrate is not uniform, as illustrated in Fig. \ref{ZZZ.ThreeCases1.eps} for an elastic block where the driving force is applied on the left-hand side. We first show that pre-slip within the contact region causes the breakloose friction force to be lower than the product of the true contact area and the static shear stress. 
For elastically soft materials, in the absence of the flash temperature effect described above, this can result in a breakloose friction force that is equal to the kinetic friction force.

Consider a ``kinetic'' frictional shear stress $\sigma_{\rm k}$ acting on the contact area $A$ between a block and substrate during sliding.
After a stationary contact duration $t_{\rm s}$, the shear stress at the new onset of sliding becomes $\sigma_{\rm s}$,
which we assume to be larger than $\sigma_{\rm k}$, e.g., due to thermally activated interfacial bond formation.
To reduce the shear stress to its kinetic value $\sigma_{\rm k}$, the slip distance must be of the order of the size $D$ of the macroasperity contact regions.
If, due to elasticity, different regions at the interface have already slipped a distance larger than $\lambda_{\rm c} \approx D$  
when full slip occurs, then the breakloose friction will be smaller than $\sigma_{\rm s} A$, which would be the friction force  
if all contact regions slipped simultaneously, as in the case of a rigid solid.

For the situation in Fig. \ref{ZZZ.ThreeCases1.eps},  
one can show that the elastic deformations are characterized by a length scale  
$\lambda_{\rm el} \approx (\sigma_{\rm s} - \sigma_{\rm k}) L^2 / E h$, where $E$ is the Young's modulus,  
$h$ is the block height, and $L$ is its length in the sliding direction.  
If $\lambda_{\rm el} > \lambda_{\rm c}$, the breakloose friction force is smaller than $\sigma_{\rm s} A$,  
and if $\lambda_{\rm el} \gg \lambda_{\rm c}$, it may vanish.  

To illustrate this, Fig. \ref{ZZZ.p.1x.2mu.last.eps} shows the ratio between the tangential (friction) force $F_x$  
and the normal force $F_{\rm N}$ as a function of sliding distance during oscillatory motion of rubber blocks  
made from a passenger car compound sliding on concrete at $T = 20^\circ {\rm C}$.  
The blocks first move forward at $v = 3 \ {\rm mm/s}$ for $20 \ {\rm cm}$,  
then backward at the same velocity, returning to the starting point. This motion is repeated four times.  
Note that the breakloose friction force equals the kinetic friction force at the start of sliding  
and also at the turning points where the slip velocity passes through $v = 0$.  

\begin{figure}
\includegraphics[width=0.35\textwidth,angle=0.0]{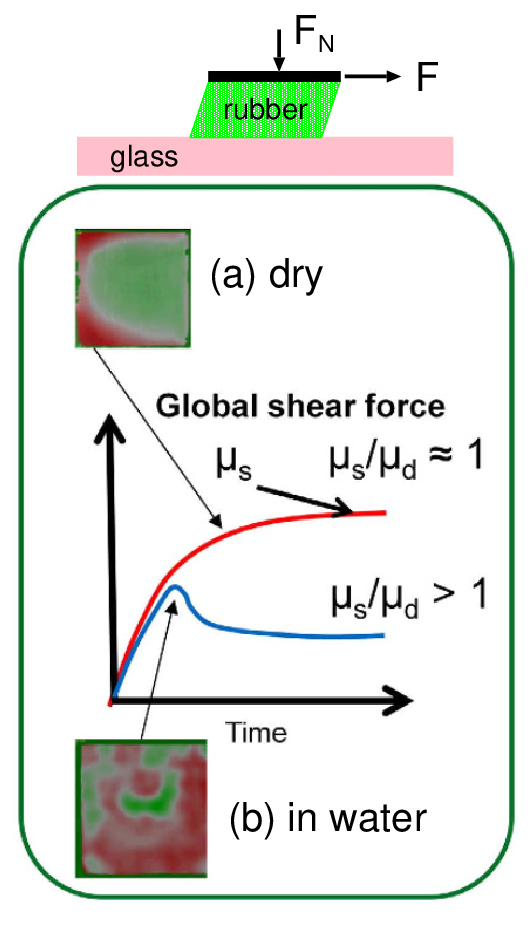}
\caption{\label{Finland.eps}
(a) In the dry state, slip initiates at the trailing edge and propagates forward, resulting in a breakloose friction force that is approximately equal to the kinetic friction force.  
(b) In the presence of water, slip initiates more uniformly along the interface, leading to a breakloose friction force that is larger than the kinetic friction force.  
In the images, green areas indicate stick regions and red areas indicate slip regions.  
Adapted from Ref. \cite{Finland}.
}
\end{figure}

\vskip 0.1cm
{\bf Pre-slip on smooth surfaces}

The study presented above involved rubber sliding on very rough substrate surfaces, but similar results have been observed for rubber in contact with smooth glass plates. 
In a very interesting study by Tuononen \cite{Finland}, a rectangular block of a tire tread compound was glued to a rigid plate and pressed against a smooth glass surface, both in the dry state and under water. 
In the dry state, when a tangential force was applied, slip initiated at the trailing-edge corners and propagated forward toward the leading edge of the block [see Fig.~\ref{Finland.eps}(a), where the green area indicates the no-slip region]. Slip began at the trailing edge because the tangential loading introduces a moment that reduces the contact pressure at the trailing edge. As a result, at the onset of global slip, where the entire interface begins to slip, the slip distance is almost everywhere longer than $\lambda_{\rm c}$, and the breakloose friction force is nearly equal to the kinetic friction force. 

On the wet glass surface, the friction is much smaller. At the onset of sliding, the motion occurs nearly simultaneously across the entire interface. As a result, slip at the interface is more uniform [see Fig.~\ref{Finland.eps}(b)], and the breakloose friction force becomes larger than the kinetic friction force. 

In Ref.~\cite{Finland}, Schallamach waves were not observed for rectangular rubber blocks. These waves are usually studied using hemispherical sliders, which may be necessary to create suitable conditions for buckling. In addition, tread compounds (which contain fillers) are elastically stiffer than the rubber types typically used in studies of Schallamach waves, which may further suppress buckling. 

\begin{figure}
\includegraphics[width=0.45\textwidth,angle=0.0]{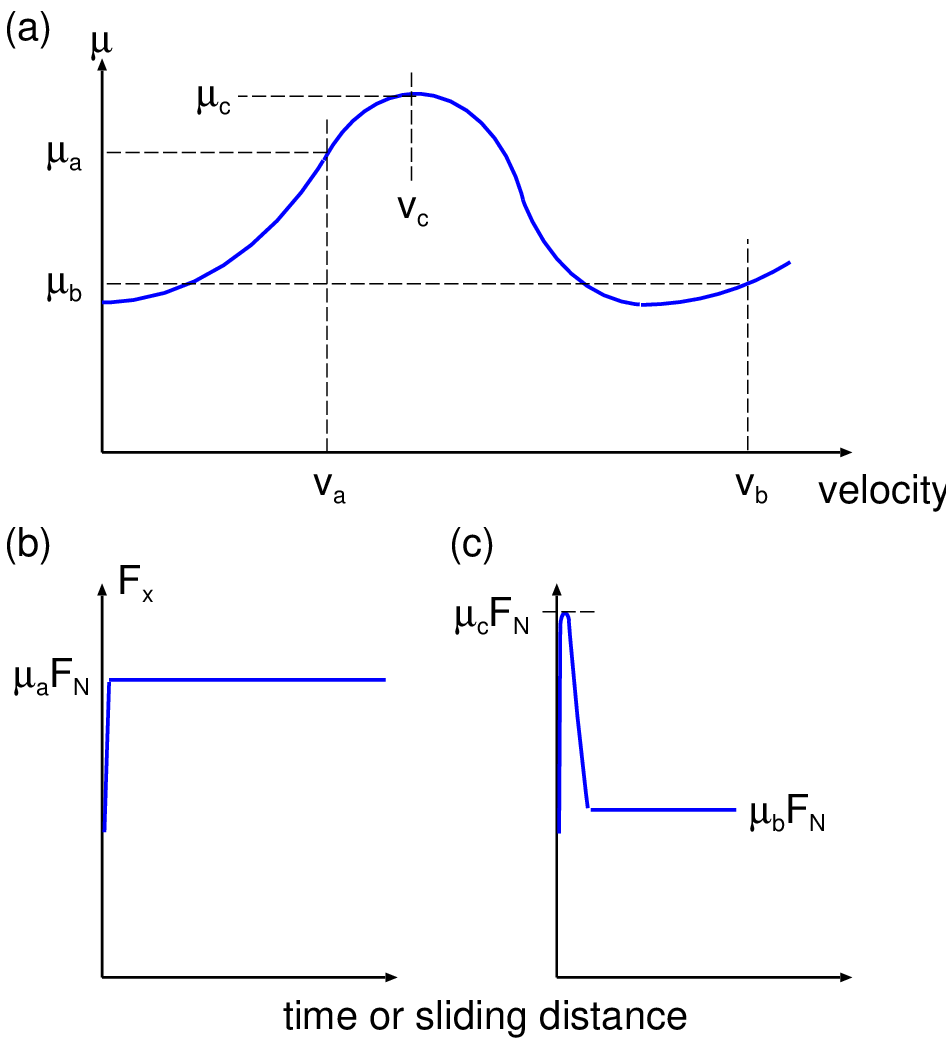}
\caption{\label{ZZZ.EffectiveMu.eps}
(a) Friction coefficient as a function of sliding speed. The $\mu_{\rm k}(v)$ curve reaches a maximum at $v = v_{\rm c}$, where $\mu_{\rm k}(v_{\rm c}) = \mu_{\rm c}$.  
(b) If a rubber block is initially stationary for $t < 0$ and its upper surface begins to move at velocity $v$ for $t > 0$, then the friction force $F_x$ increases rapidly and monotonically to $\mu_{\rm k}(v) F_{\rm N}$ if $v < v_{\rm c}$.  
(c) If $v > v_{\rm c}$, the friction force first increases rapidly to approximately $\mu_{\rm c} F_{\rm N}$ and then decreases quickly to $F_{\rm k} = \mu_{\rm k}(v) F_{\rm N}$.
}
\end{figure}

\vskip 0.1cm
{\bf Effective breakloose force}

Even when $\lambda_{\rm el} \gg \lambda_{\rm c}$,  
the breakloose friction may exceed the kinetic friction. We will now consider two 
{\it kinetic effects} which result in effective breakloose friction 
forces larger than the kinetic friction force. 

The first kinetic effect becomes important if the kinetic friction force $F_{\rm k}(v)$ has a maximum 
at a low sliding speed, say at $v = v_{\rm c}$. If the driving speed $v$ is larger than $v_{\rm c}$, 
then at the onset of slip, the friction force will first increase rapidly to approximately 
$F_{\rm k}(v_{\rm c})$, and then drop quickly to $F_{\rm k}(v)$, resulting in an effective breakloose 
friction force greater than the kinetic friction force.

The second kinetic effect is caused by the flash temperature, which requires a finite slip distance 
(typically $\sim 1 \ {\rm mm}$) to fully develop. Since the friction force tends to increase with increasing 
sliding speed at low slip velocities, and decrease with increasing temperature, the friction rises 
to a maximum before the flash temperature is fully developed.

Fig. \ref{ZZZ.EffectiveMu.eps}(a) shows the friction coefficient as a function of sliding speed acting on an elastic block, assuming that the $\mu_{\rm k}(v)$ curve has a maximum at $v = v_{\rm c}$, with $\mu_{\rm k}(v_{\rm c}) = \mu_{\rm c}$. If the block is stationary for $t < 0$ and the upper surface of the block moves at velocity $v$ for $t > 0$, then the friction force $F_x$ will rapidly and monotonically increase to $\mu_{\rm k}(v) F_{\rm N}$ if $v < v_{\rm c}$ [case (b)].  
However, if $v > v_{\rm c}$, the friction force will initially rise rapidly to approximately $\mu_{\rm c} F_{\rm N}$ and then decrease rapidly to $F_{\rm k} = \mu_{\rm k}(v) F_{\rm N}$ [case (c)]. For a rigid block, these transitions would occur instantaneously with negligible displacement. For an elastic block, the transitions are continuous, but fast enough that the peak force appears as an effective breakloose (or static) friction force. In the first case the effective breakloose friction force equals the kinetic friction force, while in the second case, it approximately equals $\mu_{\rm c} F_{\rm N}$.

As an illustration of this kinetic origin of breakloose friction,  
we consider the sliding of a racing tire compound under the same test conditions as in Fig. \ref{ZZZ.p.1x.2mu.last.eps}.  
Note that for rubber friction on rough surfaces at room temperature,  
the friction force for passenger car tires typically reaches a maximum at a sliding speed of approximately $10 \ {\rm mm/s}$  
(higher than the $3 \ {\rm mm/s}$ used in the experiment),  
while for racing tires, the maximum occurs at much lower speeds, typically around $1 \ {\rm \mu m/s}$  
(see Fig. \ref{WW.RACING.TR12masterwithTheory.1.eps}).

\begin{figure}
\includegraphics[width=0.45\textwidth,angle=0.0]{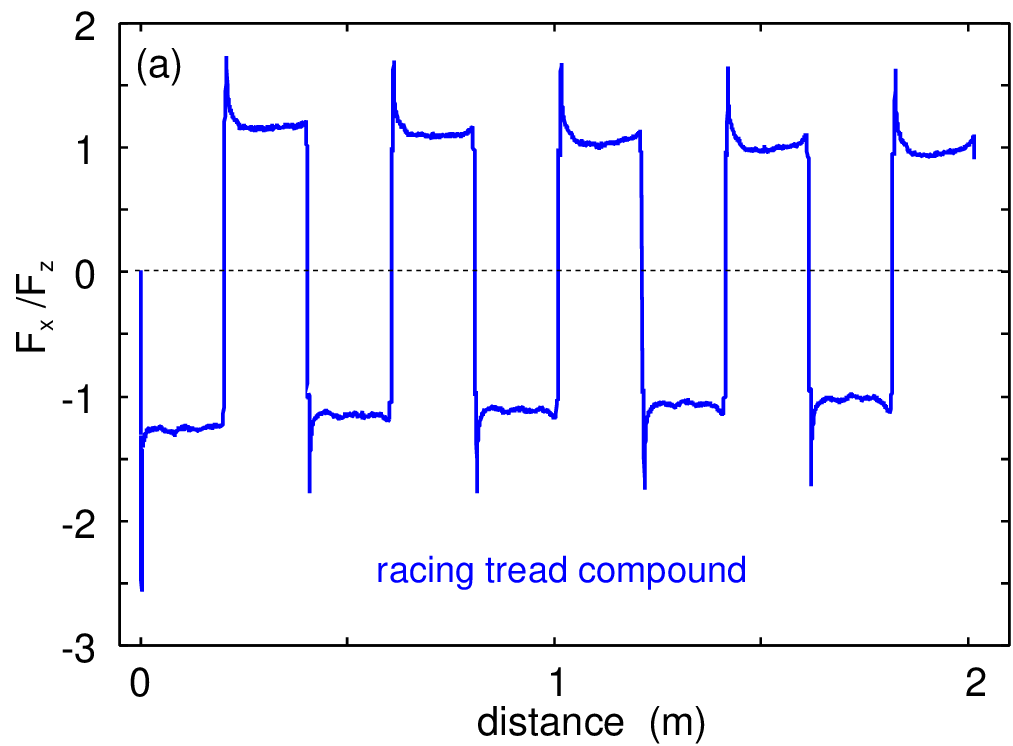}
\caption{\label{ZZZ.r.1x.2mu.last.eps}
The ratio between the tangential (friction) force $F_x$ and the normal force $F_{\rm N}$  
as a function of sliding distance during oscillatory motion of rubber blocks made from a racing compound,  
sliding on concrete at $T = 20^\circ {\rm C}$.  
The rubber blocks first move forward at $v = 3 \ {\rm mm/s}$ for $20 \ {\rm cm}$,  
then backward at the same velocity, returning to the starting point.  
This motion is repeated five times.  
The blocks are $0.5 \ {\rm cm}$ high, with lateral dimensions of $4 \ {\rm cm} \times 4 \ {\rm cm}$.
}
\end{figure}

Fig. \ref{ZZZ.r.1x.2mu.last.eps} shows the ratio $F_x/F_{\rm N}$ at $T = 20^\circ {\rm C}$  
as a function of sliding distance for a racing compound with a glass transition temperature  
$T_{\rm g} = -12^\circ {\rm C}$.  
The rubber block first moves forward at $v = 3 \ {\rm mm/s}$ for $20 \ {\rm cm}$,  
then reverses direction at the same velocity, repeating this cycle five times.  
Notably, large friction peaks occur when the velocity changes sign,  
as the bottom surface of the block undergoes a sharp (but continuous) reversal from $v = 3 \ {\rm mm/s}$ to $-3 \ {\rm mm/s}$.  
This is expected, since for the racing compound, the maximum friction occurs at $v \approx 1 \ {\rm \mu m/s}$.

For the racing compound, the breakloose friction is larger than the friction peak observed during velocity reversal.  
This may be attributed to a larger real contact area at the onset of sliding compared to that during velocity reversal,  
where the system only briefly remains near $v \approx 0$.

Note that the breakloose friction equals exactly \(\mu_{\rm c}F_{\rm N}\) only if the slip velocity at the interface is uniform when the average velocity reaches \(v_{\rm c}\). This is not always the case, as different surface regions may begin to slip at different times. In particular, if \(v_{\rm c}\) is very small, some parts of the interface may move faster than \(v_{\rm c}\) when the average interface speed equals \(v_{\rm c}\). In this case, the effective breakloose friction force will be less than \(\mu_{\rm c}F_{\rm N}\) but still greater than the kinetic friction force \(F_{\rm k}\).

An increase in the contact area with stationary contact duration is also expected for the passenger car compound.  
However, this effect is not evident in the friction curve in Fig. \ref{ZZZ.p.1x.2mu.last.eps}, for the following reason:

For the passenger car rubber compound, the frictional shear stress increases with sliding speed.  
When the velocity reaches the driving speed of $3 \ {\rm mm/s}$,  
the bottom surface of the block has already slipped far enough to renew the asperity contact regions,  
achieving the steady-state sliding contact area. At this speed, the flash temperature effect is negligible,  
and the rubber temperature remains constant.

Many other experimental and theoretical studies have shown that the breakloose (or static) friction force depends  
on the geometry of the contacting bodies and on how the driving force is applied.  
As an example, Zhang and Ciavarella~\cite{Ciav4} (see also Refs.~\cite{Ciav2,Ciav5}) used the  
Cattaneo-Mindlin friction model with Griffith friction  
(where the effect of roughness is included in a fracture energy linearly dependent  
on the local pressure), and predicted that at low normal loads, the breakloose friction coefficient depends on the geometry of the contact.

\begin{figure}
\includegraphics[width=0.2\textwidth,angle=0.0]{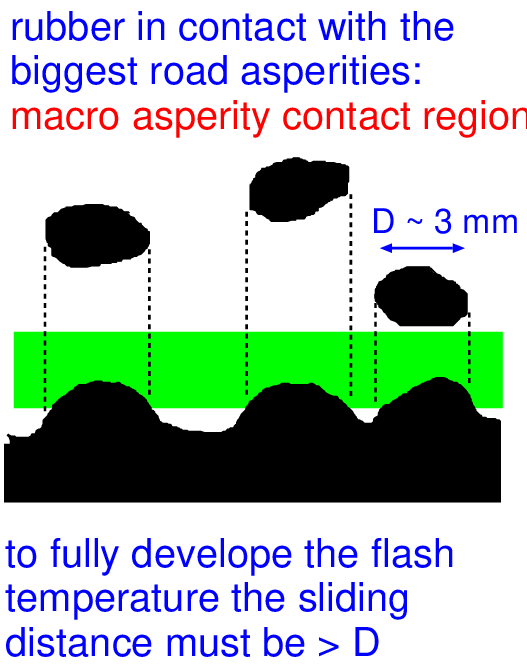}
\caption{\label{WW.macro1b.eps}
The macroasperity contact region (black area) refers to the contact regions observed at
low magnification, where the surface roughness of the large stone particles is not visible.
To fully develop the flash temperature in the asperity contact regions, the slip distance must be
at least as large as the macroasperity contact region.
}
\end{figure}

\begin{figure}
\includegraphics[width=0.45\textwidth,angle=0.0]{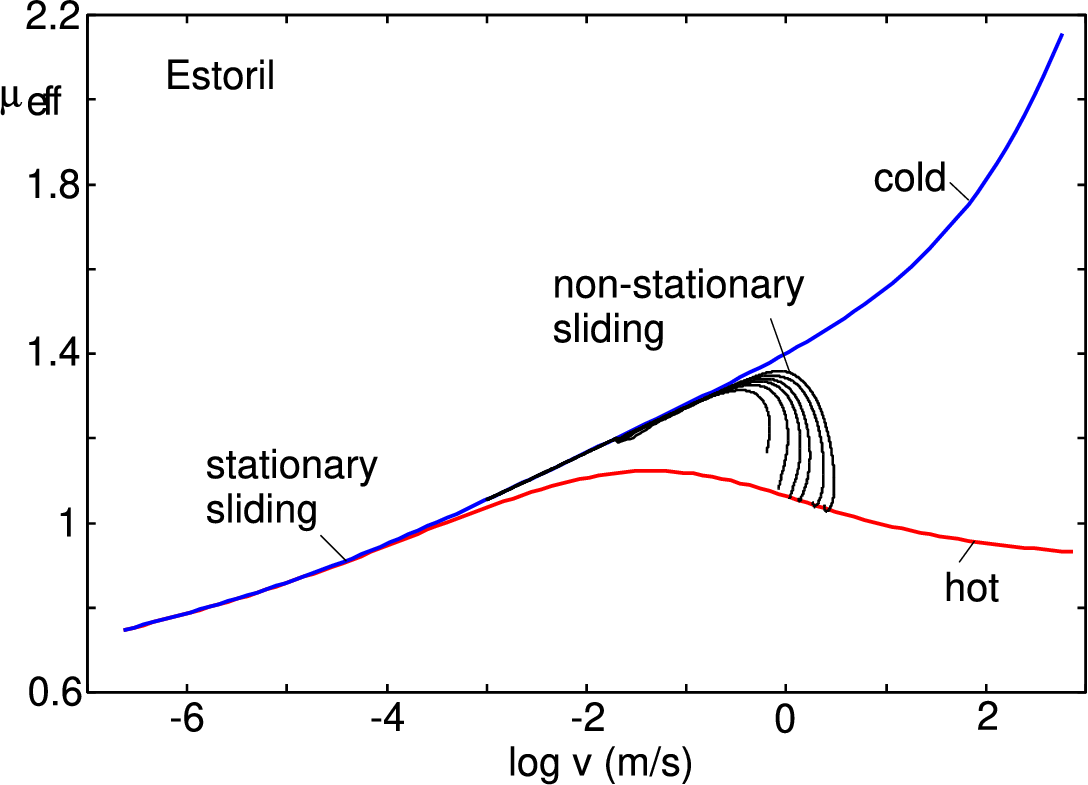}
\caption{\label{WW.Koln.Tire.muslip1.eps}
The friction coefficient as a function of sliding velocity during steady sliding,
neglecting the flash temperature (the {\it cold} branch, blue line) 
and including the flash temperature (the {\it hot} branch, red line).
The tread blocks must slip a distance comparable to the size of the macroasperity contact region 
before the flash temperature is fully developed. The black lines show the friction coefficients 
obtained from tire dynamics calculations experienced by the tread blocks 
as they pass through the tire footprint at different tire slip values.
}
\end{figure}

\begin{figure}
\includegraphics[width=0.45\textwidth,angle=0.0]{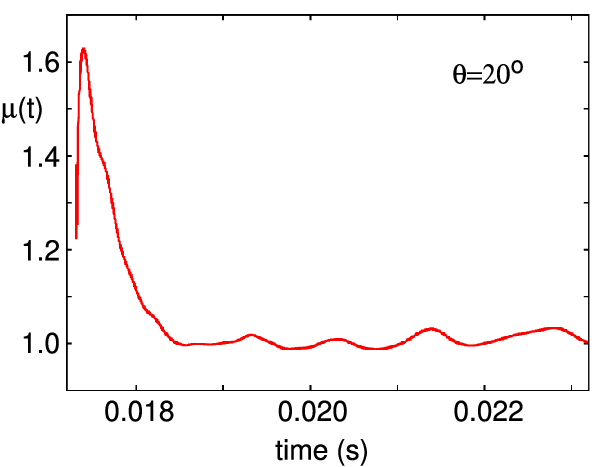}
\caption{\label{good.mu.time.slipangle.20.eps}
The effective (time-dependent) friction coefficient experienced by a tread block 
as it passes through the tire footprint during cornering.
}
\end{figure}

\vskip 0.1cm
{\bf Flash temperature effect}

A second effect which can generate an effective breakloose friction force is the flash
temperature effect \cite{tire}. A tire tread block makes contact with road asperities at many length scales.
The macroasperity contact regions are those observed at
low magnification, where the surface roughness on top of the large stone particles 
cannot be observed (black area in Fig. \ref{WW.macro1b.eps}).
The diameter $D$ of the macroasperity contact regions is typically on the order of $0.1-3 \ {\rm mm}$.
To fully develop the flash temperature in the macroasperity contact regions, the tread
blocks must slip at least a distance comparable to the size of these regions.

For slip distances much smaller than the size of the macroasperity contact regions,
the friction coefficient is determined by the theory without considering the flash temperature.
We denote this branch of the $\mu(v)$ relation as the {\it cold} branch (blue line in
Fig. \ref{WW.Koln.Tire.muslip1.eps}). As the slip distance increases, the flash temperature
begins to develop, and for large slip the friction follows the {\it hot} branch of the $\mu(v)$ relation
(red line in Fig. \ref{WW.Koln.Tire.muslip1.eps}).
The tread blocks must slip a distance on the order of the macroasperity contact region 
before the flash temperature is fully established. The black lines show the tread block velocities as they 
pass through the tire footprint at different tire slip levels, as obtained from numerical simulations of tire dynamics.

Fig. \ref{good.mu.time.slipangle.20.eps} shows the effective friction coefficient experienced by a tread block as it passes through the tire-road 
footprint during cornering (cornering angle $20^\circ$).
The result was obtained from a full tire dynamics simulation. It was found that using the following relation for the
friction coefficient in the tire model gave nearly the same tread block motion as observed when the full dynamical
friction coefficient was used:
$$\mu(t) = \mu_{\rm cold} (v(t),T_0) e^{-s(t)/s_0} $$
$$+ \mu_{\rm hot} (v(t),T_0) \left (1 - e^{-s(t)/s_0} \right )\eqno(28)$$
where $s_0 \approx 0.2 D$, and where $\mu_{\rm cold}$ and $\mu_{\rm hot}$ are the upper and lower (steady-state) friction
branches in Fig. \ref{WW.Koln.Tire.muslip1.eps}.

Including the flash temperature effect described above is crucial in tire dynamics and effectively determines the boundary in the
tire footprint between tread blocks that have undergone significant slip and those that have experienced negligible slip.

\vskip 0.3cm
{\bf 6 Rolling friction for a hard cylinder and sphere on a viscoelastic solid}

We investigate the friction force acting on a hard cylinder or spherical ball rolling on the flat surface of a viscoelastic solid. The rolling friction coefficient depends nonlinearly on both the normal load and the rolling velocity. The results of rolling friction experiments have often been analyzed using a simple model proposed by Greenwood and Tabor \cite{Tabor1}. However, this model includes an unknown factor $\alpha$, representing the fraction of the input elastic energy dissipated due to the internal friction of the rubber. Here, we review 
a simple theory for the friction force acting on a hard cylinder or spherical ball rolling on a flat rubber surface \cite{theory}.

Greenwood and Tabor \cite{Tabor1} have studied the sliding and rolling friction of a hard sphere on a well-lubricated rubber surface. 
They found that nearly the same friction force is observed during sliding as during rolling, under the condition that the interface is 
lubricated and the sliding velocity and fluid viscosity allow for the formation of a thin lubrication film. 
This film must be much thinner than the indentation depth of the sphere but thicker than the amplitude of the surface roughness  \cite{Tabor1}.

Consider a sphere or cylinder rolling with a constant velocity ${\bf v}$ on the surface of a viscoelastic solid, and assume that adhesion can be neglected. The normal stress acting on the rubber will depend on time:
$$
\sigma_z({\bf x}, t) = \sigma_z ({\bf x} - {\bf v}t)
$$
This stress results in time-dependent deformations of the rubber and leads to viscoelastic energy dissipation, 
which gives rise to the rolling friction force \cite{theory}:
$$
F_{\rm f} = {2 \left (2\pi \right )^2\over v} 
\int d^2q \ {\omega \over q} {\rm Im} {1\over E_{\rm eff} (\omega)} 
|\sigma_z({\bf q})|^2\eqno(29)
$$
Here,
$$
\sigma_z ({\bf q}) = {1\over (2 \pi)^2} \int d^2x \ \sigma_z ({\bf x}) {\rm e}^{-i{\bf q}\cdot {\bf x}}
$$
is the Fourier transform of the stress. The effective viscoelastic modulus of the rubber is defined as $E_{\rm eff} = E/(1 - \nu^2)$.

Note that (29) is also valid for a sliding object of arbitrary shape, if the frictional shear stress in the area of real contact can be neglected. 
We will apply (29) in Sec. 7 to a sliding triangular slider. The theory that leads to (29) is highly versatile and can also be applied 
to layered viscoelastic materials. Here, we first apply it to rolling cylinders and spheres.

\vskip 0.1cm
{\bf Cylinder}

Consider a hard cylinder (radius $R$ and length $L_y \gg R$) rolling on a viscoelastic solid. 
The same result is obtained during sliding if one assumes lubricated contact and if the viscous energy dissipation in the lubrication film can be neglected. 
Assuming that the stress at the interface can be approximated by the Hertz stress field for the cylinder-flat contact, one obtains from (29) the friction coefficient:
$$
\mu =  {8 f_{\rm N}\over \pi} \int_0^\infty dq_x \ {\rm Im} {1\over E_{\rm eff} (q_xv)} 
{1\over (a q_x)^2} J_1^2(q_xa)\eqno(30)
$$
The half-width of the contact area in Hertz contact theory is
$$
a = \left( \frac{4 f_{\rm N} R}{\pi E_{\rm eff}} \right)^{1/2} \eqno(31)
$$
where we take $E_{\rm eff}$ to be $|E_{\rm eff}(\omega)|$ with $\omega = {\bf q} \cdot {\bf v}$.

\begin{figure}
\includegraphics[width=0.45\textwidth,angle=0]{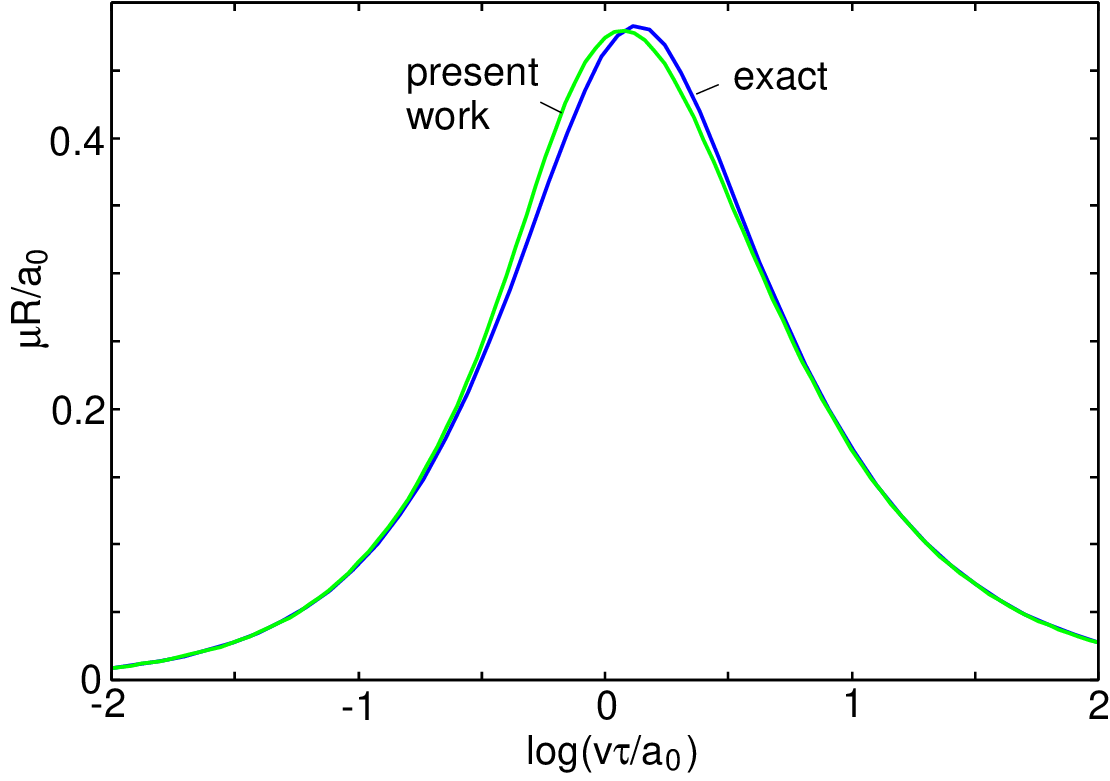}
\caption{\label{compare}
The friction coefficient (multiplied by the radius $R$ of the cylinder and divided by the
half-width $a_0$ of the static contact region) as a function of $v\tau /a_0$, where $v$ is the rolling velocity and $\tau$ the
rubber viscoelastic relaxation time (for $E_1/E_0 = 10$).
We compare the exact result (blue curve) of Hunter \cite{Hunter} with 
the prediction of (30) (green curve).
}
\end{figure}

Let us now assume the simplest possible viscoelastic modulus characterized by a 
single relaxation time $\tau$ [Eq. (4)]:
$$
\frac{1}{E} = \frac{1}{E_1} +  \left( \frac{1}{E_0} - \frac{1}{E_1} \right) \frac{1}{1 - i\omega \tau}
$$
where $E_1/E_0$ is the ratio between the high-frequency and low-frequency 
modulus. 

In Fig. \ref{compare}, we show the
friction coefficient (multiplied by the radius $R$ of the cylinder and divided by the
half-width $a_0 = a(v=0)$ of the static contact region) as a function of $v\tau /a_0$, where $v$ is the rolling velocity.
We have assumed $E_1/E_0 = 10$. 
We compare the exact result (blue curve) of Hunter \cite{Hunter} with 
the prediction of (30) (green curve). Note that some distance away from the maximum, the agreement between the
two curves is nearly perfect. This is expected because these regions correspond to small ${\rm tan} \delta$, where (13) should be essentially exact. Near the maximum, a small difference occurs between the two curves, but from
a practical point of view this is not important, since real rubber exhibits some nonlinearity,
making any linear viscoelastic theory only approximately valid.

\vskip 0.1cm
{\bf Sphere}

Consider now a hard spherical ball (radius $R$) rolling on a viscoelastic solid.
Assuming that the stress at the interface can be approximated by the Hertz stress field for the sphere-flat
contact, one obtains from (29) the friction coefficient:

$$\mu = {9 F_{\rm N}\over 2 \pi^2} \int_0^\infty dq \ q \int_0^{2\pi} d\phi \ 
{\rm cos}\phi \ {\rm Im} {1\over E_{\rm eff} (qv{\rm cos}\phi)}$$
$$ \times
{1\over (qr_c)^6}\left [{\rm sin} (qr_c)-qr_c {\rm cos}(qr_c)\right ]^2\eqno(32)$$

The radius of the contact area in the Hertz contact theory is given by:
$$
r_c = \left( \frac{3 F_{\rm N} R}{4 E_{\rm eff}} \right)^{1/3} \eqno(33)
$$
where we take $E_{\rm eff}$ to be $|E_{\rm eff}(\omega)|$ with $\omega = {\bf q} \cdot {\bf v}$.

Eqs. (30) and (32) are very general and impose no restrictions on the viscoelastic properties of the rubber or on the sliding velocity $v$
(assuming that $v$ is small enough for frictional heating to be negligible and that $v \ll c_{\rm T}$).

\begin{figure}
\includegraphics[width=0.45\textwidth,angle=0]{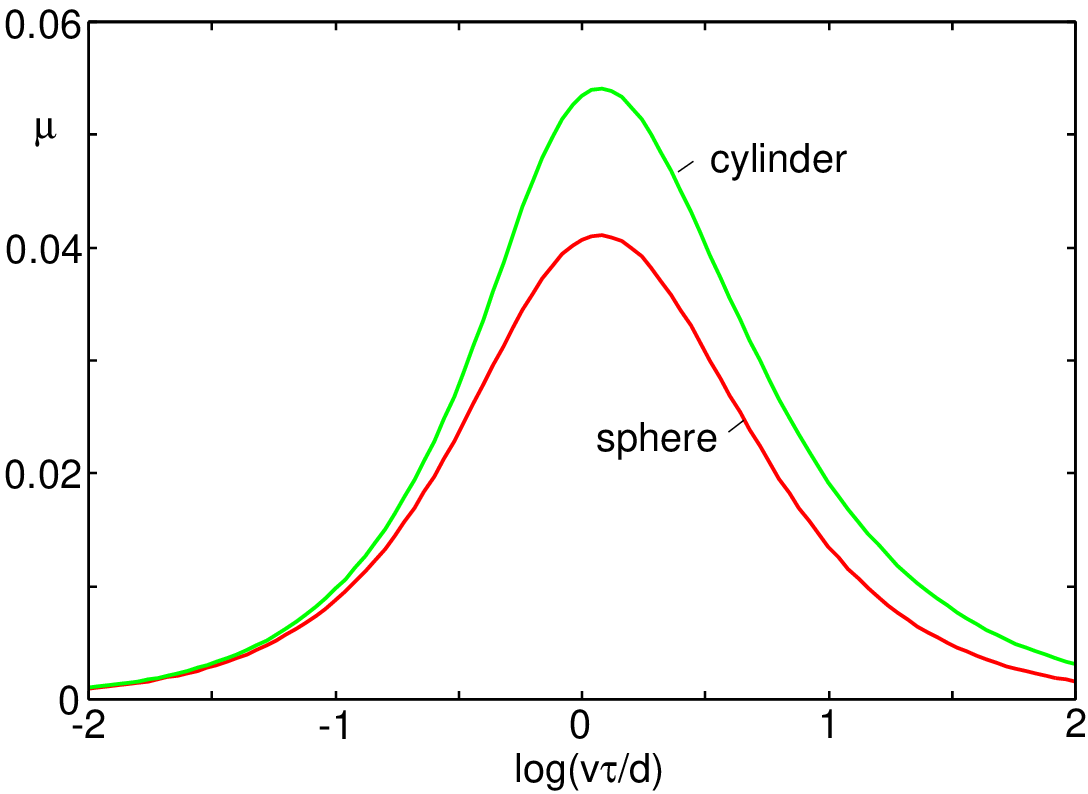}
\caption{\label{mu}
The friction coefficient as a function of $v\tau /d$, where $v$ is the rolling velocity, $\tau$ the
rubber viscoelastic relaxation time, and $d=a_0$ for the cylinder (upper curve) or $d=r_0$ for the sphere (lower curve),
rolling on a rubber substrate described by a simple
viscoelastic model [see Eq. (4)].}
\end{figure}

In Fig. \ref{mu}, we show the rolling friction coefficient 
as a function of $v \tau /r_0$ for a sphere, and as a function of $v \tau /a_0$ for a cylinder. 
Here, we use the simple rheological model described in (4), where $a_0$ is the half-width of the cylinder-substrate contact area and
$r_0$ is the radius of the contact region for the sphere, both defined in the limit of vanishing rolling velocity.
In the calculations, the radius of the sphere and the cylinder are equal, and the applied loads were
chosen such that the average contact pressure under stationary conditions was the same.

\begin{figure}
        \includegraphics[width=0.33\textwidth]{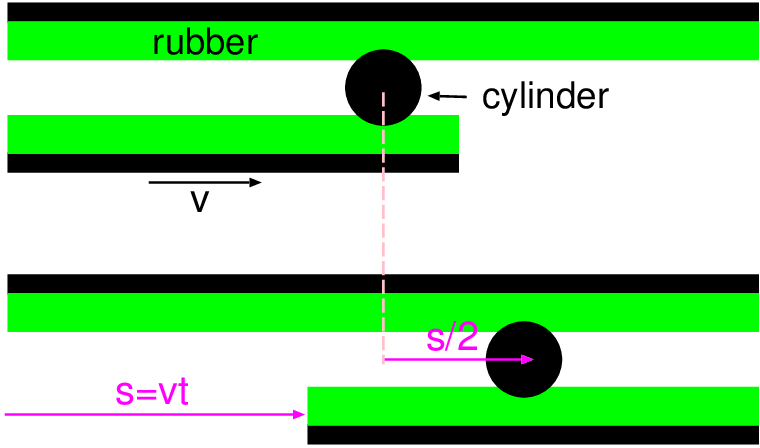}
        \caption{\label{XXX.RollingCylinder.eps}
A rigid cylinder squeezed between two rubber slabs glued to flat steel surfaces.
The lower plate moves with velocity $v$, and the roller exerts a force $F$ on the upper plate.
The rolling friction force is $F = F_{\rm R}$, and the rolling velocity is $v_{\rm R} = v/2$.}
\end{figure}

\begin{figure}[tbp]
\includegraphics[width=0.47\textwidth,angle=0]{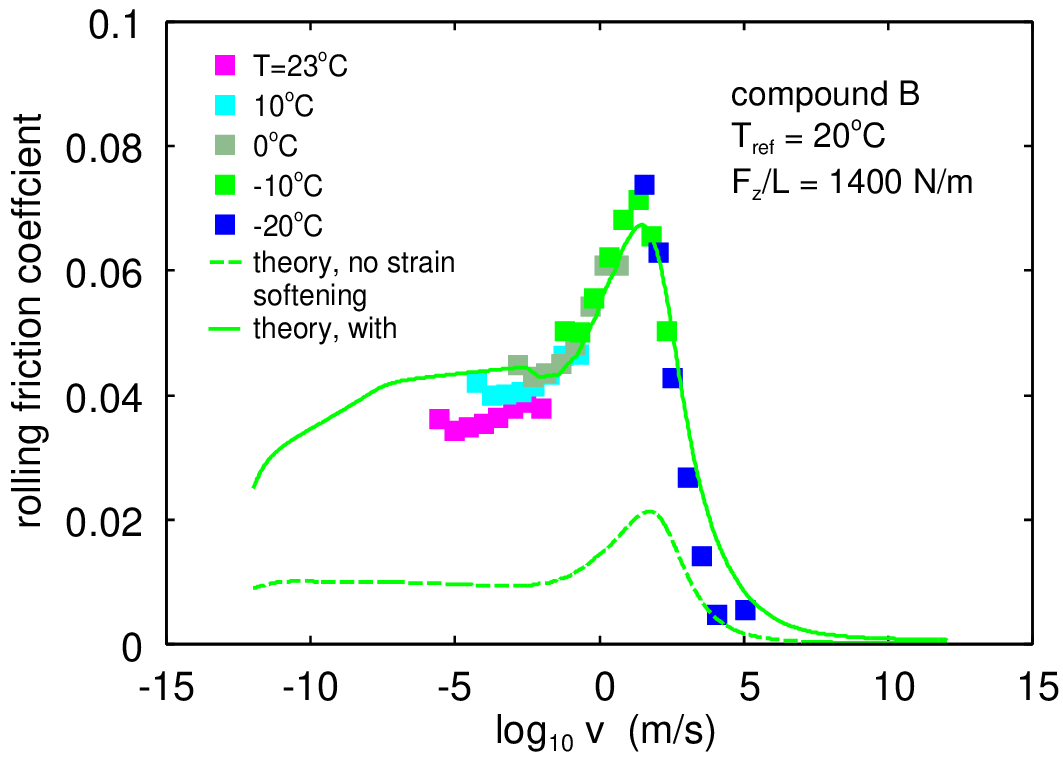}
\caption{
The rolling friction master curve for a tire tread compound. Two Plexiglas cylinder rollers
(each with 10 cm in contact with the rubber) have a diameter of 1 cm.
The squares represent measured rolling friction values at the indicated temperatures, 
shifted using the bulk viscoelastic shift factor $a_T$. 
The solid and dashed green lines represent the calculated rolling friction coefficient 
with and without including strain softening, respectively. 
}
\label{XXX.1logv.2mu.mastercurve.B.eps}
\end{figure}

\vskip 0.1cm
{\bf Comparison with experiment}

Rolling friction experiments \cite{rolling} were performed using the setup shown in Fig. \ref{XXX.RollingCylinder.eps}.
A rigid cylinder was squeezed between two rubber slabs glued to flat steel surfaces.
The lower plate moves with velocity $v$, and the roller exerts a force $F$ on the upper plate.
The rolling friction force is $F = F_{\rm R}$, and the rolling velocity is $v_{\rm R} = v/2$.
To vary the temperature, the entire setup was placed inside a deep freezer capable of cooling down to $-40^\circ {\rm C}$.

Fig. \ref{XXX.1logv.2mu.mastercurve.B.eps} shows the rolling friction master curve for a rubber compound (denoted as compound B).
The squares represent the measured rolling friction values at the indicated temperatures, 
shifted using the bulk viscoelastic shift factor $a_T$.
The solid and dashed green lines are the calculated rolling friction coefficients with and without accounting for strain softening, respectively.
The difference between the measured data (square symbols) and the theoretical prediction (solid green line) 
is likely due mainly to uncertainties in the strain softening measurements 
and the simplified way strain softening is incorporated into the theory (see Sec. 2).
Note the significant influence of strain softening on the magnitude of the rolling friction.

\vskip 0.1cm
{\bf Adhesive contribution to rolling friction}

The theory presented above for rolling friction assumes negligible
adhesion. However, for very smooth and clean surfaces adhesion is important, and for
small normal forces it can give the dominant contribution to the rolling friction force.

Consider a rigid sphere or cylinder in contact with a viscoelastic solid with a flat surface.
When adhesion occurs, the boundary line between the contact and non-contact surface area
can be considered as an adhesive crack tip.
When a cylinder (or sphere) rolls on a rubber surface, the rubber comes into
contact with the cylinder on one side (closing crack) and moves out of contact on the other
side (opening crack). If a crack moves with a finite speed $v$, the energy required to break the bonds at 
the crack tip is in general larger at the opening crack than the energy gained by bond formation
at the closing crack. This adhesion hysteresis can be understood by a simple example.

\begin{figure}
\includegraphics[width=0.40\textwidth,angle=0.0]{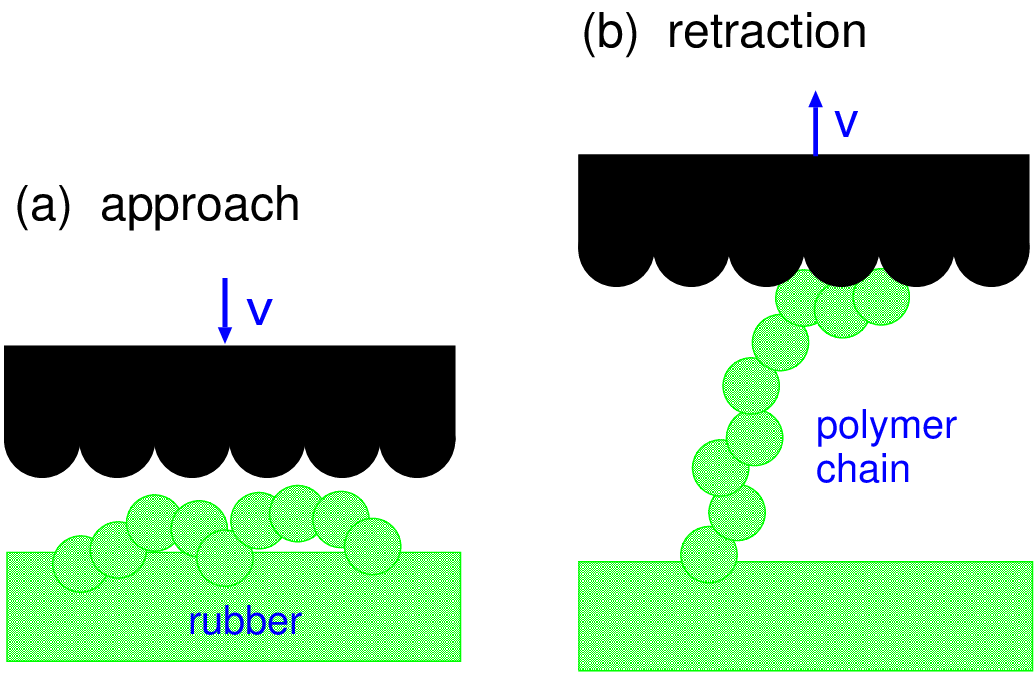}
\caption{\label{PolymerFlatInteraction.eps}
(a) A solid with a flat surface (black) approaching a solid consisting of long-chain molecules.
The solids make contact only when they are at atomic separation. (b) During separation, a chain molecule
is bonded to the upper solid, and the bond breaks only when the chain is nearly fully stretched.
}
\end{figure}

Consider a flat surface approaching a solid consisting of long-chain molecules,
see Fig.~\ref{PolymerFlatInteraction.eps}. At a finite approach
speed $v$, the solids make contact only when they are within the range of the interaction potential
between the solids (usually of order 1 nm or less). This bond formation 
gives an energy gain per unit surface area
denoted by $\gamma_0$. Next, we separate the solids. If the separation velocity is high, the 
bond between the chain molecules and the flat surface breaks only when the chain molecules
are nearly fully stretched. Hence, the work per unit surface area $\gamma_1$ required to separate 
the surfaces may be much larger than $\gamma_0$. In general, $\gamma_0(v)$ and
$\gamma_1(v)$ depend on the speed $v$. For infinitesimally slow
motion ($v \rightarrow 0$), thermal equilibrium occurs at all times and $\gamma_0(0) = \gamma_1(0)$. 

\begin{figure}
\includegraphics[width=0.40\textwidth,angle=0.0]{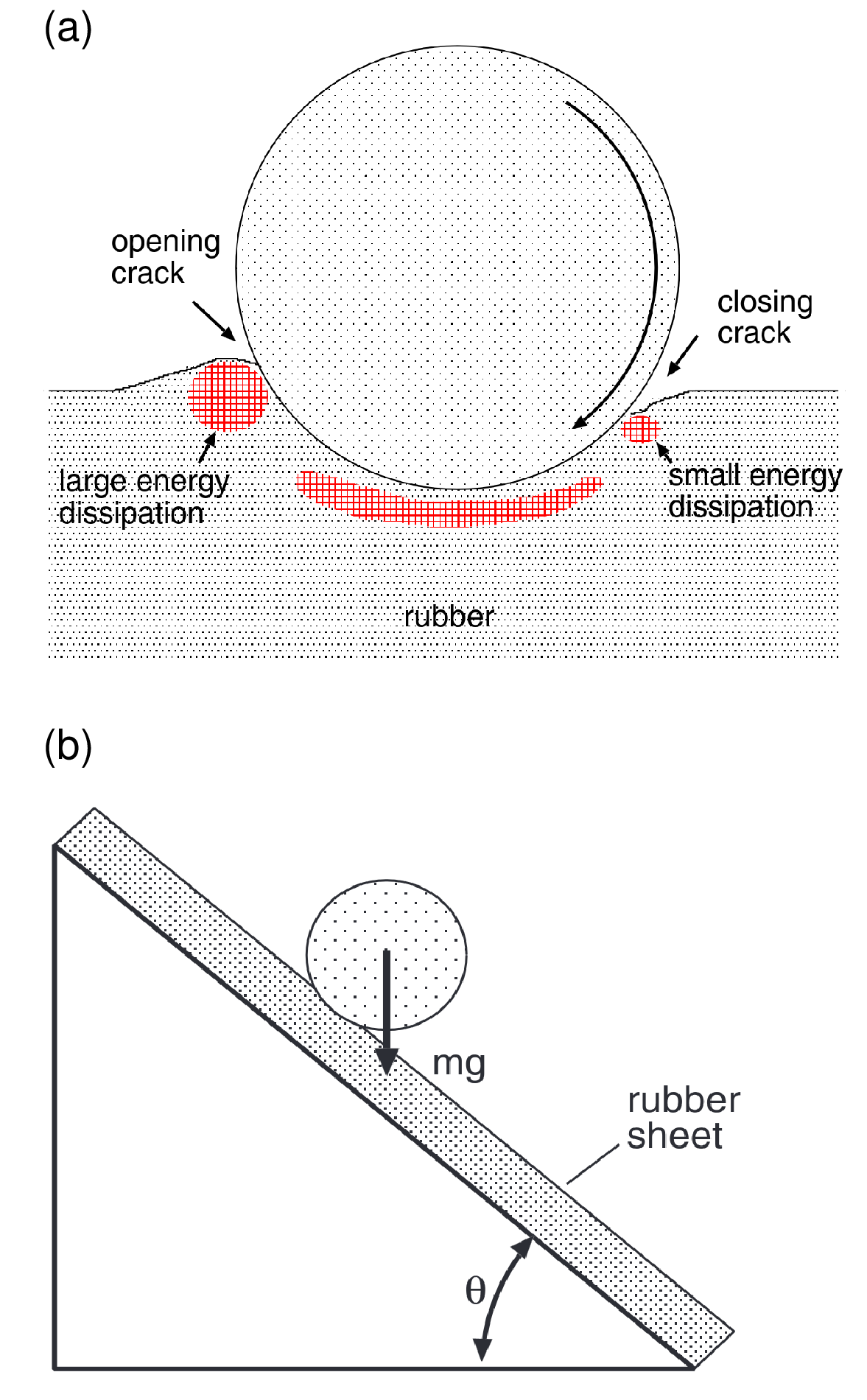}
\caption{\label{RollingBall.eps}
(a) A hard cylinder or ball rolling on a flat rubber track. The regions where most of the
viscoelastic energy dissipation occurs are indicated by the red dashed areas.
(b) A rigid cylinder rolling down a rubber track. From the tilt angle $\theta$ one can
calculate the effective interfacial energy $\gamma_{\rm eff}$. 
}
\end{figure}

\begin{figure}
\includegraphics[width=0.47\textwidth,angle=0.0]{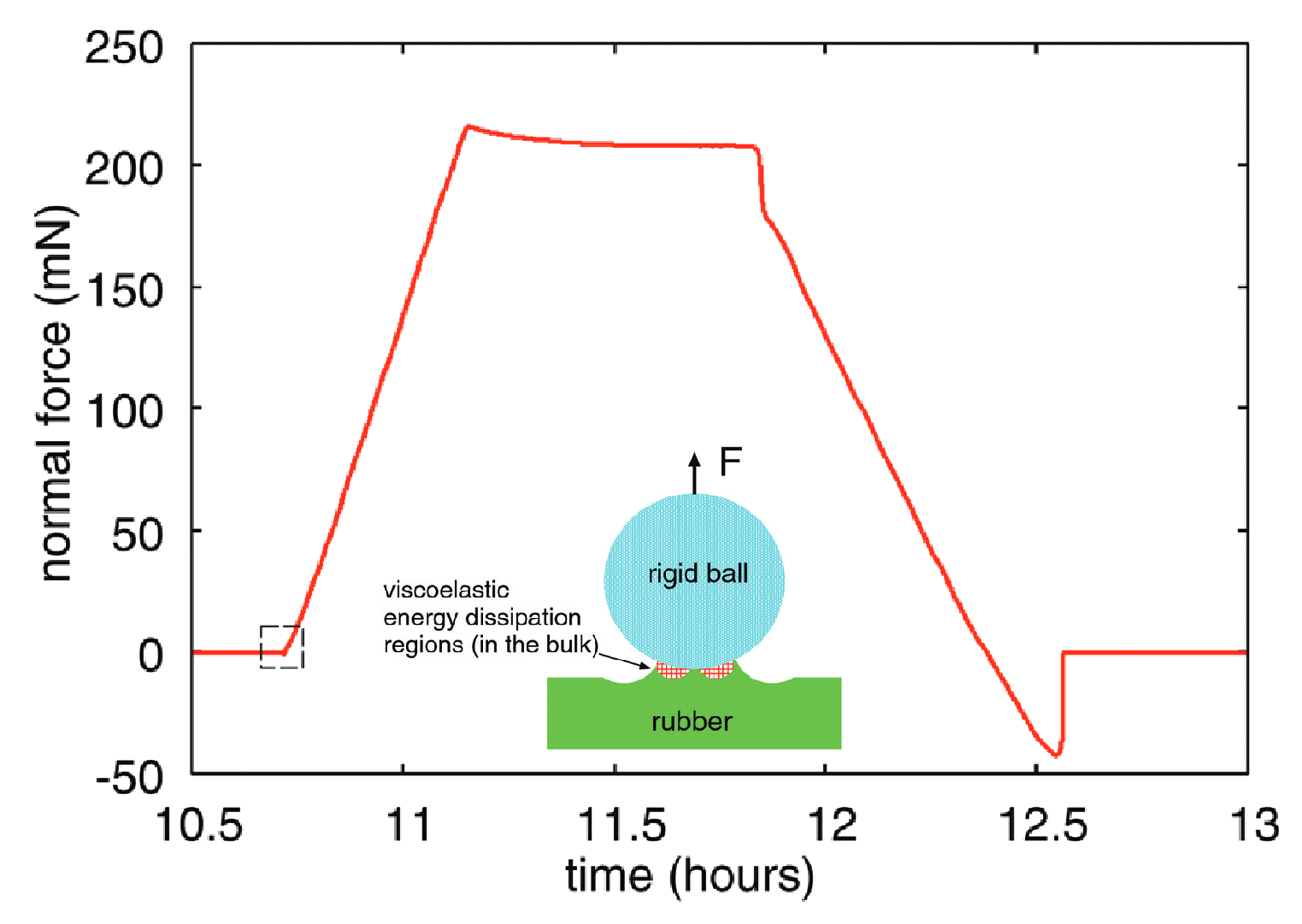}
\caption{\label{MoveBallUpDownHysteresis.eps}
A glass sphere (radius $R= 2 \ {\rm cm}$) brought into contact with a weakly crosslinked 
PDMS rubber surface, kept in contact for about 1 hour, and then removed. The approach and retraction speed
is $0.87 \ {\rm \mu m/s}$. Note that no adhesion force can be observed during contact formation (closing crack)
(dashed rectangle), while a large attractive force is observed during retraction (opening crack). 
The pull-off force is related to $\gamma_{\rm open}$
via the JKR formula $F_{\rm pull-off} = (3\pi /2) \gamma_{\rm open}$. Adapted from Ref.~\cite{WithGreg}. 
}
\end{figure}

For viscoelastic solids, the opening crack propagation energy $\gamma_{\rm open}(v)$ usually increases
rapidly with increasing crack-tip velocity due to viscoelastic energy dissipation close to the crack 
tip [see Fig.~\ref{RollingBall.eps}(a)]~\cite{Brener}.
Hence, $\gamma_{\rm open}(v)$ can be much larger than $\gamma_1(v)$. Similarly, viscoelastic energy dissipation close to the 
closing crack tip reduces the energy gain from bond formation, so that $\gamma_{\rm close}(v)$ may be much
smaller than $\gamma_0(v)$. Rolling friction depends on the difference 
$\gamma_{\rm open}(v)-\gamma_{\rm close}(v) \approx \gamma_{\rm open}(v)$
(see below), so no direct information about $\gamma_{\rm close}(v)$ can be obtained from rolling friction experiments.
However, both $\gamma_{\rm open}(v)$ and $\gamma_{\rm close}(v)$ can be deduced from adhesion experiments in which a spherical ball is moved
into and out of contact with the viscoelastic solid. One example is shown in Fig.~\ref{MoveBallUpDownHysteresis.eps}, where negligible adhesion
occurs on approach (closing crack), while strong adhesion is observed on retraction.

The adhesive contribution to the rolling friction of a cylinder on a flat viscoelastic solid
can be obtained as follows: if the cylinder has length $L$ and rolls a distance 
$\Delta x$, the opening and closing crack tips move over the surface area $A = L \Delta x$,
so the energy dissipated at the opening crack tip is $\gamma_{\rm open}(v) L \Delta x$ and the
energy gained at the closing crack tip is $\gamma_{\rm close}(v) L \Delta x$, giving the net energy
loss
$$\Delta E = [\gamma_{\rm open}(v) - \gamma_{\rm close}(v)] L \Delta x$$
If $f_{\rm R}$ is the rolling force per unit length, the rolling friction energy is $f_{\rm R} L \Delta x$,  
giving
$$f_{\rm R} = \gamma_{\rm open}(v) - \gamma_{\rm close}(v) = \gamma_{\rm eff}(v)$$
Note that $f_{\rm R}$ is independent of the normal force, so the rolling friction coefficient
$\mu = f_{\rm R}/f_{\rm N}$ increases without limit as $f_{\rm N} \rightarrow 0$. 
This differs from the viscoelastic contribution studied before [see (30)], 
where $f_{\rm R}$ depends nonlinearly on $f_{\rm N}$ (typically $f_{\rm R} \sim f_{\rm N}^{1/2}$ for a cylinder
and $F_{\rm R} \sim F_{\rm N}^{1/3}$ for a sphere), and in 
particular $f_{\rm R} \rightarrow 0$ as $f_{\rm N} \rightarrow 0$. Hence, the adhesive contribution dominates the
rolling friction for sufficiently small normal loads. In Ref.~\cite{Crack1}, an equation was derived for 
the load where the rolling friction crosses over from the adhesive to the viscous limiting case:
$$\gamma_{\rm eff} \approx \left( {\sigma_0 \over E_0} \right)^3 R {\rm Im}E (\omega_0)\eqno(34)$$
where $\sigma_0$ is the average contact pressure, and $\omega = r/d$, where $d$ is the width of the
contact regions. $\sigma_0$ and $d$ can be estimated using Hertz contact theory. (34) shows that
for large enough contact pressure the viscoelastic bulk contribution will dominate.

\begin{figure}
\includegraphics[width=0.45\textwidth,angle=0.0]{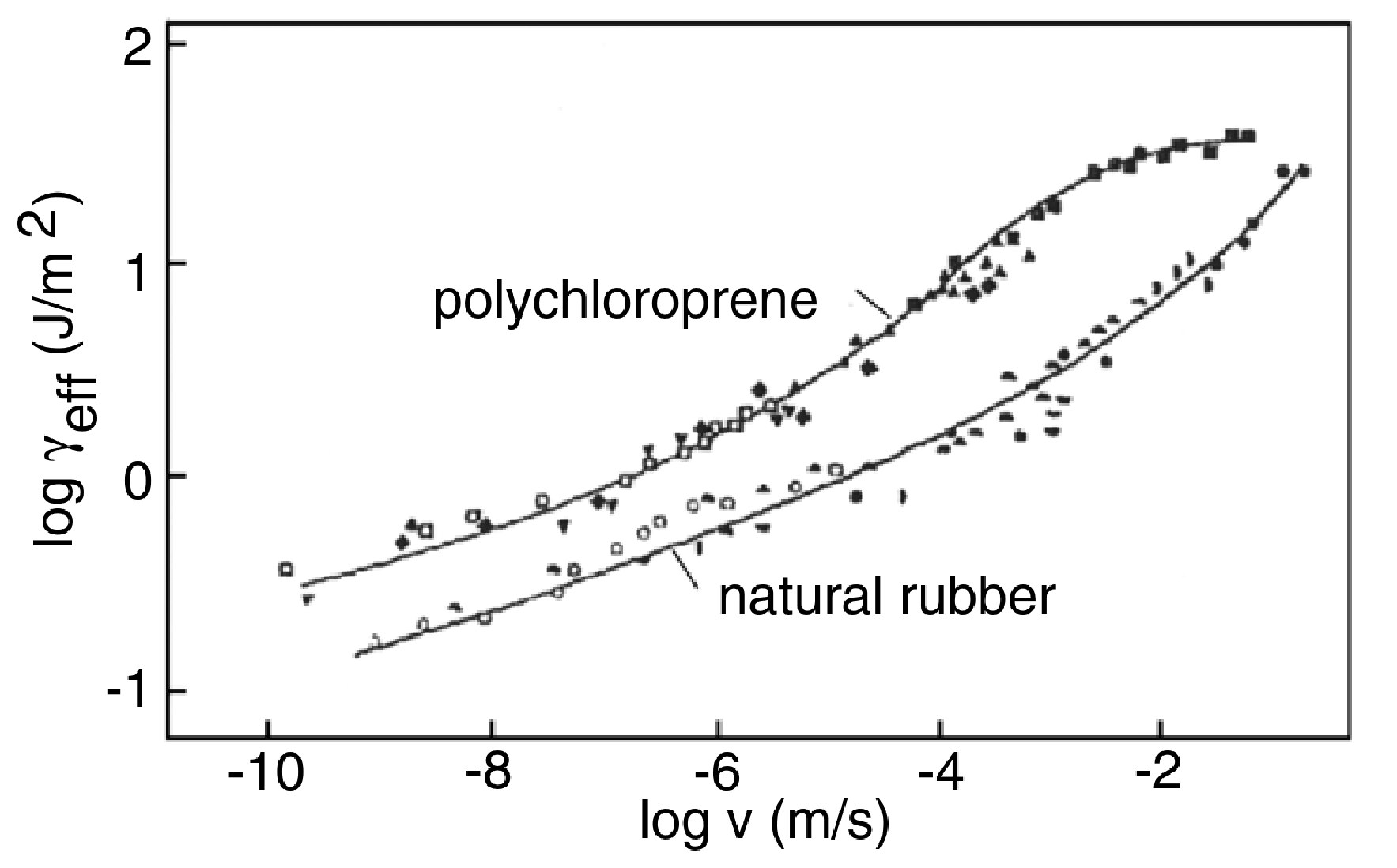}
\caption{\label{EffectiveGamma.eps}
The logarithm (base 10) of the effective interfacial energy $\gamma_{\rm eff}(v)$ as a function
of the logarithm of the crack-tip velocity $v$, deduced from rolling experiments with glass cylinders on smooth
plane rubber tracks. The different symbols correspond to different glass cylinder loads ($0$, $6$, $18$,
$28$, and $102 \ {\rm g}$), where increasing velocity corresponds on average to increasing load.
Results are shown for natural rubber and polychloroprene rubber tracks. Adapted from Ref.~\cite{Thom}.
}
\end{figure}

The rolling friction can be measured using a tilted substrate, as indicated in Fig.~\ref{RollingBall.eps}(b).
The tangential component of the gravitational force acting on the cylinder is $F = mg \sin\theta$, which
must equal the friction force $L f_{\rm R}$ during rolling at constant speed. This gives
$\gamma_{\rm eff}(v) = (mg / L) \sin\theta$. We note that steady rolling occurs only if $\gamma_{\rm eff}(v)$ 
increases with the crack-tip velocity $v$, which is almost always the case for rubber due to the strong increase
in viscoelastic energy dissipation at the crack tip with increasing velocity.

Roberts and Thomas~\cite{Thom} performed rolling friction experiments with glass cylinders on smooth
plane rubber tracks. Fig.~\ref{EffectiveGamma.eps} shows $\gamma_{\rm eff}(v)$ 
for natural rubber and polychloroprene as a function
of the crack-tip velocity $v$ (log-log scale).
The different symbols correspond to different glass cylinder loads ($0$, $6$, $18$,
$28$, and $102 \ {\rm g}$). As the cylinder mass increases, the rolling velocity increases,
and the fact that all data points collapse onto a common master curve shows that the rolling friction is independent of
the normal force and depends only on the rolling speed.

\begin{figure}[tbp]
\includegraphics[width=0.30\textwidth,angle=0.0]{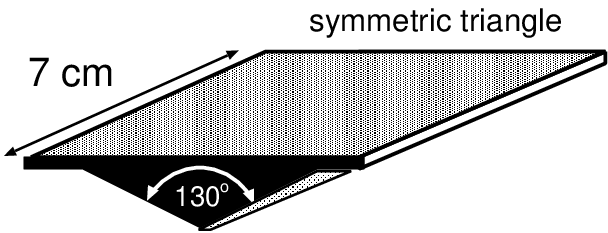}
\caption{\label{TwoTriangles.eps}
The triangular steel slider (wedge) used in the rubber friction experiments.
}
\end{figure}

\begin{figure}[tbp]
\includegraphics[width=0.45\textwidth,angle=0.0]{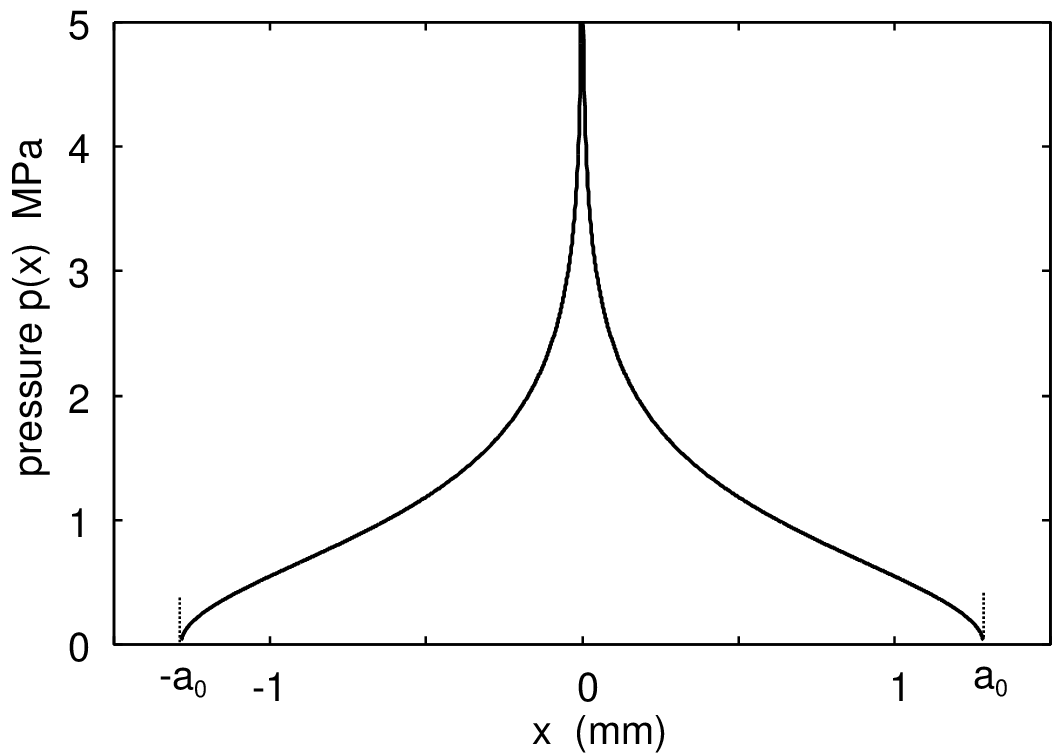}
\caption{\label{1x.2p.TRIANGLE.eps}
The pressure distribution $p(x)$ as a function of $x$ for the symmetric indenter pressed
against the rubber with a normal force per unit length $F_{\rm N}/L_y = 3000 \ {\rm N/m}$.
The effective elastic modulus, including a strain softening factor of 0.2, is taken as
$E_{\rm eff} (1/t_0,\epsilon) = 25 \times 0.2 = 5 \ {\rm MPa}$, corresponding to a contact time of approximately
$t_0 \approx 1000 \ {\rm s}$. The indenter tip is assumed to be perfectly sharp, i.e., $r_{\rm tip}=0$.
The width of the contact region is $w = 2a_0 \approx 2.5 \ {\rm mm}$.
}
\end{figure}

\vskip 0.3cm
{\bf 7 Triangular slider on rubber surfaces} 

Here we investigate the sliding friction of a rigid triangular steel slider (wedge) on a rubber 
surface under both dry and lubricated conditions \cite{triangle}. 
The substrate is a carbon-filled styrene-butadiene rubber (SBR) compound 
with a glass transition temperature of $T_{\rm g} \approx -50^\circ {\rm C}$.
We first consider lubricated friction at low sliding speeds, assuming that the frictional
shear stress in the area of real contact is negligible. 
The experimental results are compared to the theory 
developed in Ref. \cite{theory} for rolling friction,
which is also valid for lubricated sliding at low speeds 
if the frictional shear stress can be neglected. 

We use the symmetric slider shown in Fig. \ref{TwoTriangles.eps}.
The profile has a half-opening angle $\alpha = 65^\circ$ and is pressed against a flat rubber sheet with a
normal force per unit length $f_{\rm N} = F_{\rm N} /L_y$. 
The pressure distribution between the slider and the rubber is given by \cite{Johnson}
$$
p(x)= p_0 \, {\rm cosh}^{-1} \left( \frac{a}{x} \right)
$$
where $p_0 = f_{\rm N}/\pi a$. Fig. \ref{1x.2p.TRIANGLE.eps} shows the pressure distribution $p(x)$
taking into account the influence of strain softening on the elastic modulus.
The half-width of the contact region is
$$
a = \frac{f_{\rm N}}{E^* \, {\rm cot}\alpha},
$$
where $E^* = E/(1-\nu^2)$.
When used in (29), $E^*$ is replaced with $|E_{\rm eff} (\omega ,\epsilon)|$, 
with $\omega = {\bf q}\cdot {\bf v}$, so the contact 
width $w=2a$ depends on the sliding speed $v$. 

\begin{figure}[tbp]
\includegraphics[width=0.45\textwidth,angle=0.0]{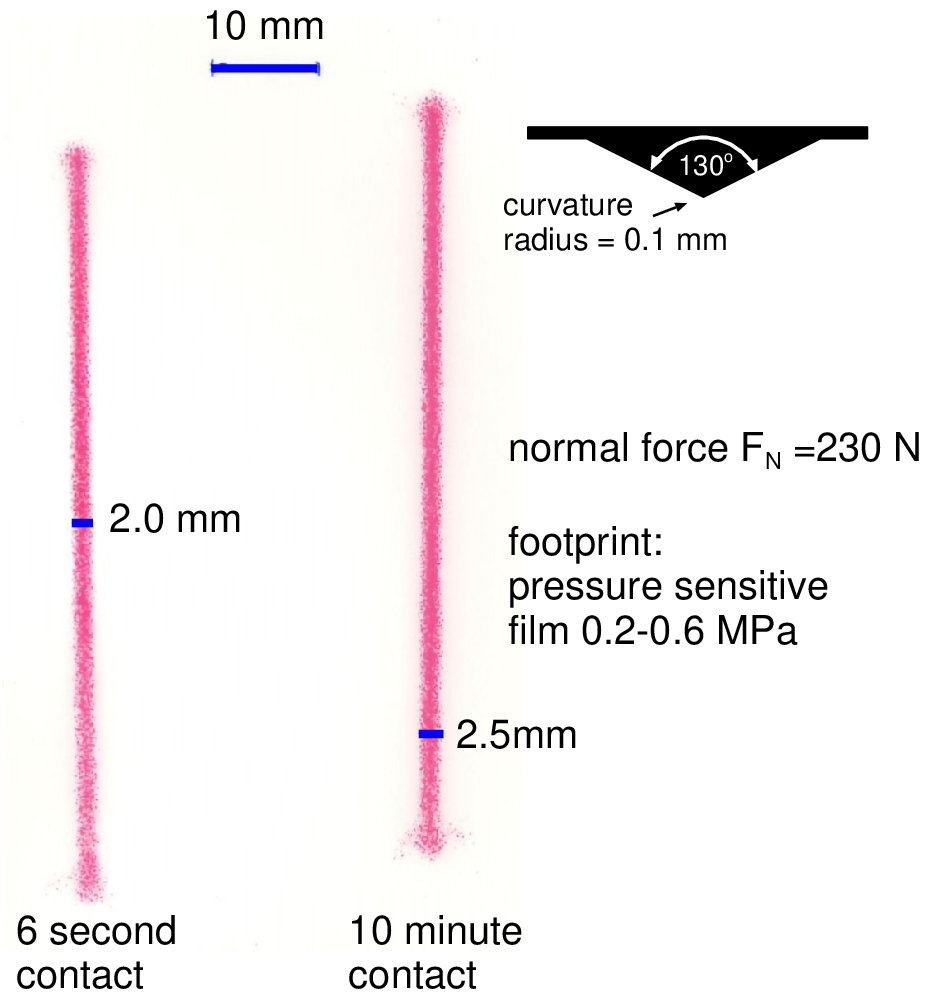}
\caption{\label{triangle.eps}
The wedge loaded against the rubber surface with a pressure-sensitive film placed in between.
The two pressure footprints correspond to contact times of $6 \ {\rm s}$ and $600 \ {\rm s}$, respectively.
The short blue bars indicate contact widths of 2 and 2.5 mm.
}
\end{figure}

Fig. \ref{triangle.eps} shows two pressure footprints of the contact between the slider
and the rubber substrate obtained using a pressure-sensitive film. 
The slider was loaded against the rubber surface 
with $f_{\rm N} \approx 3000 \ {\rm N/m}$.
The two footprints correspond to contact times of $6 \ {\rm s}$ and $600 \ {\rm s}$, 
with widths (indicated by the blue bars) of 2 and 2.5 mm, respectively.

The strain in the region where viscoelastic deformations occur varies spatially,
but it can be shown to be of the order $\epsilon \approx 1/ (2 \, {\rm tan}\alpha)$,
which in our case gives $\epsilon \approx 0.25$.
In the analysis below, we present theoretical results that include
strain softening, using the effective 
modulus (see Sec. 2) $E_{\rm eff} (\omega,\epsilon)$
for $\epsilon \approx 0.25$. 
For this relatively large strain, the nonlinear (strain softening) correction is significant,
increasing the predicted sliding friction by approximately a factor of 5.

\begin{figure}[tbp]
\includegraphics[width=0.45\textwidth,angle=0.0]{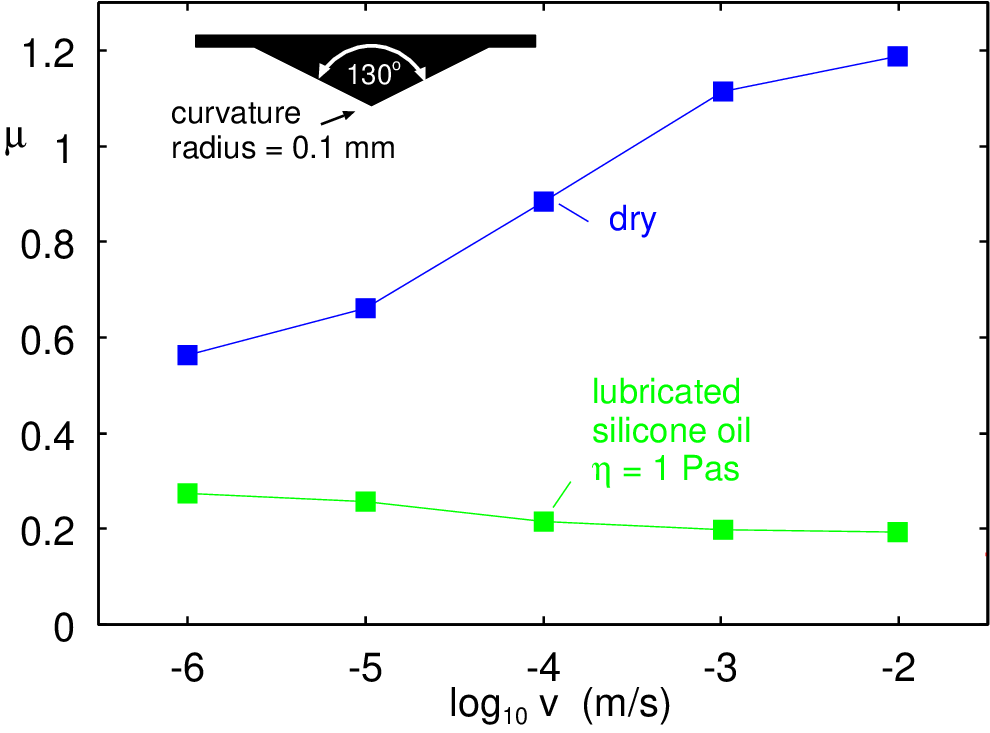}
\caption{\label{logv.2mu.WithExp1.eps}
Sliding friction for the wedge on rubber under dry (blue squares) and lubricated (green squares) conditions.
}
\end{figure}

Fig. \ref{logv.2mu.WithExp1.eps}
shows the measured sliding friction for the wedge on dry (blue squares) and lubricated (green squares) rubber.
The lubricated surface was prepared by spreading a teaspoon of silicone oil with a viscosity of $1 \ {\rm Pa \cdot s}$ over the rubber.
Due to the low sliding speed and high contact pressure, the silicone oil acts as a boundary lubricant (without hydrodynamic effects).

\begin{figure}[tbp]
\includegraphics[width=0.45\textwidth,angle=0.0]{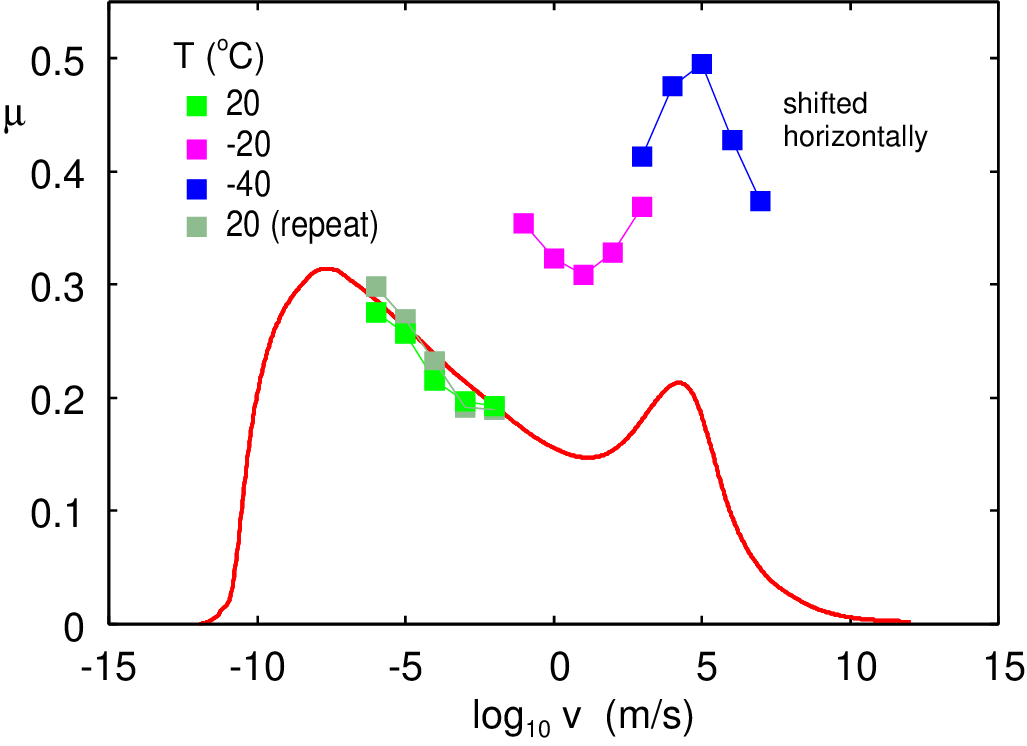}
\caption{\label{1logv.2mu.all.temp.shift1.eps}
Sliding friction for the wedge on lubricated rubber at $T = 20^\circ {\rm C}$ (green), $-20^\circ {\rm C}$ (pink), and $-40^\circ {\rm C}$ (blue), 
along with a repeated measurement at $20^\circ {\rm C}$ (dark green). 
The data at $-20^\circ {\rm C}$ and $-40^\circ {\rm C}$ are shifted horizontally using the bulk viscoelastic shift factor. 
The red line shows the theoretically predicted viscoelastic contribution, assuming a tip radius of $r = 0.1 \ {\rm mm}$.
}
\end{figure}

Fig. \ref{1logv.2mu.all.temp.shift1.eps} shows the measured sliding friction on the lubricated rubber substrate at 
different temperatures: $T = 20^\circ {\rm C}$ (green), $-20^\circ {\rm C}$ (pink), $-40^\circ {\rm C}$ (blue), and a repeated measurement at $20^\circ {\rm C}$ (dark green).
The red line represents the theoretically predicted viscoelastic contribution to the friction. 
The measured data at $-20^\circ {\rm C}$ and $-40^\circ {\rm C}$ are shifted horizontally using the viscoelastic bulk shift factor,
so that they correspond to equivalent friction values at $20^\circ {\rm C}$ but at higher sliding speeds.
However, these measured friction values lie significantly above the theoretical prediction (red curve).

We attribute the enhanced friction at $-20^\circ {\rm C}$ and $-40^\circ {\rm C}$ to the high contact pressure resulting from the increased elastic modulus of 
the rubber at low temperatures, which may lead to penetration of the lubricant film by surface asperities. 
Nevertheless, the sliding friction at $-20^\circ {\rm C}$ and $-40^\circ {\rm C}$ remains lower than expected for dry surfaces, 
indicating that the lubricant film continues to reduce friction even under these colder conditions.

Assuming, as supported by the theory, that the sliding friction on the lubricated surface 
at room temperature is dominated by viscoelastic effects, we can estimate the shear stress 
acting in the area of real contact under dry conditions using
$$\sigma_{\rm f} (v) = [\mu (v)-\mu_{\rm visc} (v)] {F_{\rm N} \over w(v)L_y}\eqno(35)$$
Here, $w(v)$ is the width of the contact region, $\mu(v)$ is the total friction coefficient, and
$\mu_{\rm visc}$ is the viscoelastic contribution, as given by the green and blue curves
in Fig. \ref{logv.2mu.WithExp1.eps}. Using this equation gives a frictional shear stress
$\sigma_{\rm f}(v)$ that increases nearly linearly with the logarithm of the sliding speed,
as expected for thermally activated processes in the low-velocity regime.

\begin{figure}[tbp]
\includegraphics[width=0.45\textwidth,angle=0.0]{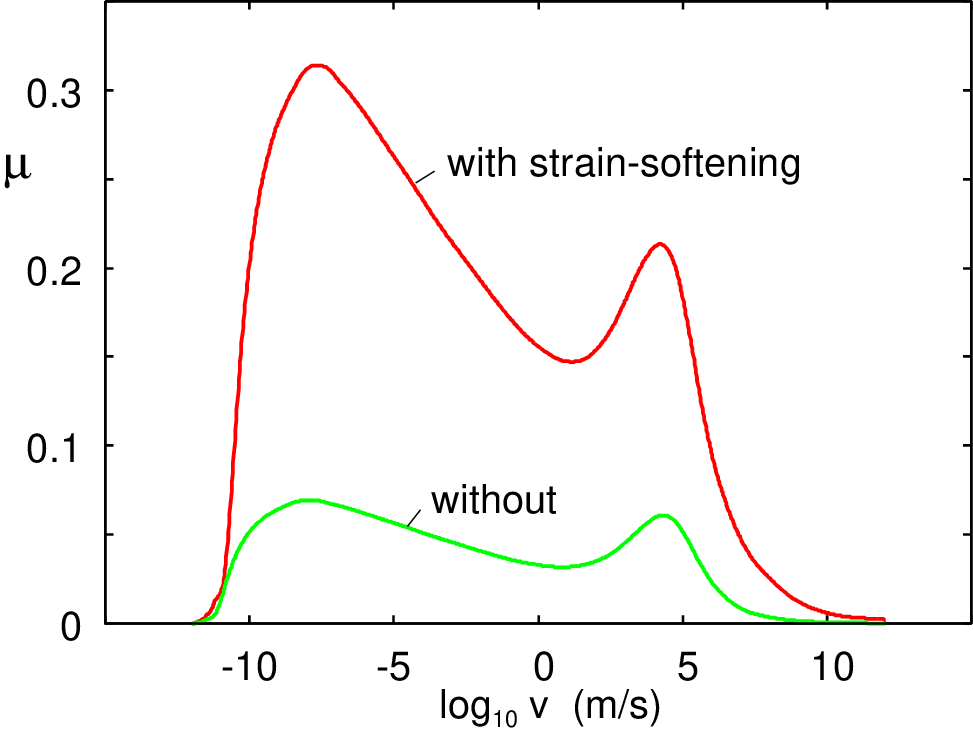}
\caption{\label{1logv.2muROLLING.eps}
Calculated sliding friction coefficient for the wedge, using the measured tip radius $r = 0.1 \ {\rm mm}$. 
The red curve includes the effect of strain softening (nonlinear modulus), while the green curve uses the linear viscoelastic modulus (low-strain response).
}
\end{figure}

Rubber filled with reinforcing particles exhibits strong strain softening, as shown in Fig. \ref{1strain.2reduction.eps} for the compound used in this study.
It is crucial in studies of rolling and sliding friction to account for the rapid decrease in effective modulus with increasing strain. 
This has been emphasized previously in the context of rolling friction and sliding friction. 
In the present study, the strain in the contact region is approximately $0.25$, which is similar to the strain encountered in asperity contacts when a tire is in contact with a road surface. 
Fig. \ref{1logv.2muROLLING.eps} compares the calculated friction using the linear response modulus $E(\omega)$ 
with the result obtained using the effective modulus $E_{\rm eff} (\omega, \epsilon)$ for the strain $\epsilon=0.25$.
The calculated result that includes strain softening yields a friction value approximately five times higher than the result without softening. 
A similarly strong effect of strain softening is observed in rolling friction on rubber with filler particles. 
Strain softening is equally important for static contacts, and the width of the wedge footprints in Fig. \ref{triangle.eps} can only be reproduced
by the theory if both strain softening and the frequency dependence of the effective viscoelastic modulus are taken into account.

In Ref.~\cite{Ciav1}, the authors studied the sliding of 
non-cylindrical wedges on a viscoelastic half-space.  
Using the boundary element method, they found that for wedges with the
height profile $h(x) \sim |x|^k$, with $k = 4, 6,$ and $8$, the viscoelastic contribution
to the friction force differs from the prediction
of (30). However, if the friction coefficients are scaled by
$E_0/p_0$, where $E_0$ and $p_0$ are the modulus and mean pressure at zero speed, 
nearly identical results are obtained for all wedges, including the cylindrical
wedge (where $h(x) \sim x^2$).  
Hence, the friction coefficients as a function of sliding
speed can be obtained from the cylindrical case, where (30) is accurate, by scaling.

\begin{figure}[tbp]
\includegraphics[width=0.45\textwidth,angle=0]{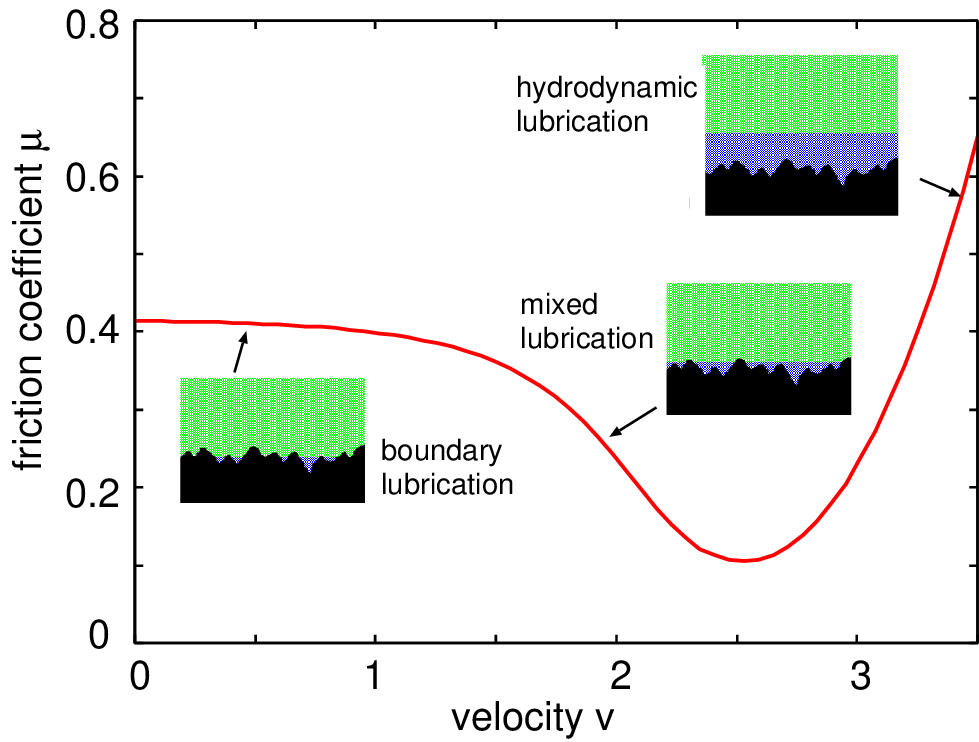}
\caption{
Stribeck curve: The typical relation between the friction coefficient and the sliding speed.
In the boundary lubrication regime, the rubber (green) is in contact with the substrate asperities,
and there is no hydrodynamic lift-off pressure from the fluid, except possibly from some trapped (sealed-off)
fluid islands with pressurized fluid. In the mixed lubrication regime, the fluid pressure increases the separation
between the surfaces, although some asperity contacts still occur. In the (elasto)hydrodynamic regime, the fluid pressure is sufficient
to fully separate the surfaces so that no asperity contact remains.
}
\label{STRIBECK.1logv.2mu.eps}
\end{figure}

\vskip 0.3cm
{\bf 8 Lubricated rubber friction} 

This review mainly focuses on rubber friction on dry surfaces. However, rubber friction on wet surfaces, or on surfaces covered by lubrication or contamination films, is equally important. Applications include tires operating on wet or contaminated road surfaces and dynamic seals in which the sliding surfaces are covered by oil films. Other examples include the friction between shoes and contaminated ground surfaces, and syringes, where the interface between the rubber stopper and the glass or polymer barrel is often lubricated with silicone oil. 

Here we do not present a comprehensive study of rubber friction on lubricated surfaces, but we highlight a few basic facts.

When an elastic body slides on a substrate in the presence of a fluid, one can distinguish three regimes, as illustrated in Fig.~\ref{STRIBECK.1logv.2mu.eps}, which schematically shows the friction coefficient as a function of sliding speed. Here, we focus on the behavior of thin fluid films that may exist in the boundary and early mixed lubrication regions of the Stribeck curve. 

The hydrodynamic region prevails when there is no direct contact between the solids. For randomly rough surfaces with a roll-off, the highest asperities are typically on the order of a few times $h_{\rm rms}$. However, if the surfaces have already been in contact in the boundary lubrication region, the highest asperities may be significantly reduced due to deformation or wear.

In the boundary lubrication region, most of the load is carried by the area of real contact. Nevertheless, a small fraction of the load—typically around 10\%—may be supported by pressurized fluid trapped in sealed-off regions.

Thin films between solid surfaces may be stable or unstable depending on interfacial energies and the nominal contact pressure. We first consider smooth surfaces as studied in Refs.~\cite{dewet1,dewet2,dewet3,dewet4,dewet5}.

Interfacial dewetting in water is observed for
surfaces of hydrophobic solids. A fluid film between two solids with flat surfaces
is unstable and will be removed by dewetting if this results in a reduction of
the free energy. If the interfacial energies (per unit surface area) between the solids
and the fluid are denoted by $\gamma_{01}$ (fluid 0, solid 1) and  $\gamma_{02}$ (fluid 0, solid 2),
and the interfacial energy for the dry solid-solid contact by $\gamma_{12}$, then
the condition for dewetting is that the interfacial (one-dimensional or line) spreading pressure
$$S = \gamma_{12} - \gamma_{01} - \gamma_{02} \eqno(36)$$
is negative. Note that $S$ represents the change in free energy per unit surface area
as the two solids come into contact in the fluid. If $\theta_1$ and $\theta_2$ are the
fluid (thermal equilibrium) contact
angles on the surface of solids 1 and 2, respectively, and $\gamma_{1}$ and $\gamma_{2}$ the corresponding
solid-vapor surface energies, then by Young's equations
$$\gamma_{1} = \gamma_{01} + \gamma_0 \cos \theta_1, \ \ \ \ \gamma_{2} = \gamma_{02} + \gamma_0 \cos \theta_2$$
where $\gamma_0$ is the surface tension of the fluid. Hence, we can also write (36) as
$$S = \gamma_{12} - \gamma_{1} - \gamma_{2} + \gamma_0 (\cos \theta_1 + \cos \theta_2) \eqno(37)$$
The term
$$w = \gamma_{1} + \gamma_{2} - \gamma_{12}$$
is the work of adhesion, that is, the energy per unit surface area required to separate the solids
in dry conditions (no fluid). Thus,
$$S = -w + \gamma_0 (\cos \theta_1 + \cos \theta_2)$$
If $\theta_1$ and $\theta_2$ are larger than $90^\circ$, then $S$ will be negative regardless of the
work of adhesion, which is always positive for neutral solids.

\begin{figure}[tbp]
\includegraphics[width=0.47\textwidth,angle=0]{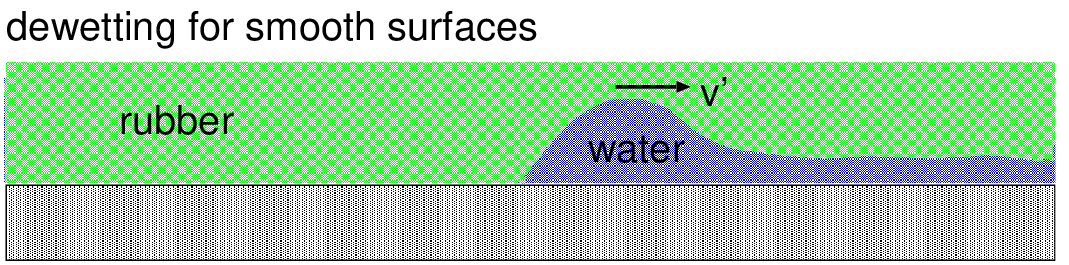}
\caption{
	Dewetting for a soft solid (rubber) initially separated from the substrate
	by a thin fluid film. During dewetting, the fluid accumulates in a rim at the front of the
	expanding dry surface area.
}
\label{DewettinAndAdhesion.eps}
\end{figure}

If $S$ is positive, then in the
absence of external forces a thin fluid film will separate the surfaces of the solids.
If $S$ is negative, the fluid film is unstable and dewetting will occur. For soft solids such as rubber,
this may take place by the fluid accumulating in a rim at the front of the expanding dry surface area,
as illustrated in Fig.~\ref{DewettinAndAdhesion.eps}.

Here we are interested in rubber friction on solids in a fluid environment, so that $\gamma_{12} = \gamma_{\rm rs}$ (rubber-solid), 
$\gamma_{01} = \gamma_{\rm rf}$ (rubber-fluid), and $\gamma_{02} = \gamma_{\rm sf}$ (solid-fluid).

\begin{figure}[tbp]
\includegraphics[width=0.47\textwidth,angle=0]{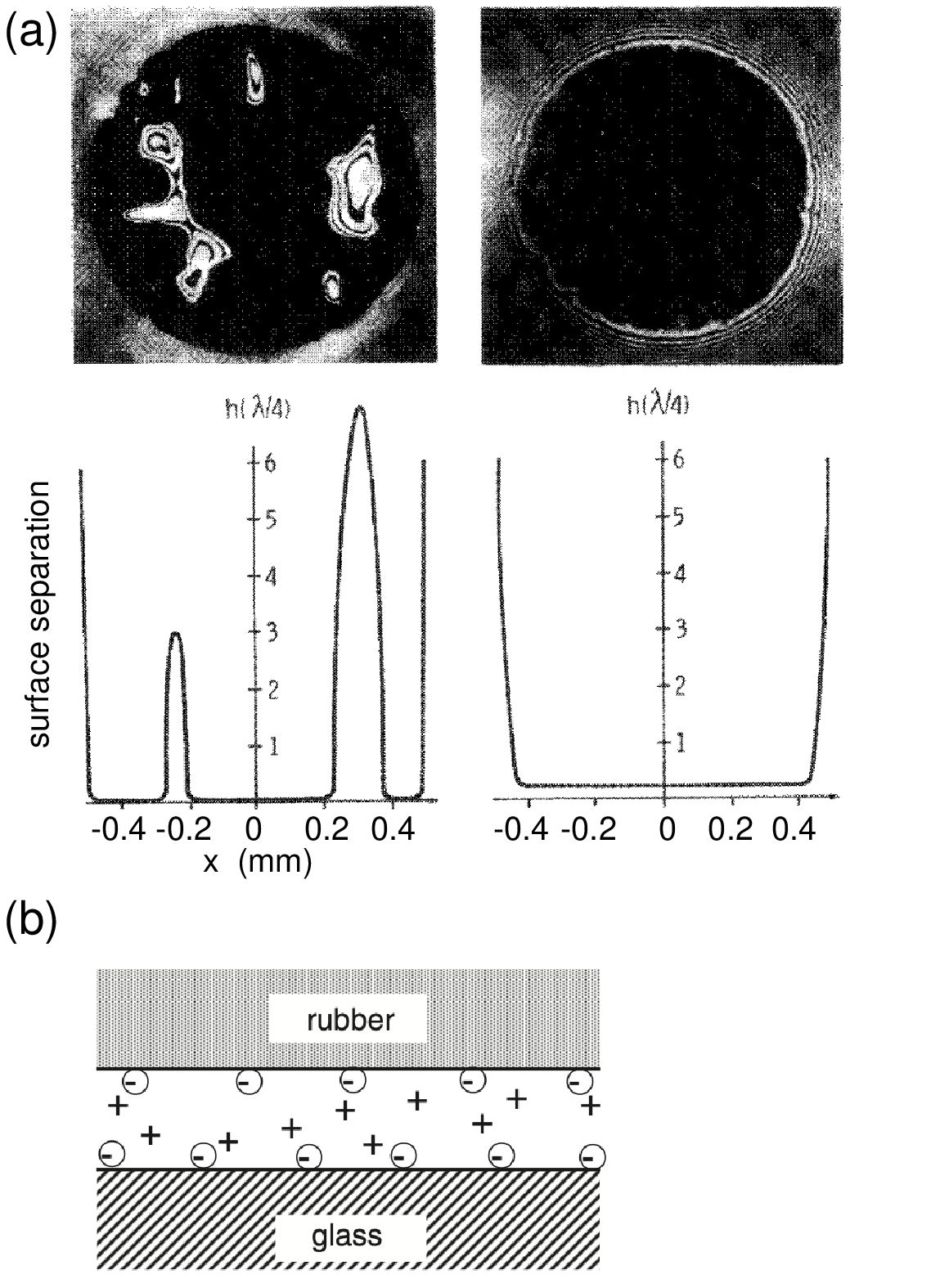}
\caption{
(a) Interferograms (top) and deduced surface separation profiles (bottom) showing the marked difference  
between distilled water (left) and water containing 0.01 $\rm mol/L$ SDS. Distilled water becomes trapped  
in pockets, while the SDS solution forms a thin equilibrium film of approximately $20 \ {\rm nm}$ uniform thickness.  
(b) Water with positive ions (hydrated protons) confined between a rubber surface and a glass surface  
with negatively charged atomic groups. As the water is squeezed out, the concentration of  
positive charges between the surfaces increases (the total number of positive and negative charges remains constant),  
resulting in an increase in entropy and a repulsive force between the solid walls.  
Adapted from Ref.~\cite{dewet5}.
}
\label{ROBERTSpic.eps}
\end{figure}

\vskip 0.1cm
{\bf Wetting and dewetting experiments}

In a series of pioneering studies, Roberts and Tabor \cite{RT1,RT2} investigated the squeezing of liquid films between rubber balls and flat hard substrates.
The thickness of the liquid layer was deduced by analyzing the optical interference pattern in the junction.
During squeezing of the rubber ball against the substrate in a liquid, the profile of the rubber surface in the contact zone bends upward,
and the rubber-substrate separation becomes smallest at the periphery of the contact area, as expected from fluid squeeze-out theory.
When the thickness of the film falls below $40 \ {\rm nm}$, the surfaces suddenly spring together at various points, leading to adhesive contact
over a major part of the contact region. The contact first occurs at some protrusion (defect) near the edge of the contact zone, where the
film thickness is smallest. Once a point of contact is established, it pulls the rest of the rubber into contact.
This is referred to as a dewetting transition.

With a low-viscosity fluid such as water as the lubricant, at the stage of film collapse, the surfaces seal together exceedingly quickly and trap small
islands of fluid. After a few hours, these islands disappear, possibly due to a slow drift toward the periphery of the contact area,
driven by the spatial (Hertzian) pressure distribution in the contact zone. If instead of distilled water a dilute solution of a negatively charged soap 
(sodium dodecyl sulfate, SDS) is used, a drastically different result is observed; see Fig.~\ref{ROBERTSpic.eps}(a) (right). The surfaces no longer spring together, trapping islands of liquid,
but instead remain separated by an almost uniform distance of about $20 \ {\rm nm}$. The film does not collapse over time. In this case, the film is
stabilized by electrical double-layer repulsive forces, which support the normal load even when the squeezing pressure in the
contact region reaches $0.1 \ {\rm MPa}$. The SDS is adsorbed onto the rubber surface [see Fig.~\ref{ROBERTSpic.eps}(b)] with its negatively charged polar end-groups immersed in the water.
The glass surface most likely acquires a negative charge due to the reaction of water molecules with Si=O groups on the surface, forming HOSi. 
The two negatively charged surfaces attract positive ions from the solution, establishing a double layer of charge,
which results in repulsive forces between the surfaces at small wall-to-wall separations. One important manifestation of this thin liquid layer is its lubricating effect:
sliding at 1 m/s results in a friction coefficient of 0.001, compared to 10 under dry conditions or when using distilled water.

Roberts \cite{RT2} also performed experiments in which positively charged soap was adsorbed onto the rubber surface, while the glass surface remained negatively charged.
In this case, the rubber and the glass surfaces snapped together (dewetting transition), and the friction was much higher.
The ability to affect adhesion and sliding friction of solids in liquids by adsorbing molecules with ionic groups on solid walls
has great practical relevance, for example in conditioners for hair-care applications. The ability to modulate the contact between
two soft solid bodies by changing the ionic composition of the surrounding fluid may also be important in many
biological processes.

The discussion above has not considered the influence of external pressures or shear stresses on the
lubrication film thickness. However, if the pressure due to an external load is sufficiently large, a fluid film
can be removed from the interface even if this increases the interfacial energy, that is, even if the spreading pressure $S > 0$.
We refer to this as forced dewetting. Conversely, even if interfacial energies favor removing a fluid film from the interface
(i.e., $S < 0$), sliding motion may force a fluid film to enter the interface.
We refer to this as forced wetting.

\begin{figure}[tbp]
\includegraphics[width=0.47\textwidth,angle=0]{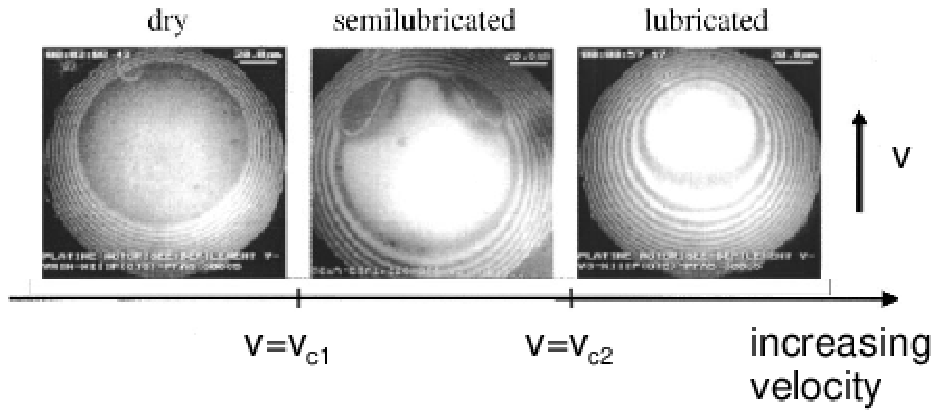}
\caption{
Regimes of forced wetting observed in the sliding rubber/liquid/glass contact as a function of increasing sliding speed $v$.
The arrow indicates the direction of motion of the glass plate.
}
\label{ForcedWettinParis.ps}
\end{figure}

\vskip 0.1cm
{\bf Forced wetting}

In a pioneering work, Martin et al. \cite{dewet1} studied sliding-induced (forced) wetting of a contact 
that was initially dry. They observed (using optical interferometry) the contact of a rubber cap squeezing a nonwetting liquid against a plate moving 
at velocity $v$, as shown in Fig. \ref{ForcedWettinParis.ps}. At low velocities, the contact remains dry [Fig. \ref{ForcedWettinParis.ps}(a)]. 
It becomes partially wetted above a threshold velocity $v = v_{\rm c1}$, 
with two symmetrical dry patches remaining at the rear of the contact. At a second critical velocity $v = v_{\rm c2}$, the contact becomes fully wet. 
The regime $v > v_{\rm c2}$ corresponds to viscous hydroplaning. 
The transitions at $v_{\rm c1}$ and $v_{\rm c2}$ result from the competition between 
liquid invasion induced by shear and spontaneous dewetting of the liquid (between nonwettable surfaces).

For a smooth spherical lens (with radius $R$) sliding on a smooth rubber substrate, 
the critical velocity $v_{\rm c}$ was found to scale with the parameter \cite{dewet1}
$$v_{\rm c} \sim {|S| \over \eta} \left ({|S| \over E R}\right )^{1/3}, \eqno(38)$$
where $S < 0$. This relation was derived using simple and approximate arguments, 
and the experiments performed in Ref. \cite{dewet1} showed a slightly weaker dependence on the fluid viscosity, 
namely $v_{\rm c} \sim \eta^{-3/4}$.

\begin{figure}[tbp]
\includegraphics[width=0.47\textwidth,angle=0]{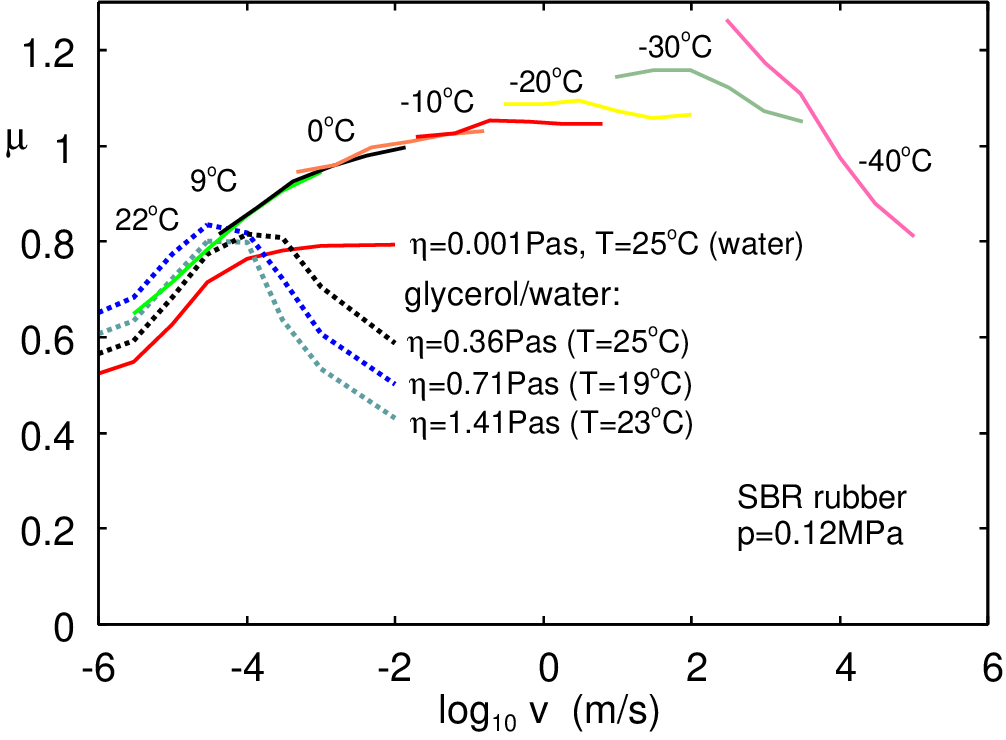}
\caption{
The sliding friction master curve for a SBR rubber with carbon filler
sliding on a dry concrete surface, obtained by shifting
velocity segments acquired at the indicated temperatures
using the bulk viscoelastic shift factor. Also shown are
the friction coefficients at room temperature for the concrete surface lubricated
by glycerol-water mixtures. Adapted from Ref. \cite{Miy}.
}
\label{1logv.2mu.C.all.1.eps}
\end{figure}

\vskip 0.1cm
{\bf Forced dewetting}

If $S$ is positive, a thin fluid film will separate the surfaces of the solids in the absence of external forces. 
This remains true even under an applied squeezing pressure $p_0$, provided it is not too high. 
However, if $S$ is small and the squeezing pressure is large, the fluid film may be removed by what we refer to as {\it forced interfacial dewetting}. 
If $p^*$ denotes the minimum pressure needed to squeeze out the fluid film, then $p^* d^* = S$, 
where $d^*$ is expected to be a molecular-scale distance, approximately $1 \ {\rm nm}$. 
We define the effective spreading pressure as $S^* = S - p d^*$, where $p$ is the pressure in the asperity contact region. 
If $S^* < 0$, then the fluid is removed from the contact regions even when $S > 0$.

We now present a case where forced dewetting may occur at low sliding speeds. 
Consider rubber sliding on a concrete surface lubricated with water-glycerol mixtures. 
Fig. \ref{1logv.2mu.C.all.1.eps} shows the sliding friction master curve for an SBR rubber with carbon filler 
sliding on a dry concrete surface, obtained by shifting velocity segments at the indicated temperatures 
using the bulk viscoelastic shift factor. Also shown are the friction coefficients at room temperature 
for the concrete surface lubricated by glycerol-water mixtures. For $v < 0.1 \ {\rm mm/s}$, 
the friction force is nearly the same in the dry state and in the presence of the water-glycerol mixtures.

The surface tension of water (and glycerol) is approximately $0.07 \ {\rm J/m^2}$. 
If $S$ is positive, its magnitude is expected to be at most a few times $0.01 \ {\rm J/m^2}$. 
Using $S = 0.01 \ {\rm J/m^2}$ and $d^* = 1 \ {\rm nm}$ gives $p^* = 10 \ {\rm MPa}$, 
which is also the typical pressure in the rubber-concrete real contact areas. 
Thus, assuming that the relative asperity contact area $A/A_0$ is approximately $0.01$ (as estimated in Ref. \cite{Miy}), 
and taking $p_0 = 0.12 \ {\rm MPa}$, we find that $p = p_0 A_0/A = 12 \ {\rm MPa}$. 
Hence, the fluid film may be removed from the asperity contact regions even if $S > 0$, 
provided that $S$ is small enough.

For sliding speeds $v > 0.1 \ {\rm mm/s}$, the friction in the fluids is smaller than in the dry state. 
However, a very thin fluid film is sufficient to explain the observed drop in the friction coefficient. 
For example, if the fluid film is approximately $3 \ {\rm nm}$ thick, then for pure glycerol 
($\eta \approx 1.4 \ {\rm Pa \cdot s}$, not accounting for shear thinning) at the sliding speed $v = 1 \ {\rm cm/s}$, 
the shear stress is expected to be $\sigma_{\rm f} \approx \eta v/d \approx 5 \ {\rm MPa}$. 
This gives a friction coefficient of approximately $\sigma_{\rm f} A / p_0 A_0 \approx 0.2$, 
which is consistent with experimental observations.

The drop in the friction coefficient in glycerol, starting at the sliding speed $v \approx 0.03 \ {\rm mm/s}$, is likely due to forced wetting. Assuming $|S^*| \approx 0.01 \ {\rm J/m^2}$, a fluid viscosity of $\eta \approx 1 \ {\rm Pas}$, an elastic modulus $E = 10^7 \ {\rm Pa}$, and using a macroasperity radius of curvature $R = 26 \ {\rm mm}$—as obtained from the concrete surface roughness power spectrum (see Ref.~\cite{Miy})—we obtain from (38) a critical velocity $v_{\rm c} = 0.03 \ {\rm m/s}$, which is consistent with the results shown in Fig.~\ref{1logv.2mu.C.all.1.eps}.

\begin{figure}[tbp]
\includegraphics[width=0.47\textwidth,angle=0]{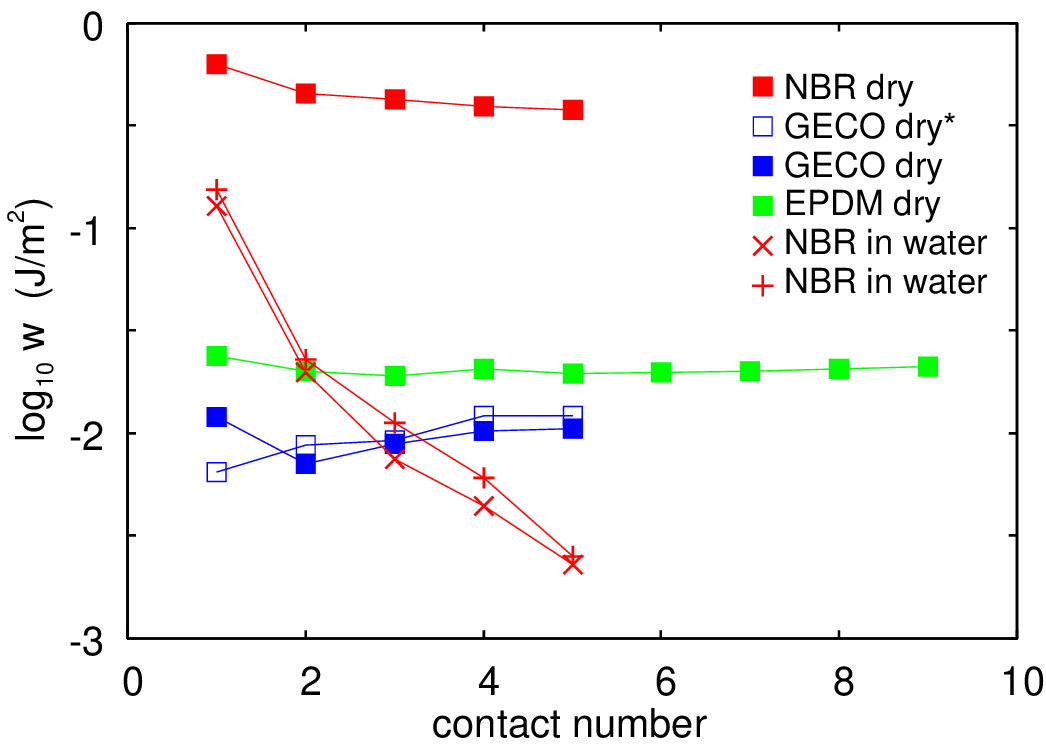}
\caption{
The logarithm of the work of adhesion $w$ during retraction measured during repeated contact 
(drive speed $\approx 1 \ {\rm \mu m/s}$) between a glass ball and 
three types of rubber (without filler): NBR, GECO and EPDM. In water no adhesion is observed for GECO and EPDM.
}
\label{1i.2w.all.dry.wet.1.eps}
\end{figure}

\begin{figure}[tbp]
\includegraphics[width=0.47\textwidth,angle=0]{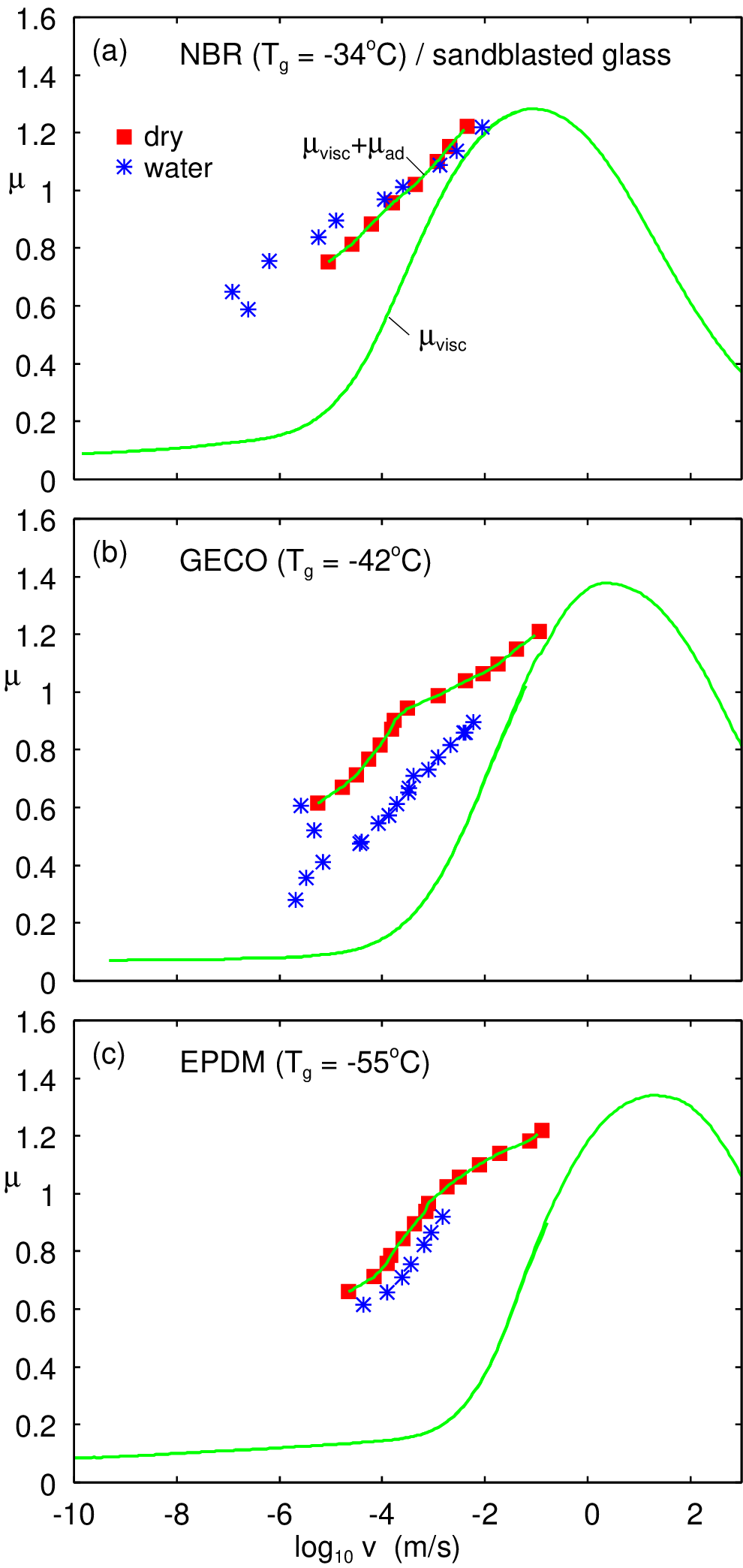}
\caption{
The friction coefficients as a function of the logarithm of the sliding speed 
for the NBR, GECO and EPDM rubbers on dry and wet sandblasted glass. The nominal contact
pressure $p\approx 0.09 \ {\rm MPa}$. The green lines are the calculated viscoelastic contributions.
}
\label{EPDM.GECO.NBR.last.eps}
\end{figure}

\vskip 0.1cm
{\bf On the adhesive contribution to sliding friction}

As discussed in Sec. 5, there are several ways in which adhesion can affect the sliding friction force.
First, it can increase the area of real contact $A$, which would influence both the viscoelastic contribution
and the real contact area contribution to friction \cite{6}. 
There may also be a contribution from the edges of the 
contact regions due to opening cracks\cite{Crack1,Carb,many,Muser,WithGreg}. 
Finally, and most importantly, rubber molecules at the sliding interface may undergo
stick-slip motion [see Fig. \ref{ShearStress.eps}(a)-(c)], as first suggested by Schallamach. This results in a
friction force contribution proportional to the real contact area $A$.

Here we compare adhesion and friction for three types of rubber in the dry state and in water.
Experimental observations indicate that, for rough surfaces, the increase in contact area due to adhesion 
and the crack-opening mechanism appear to be unimportant for rubber friction on rough surfaces \cite{water}.

We present results for three types of rubber: NBR, GECO, and EPDM. 
The compounds were produced without oil and filler particles and are therefore ``clean'' compounds,
suitable for fundamental studies. The GECO compound is strongly hydrophilic and absorbs water 
(swelling) in the wet state. All rubbers were produced in the same mold and have the same (small) roughness.

Fig. \ref{1i.2w.all.dry.wet.1.eps} shows 
the logarithm of the work of adhesion $w$ during retraction, measured during repeated contact 
(drive speed $\approx 1 \ {\rm \mu m/s}$) between a glass ball and the rubber samples. 
In water, no adhesion is observed for GECO and EPDM, while for NBR the adhesion in water is 
(depending on the number of contacts) approximately 5 to 100 times weaker than in the dry state. 
Thus, water either completely eliminates adhesion (for GECO and EPDM) or strongly reduces it.

For the same three rubbers, Fig. \ref{EPDM.GECO.NBR.last.eps} 
shows the friction coefficients as a function of the logarithm of the sliding speed. 
The green lines indicate the calculated viscoelastic contributions. Note that for the EPDM rubber, 
the friction in the dry state and in water is nearly the same up to a sliding speed of approximately $1 \ {\rm mm/s}$, 
despite the absence of adhesion in water. Therefore, for this compound, neither crack opening nor 
increased contact area in water contributes to the sliding friction. We conclude that these effects 
make a negligible contribution to the sliding friction of EPDM.

The figure also suggests that the pressure in the asperity contact regions is high enough to squeeze out the fluid film.
Since no adhesion is observed in water, the spreading pressure $S$ is positive, but clearly, at low sliding speed, 
forced dewetting must occur such that $S^* < 0$. For sliding speeds higher than approximately $1 \ {\rm mm/s}$, 
forced wetting takes place and the friction decreases, but this cannot be measured using the Leonardo da Vinci set-up 
employed in this study, as it results in run-away motion. 
With the Leonardo da Vinci set-up, the driving force is constant, 
so if the friction decreases with increasing sliding speed, accelerated motion occurs.

For the NBR rubber, the friction in water is nearly the same or even higher than in the dry state, again indicating
that the increase in contact area and crack-opening effects are negligible. For GECO, some change in the friction 
between the dry state and the wet state is observed. This can be attributed to the fact that GECO is strongly hydrophilic 
and absorbs water, resulting in swelling and modification of the bulk viscoelastic and other material properties.

\vskip 0.3cm
{\bf 9 Rubber friction on ice and snow} 

Rubber friction on ice and snow has attracted long-standing interest due to its critical relevance to mobility and safety in cold environments \cite{gent2006,robert1, robert2, higgins, ice1, tounonen, klapproth}. 

\vskip 0.1cm
{\bf Friction on ice}

Friction on ice is a fascinating and long-studied subject \cite{rosenberg, kietzig, nye, glacier, baurle}, with pioneering work by Faraday \cite{faraday} over 150 years ago, 
who proposed that the surface of ice is covered by a liquid-like film of water. This phenomenon is now known as \textit{premelting}. 
The presence of a premelted layer depends on both temperature and the hydrophobic or hydrophilic nature of the counter surface \cite{ice2, baran}. 
For inert hydrophobic materials, such as polytetrafluoroethylene (Teflon), such a layer may form well below the melting point of ice, while for hydrophilic surfaces such as silica it may not occur at all, or only very close to the melting point. 

The origin of this water film has long been debated. 
One explanation, proposed by Thomson \cite{thomson}, is pressure-induced melting, but this mechanism is unlikely since plastic deformation of ice usually occurs before the pressure is sufficient to induce melting. 
A more widely accepted explanation is frictional heating, suggested by Bowden and Hughes \cite{bowden}, who argued that sliding at sufficiently high velocity produces local melting of the ice surface. 
This view has been supported by experimental studies \cite{vanD}. 
An important observation for friction on ice is that when the sliding speed increases the friction drops, which implies reduced frictional heating. 
This shifts the onset of ice melting to higher sliding speeds and can explain the velocity dependence of friction on ice \cite{BOjcp}. 

More recent work has challenged the idea that interfacial water films alone are responsible for the low friction of ice. 
Molecular dynamics simulations by Atila et al. \cite{Martin}, inspired by the hypothesis of Moras et al. \cite{Moras} on amorphization in incommensurate contacts, showed that ice surfaces form local ``cold-welded” sites during sliding. 
These sites rapidly transform into amorphous layers under lateral displacement, without requiring significant heating or reaching melting conditions. 
The structure of the amorphous layer resembles supercooled water, providing a lubricating film that reduces shear stress. 
Its thickness scales with the square root of the sliding distance, indicating a displacement-driven process, while local pressure gradients further promote amorphization. 
In this framework, frictional heating plays a secondary role by lowering the effective viscosity of the amorphous layer, thereby reducing shear stress, while amorphization itself remains the dominant mechanism. 
The simulations further reveal that the mechanism operates similarly at hydrophilic and hydrophobic interfaces, although capillary and adhesion-enhanced dissipation increase friction for the former. 
This new perspective helps to explain observations that traditional theories could not account for, such as the persistence of interfacial water well below $0^\circ$C and the pronounced frictional contrast between hydrophilic and hydrophobic counterfaces. 
It also clarifies the dual temperature effect: at lower temperatures, amorphous layers form more rapidly but exhibit much higher viscosity, resulting in higher macroscopic friction.

Rubber friction on ice can be categorized according to temperature and sliding speed:

\begin{itemize}
    \item At temperatures close to the bulk melting point of ice, 
or at high sliding speeds, a water film or a low-viscosity amorphous layer is present at the interface. 
In this regime, the friction is dominated by the viscoelastic deformation of the rubber induced by the ice roughness \cite{ice1, ice2, ice3, Martin}.

    \item At lower temperatures and lower sliding speeds, the interfacial water film is absent or the amorphous layer exhibits a much higher effective viscosity. 
In this case, the friction arises from both the viscoelastic response of the rubber and the shear in the rubber-ice contact regions, 
or between ice fragments attached to the rubber surface and the underlying ice, which corresponds to an ice-ice interaction \cite{ice2, Martin}.

    \item At even lower temperatures (below $T_{\rm g}$), the viscoelastic contribution becomes negligible. 
As the rubber stiffens, it can cause wear of the ice asperities upon contact, leading to increased friction through a plowing-type process \cite{ice3, liefferink}.
\end{itemize}

In Ref. \cite{ice1},  the Persson rubber friction theory was applied to rubber sliding on ice. The study incorporates both the adhesive contribution from the real area of contact 
and the viscoelastic contribution from the rubber. A temperature-dependent large wavenumber cutoff, $q_1(T)$, was introduced to account for the smoothing of short-wavelength roughness 
by plastic deformation of the ice. Since the ice penetration hardness depends on the temperature and indentation speed dependent, the cut-off 
depends on the ice surface temperature and on the rubber sliding speeds. The model also includes the effect of frictional 
heating in the rubber, which leads to a shift in the viscoelastic modulus to higher frequencies. 
The model offers a consistent explanation of the measured friction behavior for three very different tread compounds.

\begin{figure}[htbp]
    \centering
    \includegraphics[width=0.47\textwidth,angle=0]{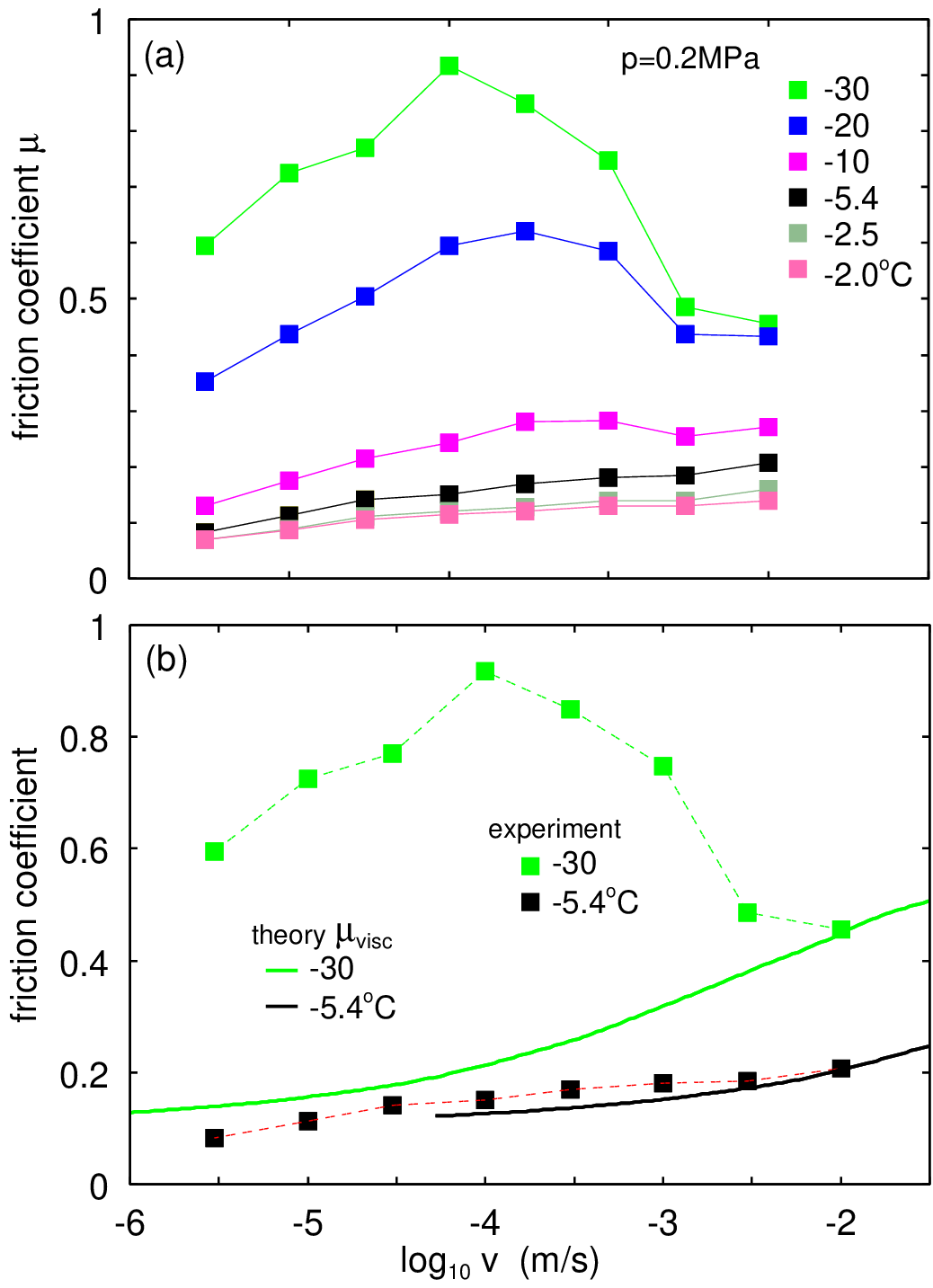}
    \caption{
(a) The measured rubber-ice friction coefficient as a function of the logarithm of the sliding speed
for several indicated temperatures. At low temperatures, the friction is very similar to that of ice sliding on ice, suggesting that the
friction originates from ice wear particles on the rubber surface sliding against the ice surface.  
(b) Comparison between experimentally measured friction coefficients (symbols) and theoretical predictions of the viscoelastic contribution (solid lines), based on the theory developed in Ref.~\cite{ice1}. At the higher temperature (black), the measured friction closely matches the viscoelastic prediction, indicating that a thin premelted water film forms at the rubber-ice interface and that friction is dominated by viscoelastic deformation. At the lower temperature (green), the measured friction is significantly higher than the theoretical prediction, suggesting additional contributions from interfacial shear. Adapted from Ref.~\cite{ice2}.
}
    \label{1logv.2mu.TADA.ice.eps}
\end{figure}

\begin{figure}[htbp]
    \centering
    \includegraphics[width=0.47\textwidth,angle=0]{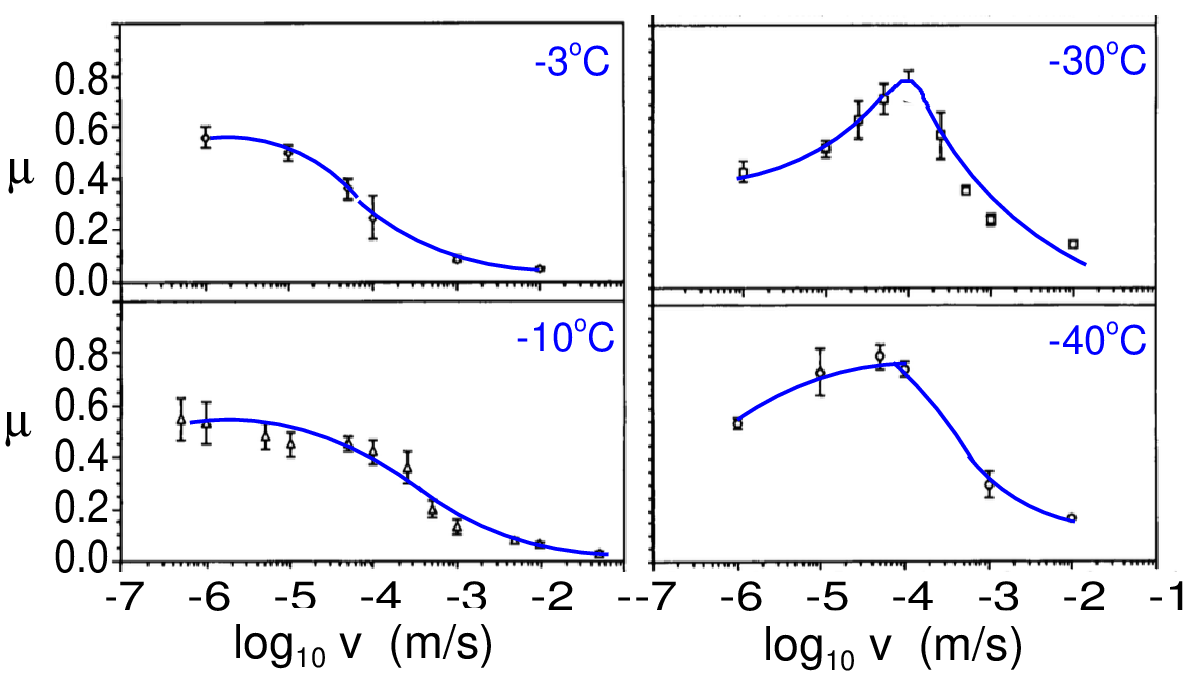}
    \caption{
The friction coefficient as a function of the logarithm of the sliding speed for ice sliding on ice at different temperatures. Adapted from Ref.~\cite{IceIce}.
}
    \label{IceOnIceFriction.eps}
\end{figure}

Fig. \ref{1logv.2mu.TADA.ice.eps}(a) 
shows the rubber friction on ice as a function of the sliding speed for several different temperatures.  
In Fig. \ref{1logv.2mu.TADA.ice.eps}(b) the experimental measurements are compared with theoretical predictions based on the viscoelastic 
contact mechanics theory developed in \cite{ice1}. At the higher temperature ($T = -5.4^\circ {\rm C}$), the experimental data nearly 
coincide with the predicted viscoelastic contribution, indicating that a thin interfacial film is present, which may consist of a premelted water layer 
or a low-viscosity amorphous layer. In this regime, the friction is dominated by viscoelastic deformation of the rubber induced by the ice roughness.  
In contrast, at the lower temperature ($T = -30^\circ {\rm C}$), the measured friction coefficient significantly exceeds 
the viscoelastic prediction. This suggests the absence of a liquid-like film or the presence of a highly viscous amorphous layer, so that 
friction arises not only from the viscoelastic response but also from interfacial shear in the rubber-ice contact regions. 
For $T<-20^\circ {\rm C}$ the friction of rubber on ice approaches that of ice on ice (see Fig. \ref{IceOnIceFriction.eps}), 
indicating a substantial contribution from shear between ice fragments and the underlying ice surface.

\begin{figure}[htbp]
    \centering
    \includegraphics[width=0.3\textwidth,angle=0]{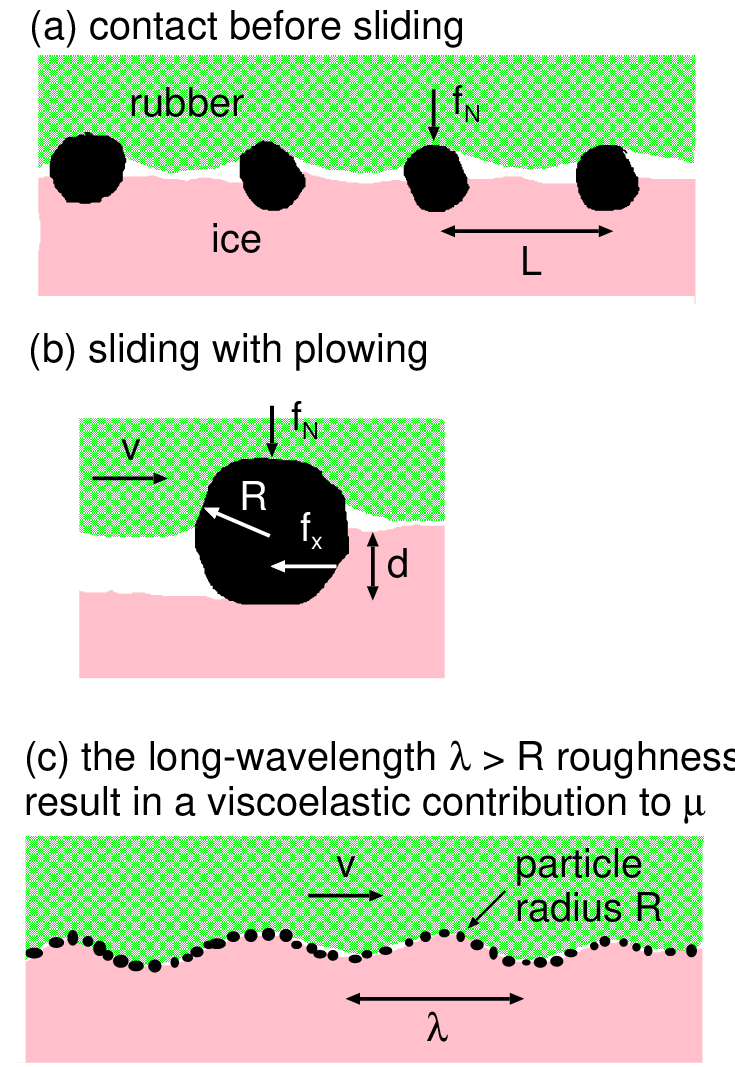}
    \caption{
    (a) A rubber block squeezed in contact with an ice surface covered by stone particles.  
    (b) During sliding, ice plowing occurs, resulting in a friction force that is maximal when
    the projected area of the particles occupies the fraction $\psi \approx 2 \sigma_0/\sigma_{\rm P}$ of 
    the nominal surface area, where $\sigma_0$ is the nominal contact pressure squeezing the rubber block against the substrate, and
    $\sigma_{\rm P}$ is the ice penetration hardness (temperature dependent). The maximum plowing friction coefficient
    is $\mu \approx 1$.  
    (c) The ice roughness with wavelengths larger than the sand particles contributes a viscoelastic component
    to the friction force.} 
    \label{PLOWING.eps}
\end{figure}

\vskip 0.1cm
{\bf Friction on ice with sand}

In many practical situations, ice surfaces are not clean but are covered with contamination. For example, sand is often spread on ice surfaces to increase friction. 
The rubber bends around the sand particles, and most of the slip occurs between the sand particles and the ice, resulting in a plowing contribution to the friction force (see Fig.~\ref{PLOWING.eps}), which can be estimated as follows:

Suppose there are $N$ sand particles in the tire-road footprint. If the concentration of sand particles is high enough, the full normal load of the tire will act on the sand particles, giving the force $f_{\rm N} = F_{\rm N}/N$ that squeezes each sand particle against the ice surface. If the sand particles are spheres with radius $R$, and if $\sigma_{\rm P}$ is the ice penetration hardness, then the sand-ice 
contact area projected onto the horizontal $xy$-plane is $A = f_{\rm N}/\sigma_{\rm P}$.

The maximum friction occurs when the force $f_{\rm N}$ is large enough to squeeze half of the particle into the ice. If $f_{\rm N}$ exceeds the value required to embed half of the sand particle, the entire particle may be pressed into the ice, and the contact would become rubber-ice contact. For the stationary case, the contact area projected onto the horizontal $xy$-plane would be $\pi R^2$, and the maximum penetration resistance is $\pi R^2 \sigma_{\rm P}$. However, for the sliding (plowing) case, we assume the contact area is half of this, $A = \pi R^2 / 2$, since the sand particle only contacts the ice on one side. In this case, the ice wear track will have a width of $2R$, and the vertical cross-sectional area of the plowing track will be $A_{\rm plow} = \pi R^2 / 2$.

If the plowing stress (the stress acting on the sand particle from the ice it displaces during sliding) is similar to the ice penetration hardness $\sigma_{\rm P}$, we obtain the friction force $F_{\rm f} = N A_{\rm plow} \sigma_{\rm P}$ and the normal force $F_{\rm N} = N A \sigma_{\rm P}$, leading to the friction coefficient
$$\mu_{\rm max} = \frac{F_{\rm f}}{F_{\rm N}} = \frac{A_{\rm plow}}{A} = 1 \eqno(39)$$
This equation is only valid if the force squeezing a particle equals $f_{\rm N} = (\pi R^2/2)\sigma_{\rm P}$.

We now apply (39) to tires and ask for which particle concentration this relation holds. Assume the tire footprint area is $A_0$, and that the projected area of the particles occupies a fraction $\psi < 1$ of the footprint area. The number of particles in the tire footprint is $N = \psi A_0/\pi R^2$, and the normal force acting on a particle is
$$f_{\rm N} = \frac{F_{\rm N}}{N} = \frac{\pi}{\psi} R^2 \sigma_0 \eqno(40)$$
where $\sigma_0 = F_{\rm N}/A_0$ is the average tire-road footprint pressure.

Using (40) and $f_{\rm N} = \sigma_{\rm P} \pi R^2/2$ gives $\psi = \psi_0$ where
$$\psi_0 = 2 \frac{\sigma_0}{\sigma_{\rm P}} \eqno(41)$$
Only for this particle concentration will the friction coefficient equal (39). We also note that (41) is the minimum possible particle concentration for which plowing occurs, since if $\psi < \psi_0$, then the normal force will be larger than $(\pi R^2/2)\sigma_{\rm P}$, and the particles will be fully squeezed into the ice surface during plowing.

The penetration hardness depends on temperature (see below), but since rubber friction on ice is lowest near the ice melting temperature, we estimate the particle concentration for this case. Close to the ice melting temperature \cite{TaborIce1,TaborIce1,BonnIce}, $\sigma_{\rm P} \approx 10 \ {\rm MPa}$, and using the average tire-road footprint pressure $\sigma_0 \approx 0.2 \ {\rm MPa}$, we obtain $\psi \approx 0.04$. 

If $L$ is the average distance between the centers of the particles, using $\psi \approx 2 \pi R^2/(\surd 3 L^2)$ gives $L/R \approx 10$. For example, if $R = 1 \ {\rm mm}$, this gives $L \approx 1 \ {\rm cm}$. At lower temperatures, $\sigma_{\rm P}$ increases; for example, $\sigma_{\rm P} \approx 30 \ {\rm MPa}$ at $T = -20^\circ {\rm C}$, and the optimum concentration of sand particles decreases to $\psi \approx 0.01$. However, at low temperatures, rubber friction on clean ice is not so low unless the sliding speed is very high. Very close to the ice melting temperature, a thin water film may form on the ice surface, and in this case, without sand particles, viscous hydroplaning effects could result in very low sliding friction.

There is a second condition that must also be satisfied for (39) to be valid. If a single spherical particle is squeezed against a rubber surface, it will penetrate into the rubber by a distance (Hertz’s theory):
$$\delta = \left( \frac{9 f_{\rm N}^2}{16 E_*^2 R} \right)^{1/3}$$
where $E_* = E/(1-\nu^2)$.
If this distance is larger than $R$, the rubber may contact the ice, and part of the normal force (the load) would then be carried by direct rubber-ice contact. As a result, the force $f_{\rm N}$ acting on the sand particles would be reduced. Hence, for (39) to be valid, we must have
$$\delta = \left( \frac{9 f_{\rm N}^2}{16 E_*^2 R} \right)^{1/3} < R$$
or 
$$f_{\rm N} < \frac{4}{3} E_* R^2 \eqno(42)$$
This is a conservative estimate of the maximum force, as the result is obtained in the limit of a low concentration of particles. For a finite concentration of particles, the upward displacement of the rubber at a given point between the particles will have contributions from all nearby particles. In general, a weaker condition than (42) (i.e., a larger $f_{\rm N}$) will prevail.

Using (40) and (42), we conclude that the rubber will not make direct contact with the ice if $\psi > \psi_1$, where
$$\psi_1 = \frac{3 \pi}{4} \frac{\sigma_0}{E_*} \approx 2 \frac{\sigma_0}{E_*} \eqno(43)$$

Like the viscoelastic modulus $E_*$, the penetration hardness $\sigma_{\rm P}$ is determined by stress-aided, thermally activated processes and depends strongly on temperature and deformation rate. The penetration hardness of ice was studied by Barnes and Tabor \cite{TaborIce1,TaborIce2} and other groups \cite{BonnIce}. In particular, both $\sigma_{\rm P}$ and $E_*$ increase as the temperature decreases and are of similar magnitude for most relevant temperatures and deformation frequencies.

Let us now consider the limit $d \ll R$. In this case, if the width of the plowing track is $2r_0$, the projected contact area is $A = \pi r_0^2 / 2$. The plowing area $A_{\rm plow}$ during slip is the area of the segment of the circle penetrating below the ice surface, and is given approximately by $A_{\rm plow} \approx 4 r_0 d / 3$. Thus, the friction coefficient becomes
$$\mu = \frac{A_{\rm plow}}{A} = \frac{4 r_0 d / 3}{\pi r_0^2 / 2} = \frac{8}{3\pi} \frac{d}{r_0}$$
For $d \ll R$, the penetration $d \approx r_0^2 / 2R$, so that
$$\mu = \frac{4}{3\pi} \frac{r_0}{R}$$
The normal force is $f_{\rm N} = (\pi r_0^2 / 2)\sigma_{\rm P}$, and using this in (39) gives $r_0 / R = \left(2 \sigma_0 / \psi \sigma_{\rm P} \right)^{1/2}$, resulting in
$$\mu = \frac{4 \surd{2}}{3 \pi \surd{\psi}} \left( \frac{\sigma_0}{\sigma_{\rm P}} \right)^{1/2} \approx \frac{0.6}{\surd{\psi}}  
\left( \frac{\sigma_0}{\sigma_{\rm P}} \right)^{1/2} \eqno(44)$$
Note that the friction coefficient is independent of the particle radius and decreases with increasing particle concentration as $1/\surd{\psi}$. This result is only valid as long as $\psi \gg \psi_0$.

Sanding of road surfaces typically involves adding $150 \ {\rm g/m^2}$ of sand particles with a typical particle radius of $1 \ {\rm mm}$. Using the sand mass density $2.6 \ {\rm g/cm^3}$, this corresponds to a volume per surface area $\rho \approx 58 \ {\rm cm}^3/{\rm m^2}$. Using $\psi = 3 \rho / 4R$, this gives $\psi \approx 0.043$. If $\sigma_0 = 0.2 \ {\rm MPa}$ and $\sigma_{\rm P} = 10 \ {\rm MPa}$, this is close to the optimum particle coverage where the friction $\mu \approx 1$. 

\begin{figure}[htbp]
    \centering
    \includegraphics[width=0.47\textwidth,angle=0]{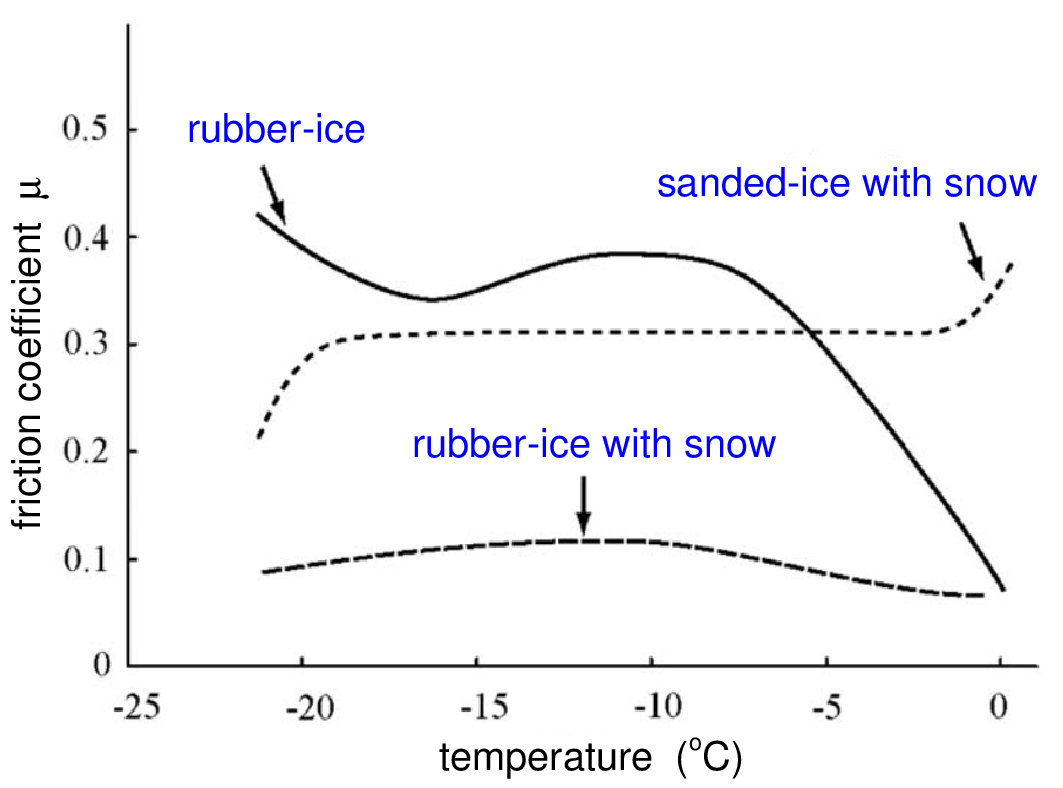}
    \caption{
    Measured rubber friction coefficient as a function of temperature for rubber sliding on clean ice, 
    on ice with a thin snow film, and on ice covered by a thin snow film and sand.
    The friction was measured using the British Pendulum Tester (sliding speed on the order of $\sim 1 \ {\rm m/s}$).
    Adapted from Ref.~\cite{sand}    
} 
    \label{RubberSandIce.eps}
\end{figure}

However, if the friction were this large, it cannot be excluded that the rubber would slip relative to the sand particles instead of maintaining the stick condition assumed in deriving (39). In any case, the exact conditions in terms of particle concentration required for (39) to hold will never be perfectly satisfied in practical applications, where the particle concentration varies along the sliding track. In particular, if $\psi$ is close to $\psi_0$, stochastic fluctuations in the number of particles in the tire-road footprint may result in some particles being exposed to squeezing forces $f_{\rm N}$ large enough to be pressed fully into the ice surface rather than contributing to plowing. This can explain why the friction in most cases is lower than unity, e.g., $\approx 0.3$ in the study reported in Fig.~\ref{RubberSandIce.eps}.

In addition to the plowing contribution to the friction force, there will be a viscoelastic contribution from ice height fluctuations on length scales longer than the size of the sand particles [see Fig.~\ref{PLOWING.eps}(c)]. This arises from the pulsating fluctuations that such long-wavelength roughness exerts on the rubber surface. This contribution to the friction force may be rather small, particularly if the sand particles are large.

Sanding with a high concentration of sand particles $\psi > 0.2$ may result in friction smaller than that on a clean surface, unless the temperature is close to the ice melting temperature (see Fig.~\ref{RubberSandIce.eps}). This is consistent with some experimental observations, where the friction level of rubber on sanded ice was found to be lower than that on clean ice \cite{sand1,sand2}. Norwegian airport maintenance personnel have also pointed out that, in some cases, sanding has a negative effect on the friction level, and they commonly refer to it as a ``ball-bearing'' effect \cite{sand,Paste}. (However, there is no general reason for the sand particles to roll on the ice surface.) 

This behavior is expected when the sand particle concentration is high, because in this case (44) predicts $\mu < 0.2$ if $\psi > 9 (\sigma_0 / \sigma_{\rm P})$, and using $\sigma_0 = 0.2 \ {\rm MPa}$ and $\sigma_{\rm P} = 10 \ {\rm MPa}$ gives $\psi > 0.18$. Since sanding removes the adhesive contribution to friction and the viscoelastic contribution from roughness with wavelengths shorter than the sand particles, this may result in a friction coefficient smaller than that observed on clean ice for $T < -5^\circ {\rm C}$.

Finally, we note that ice wear tracks have been observed even when rubber slides on clean ice surfaces \cite{scratch1,scratch2}. In Ref.~\cite{scratch1}, it was suggested that the narrow wear tracks observed may have resulted from contamination particles (dust), which cannot be avoided when performing experiments in the normal atmosphere. However, in Ref.~\cite{scratch2}, it was suggested that hard filler particles on the rubber surface scratch the ice surface and result in the strongly modified (smoothed) ice surface observed after repeated sliding on the same area. In addition, the stress in rubber asperities that are strongly deformed (strain of order unity) is on the order of the modulus $E_*$, which can be similar to the yield stress of ice. Hence, wear tracks may arise if large enough rubber asperities occur, even without considering the fillers.

\vskip 0.1cm
{\bf Friction on snow}

The friction of rubber on snow shares many similarities with that on ice, as snow can be regarded as loosely packed ice particles.
However, snowflakes can have a complex dendritic structure or consist of small, nearly spherical ice grains, and these two types of snow may exhibit different
frictional properties against rubber or other solids. 
Although dedicated studies on the frictional behavior of rubber on snow surfaces remain limited (mostly for tires) \cite{giessler2007,giessler2010,ripka2009,snow1}, 
there are many studies of the friction of polymer sliders (skis) on snow \cite{snow2,snow3,lulea}. This field was pioneered by Bowden and Hughes \cite{bowden2}, 
who used skis to investigate friction on snow and found characteristics similar to friction on ice, 
although they did not propose a specific model for snow friction. For stiff solid sliders such as skis,
Colbeck \cite{colbeck1988,colbeck1992,klein1947,lever2021,lever2018} identified the key components of snow friction as plowing, 
direct solid-solid contact (i.e., adhesive friction), meltwater lubrication due to frictional heating, and capillary forces from liquid bridges in partially melted snow.

\begin{figure}[htbp]
    \centering
    \includegraphics[width=0.47\textwidth,angle=0]{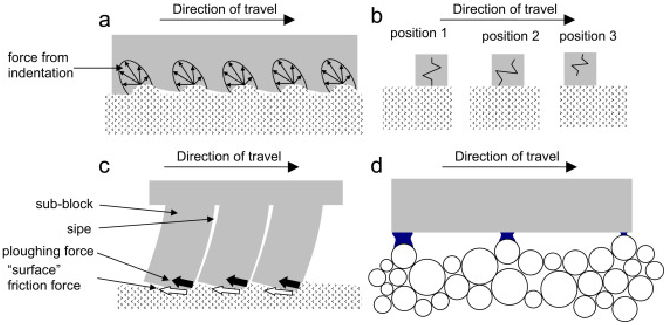}
    \caption{Schematic illustration of the primary friction mechanisms of rubber on snow: (a) indentation-induced resistance due to the normal force from local deformation, (b) molecular adhesion through the formation and rupture of van der Waals bonds at the interface, (c) plowing force and interfacial (surface) friction generated by tread blocks and sipes interacting with snow, and (d) lubricating effect of a meltwater film, which increases in thickness with the residence time of snow grains in contact with the rubber. Adapted from Ref. \cite{snow2}.
    }
    \label{snowfriction.png}
\end{figure}

Some of these mechanisms for friction between snow and hard polymers may also apply to rubber materials, 
but additional considerations are required due to their viscoelastic nature (see Fig.~\ref{snowfriction.png}). 
Any fluctuation in the height of the compacted snow layer that is not flattened in the tire footprint could generate a viscoelastic contribution
to the friction during slip. However, if the frictional shear stress exceeds the shear yield strength of ice, the 
compacted snow may instead undergo internal shear deformations, which would also contribute to the friction force.

\begin{figure}[htbp]
    \centering
    \includegraphics[width=0.40\textwidth,angle=0]{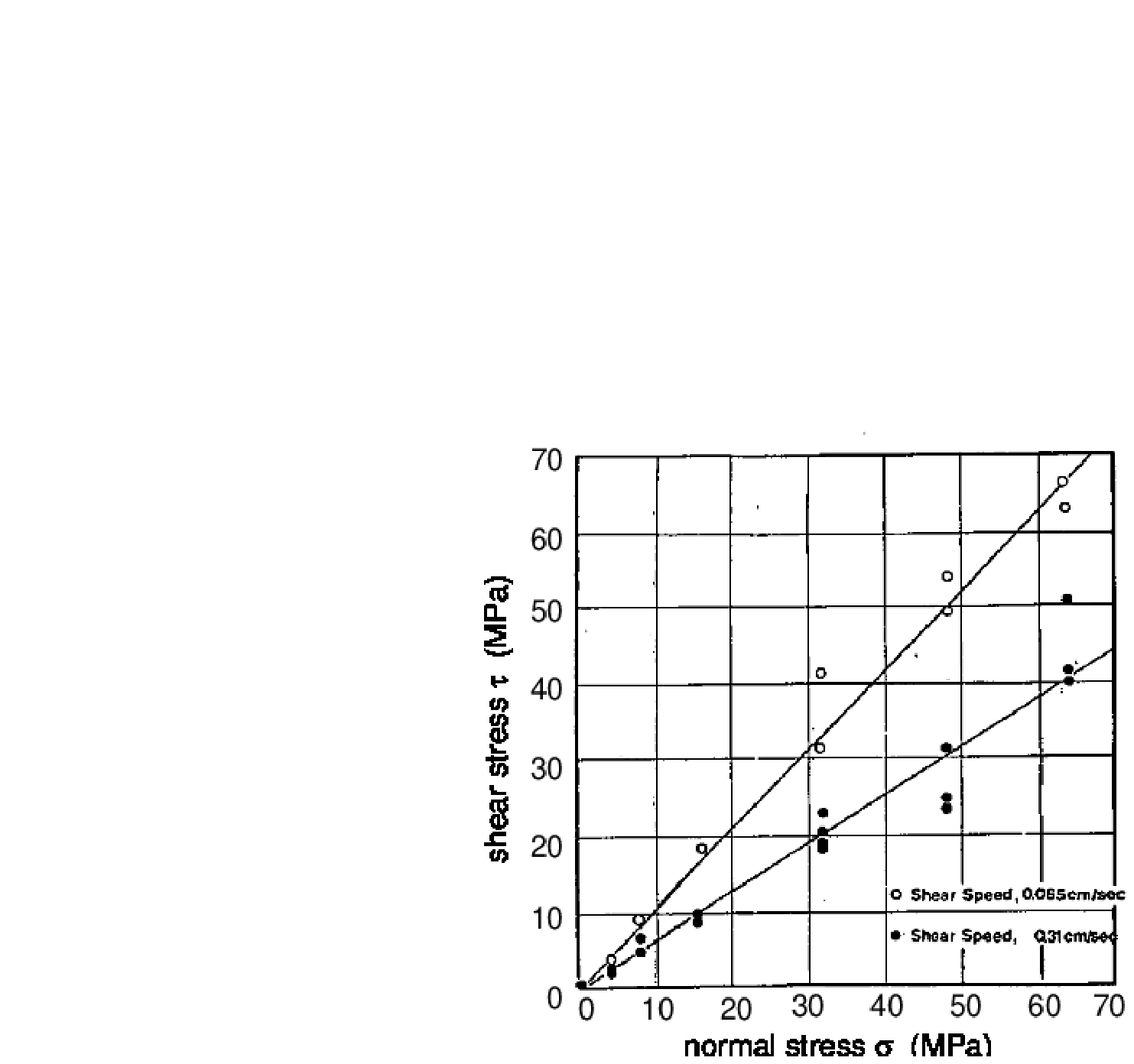}
    \caption{The shear stress of snow as a function of the normal stress for the 
sliding speeds $0.65 \ {\rm mm/s}$ and $3.1 \ {\rm mm/s}$. Adapted from Ref. \cite{ShearSnow}
    }
    \label{ShearStrethpic.ps}
\end{figure}

\begin{figure}[htbp]
    \centering
    \includegraphics[width=0.47\textwidth,angle=0]{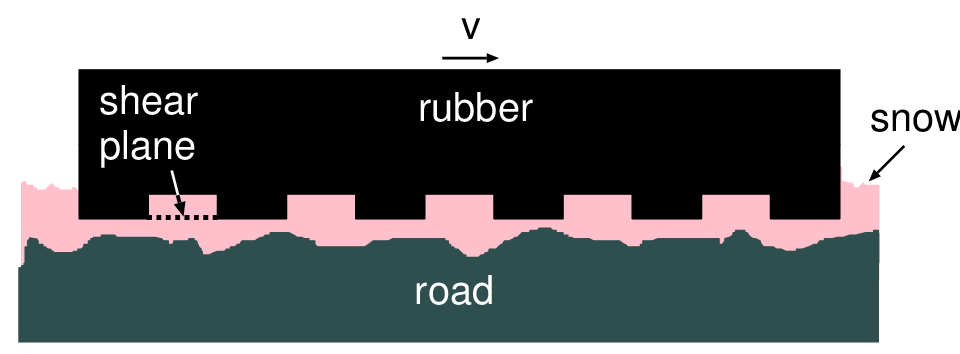}
    \caption{If the normal pressure acting on the snow in the rectangular
cavities of the rubber surface is similar to the nominal pressure acting in the tire-road footprint, then
shearing the snow-snow interface as indicated by the dashed line could contribute to the friction coefficient. 
Depending on the slip velocity, this contribution could be of the order of the fraction of the nominal contact area covered 
by the cavities, e.g., 0.5 if the cavities cover $50\%$ of the total area.
    }
    \label{ShearCompactSnow.eps}
\end{figure}

The shear strength of snow compressed by a normal stress $\sigma$ has been studied experimentally. Fig.~\ref{ShearStrethpic.ps} shows results for the shear stress
immediately after compression (without aging), as relevant for tire applications. The shear stress increases nearly linearly with the applied pressure
and is of similar magnitude to the normal stress. This implies that if the sheared area were similar in size to the nominal tire-road contact area, a friction
coefficient of order unity could be expected. In reality, the area where the shear plane lies inside the compacted snow rather than at the
rubber-snow interface may be much smaller than the nominal contact area. However, if the tire had a large concentration of cavities as in Fig.~\ref{ShearCompactSnow.eps},
and if the pressure between the snow and the rubber in the cavities were comparable to that in the nominal tire-road footprint elsewhere,
then a large friction coefficient could be expected.

Overall, the friction of rubber on snow depends not only on the material properties but also on the snow morphology and water content (which is temperature dependent):

\begin{itemize}
    \item Wet snow or slush (snow mixed with water, salt, mud, etc.) can act as a high-viscosity fluid that cannot be squeezed out of the tire-road footprint
even at relatively low speeds. Hence, it may act as a high-viscosity lubricant, which could result in viscous hydroplaning and low friction.

    \item On dry snow, friction arises in part from shearing snow contacts and in part from viscoelastic deformation of the rubber by inhomogeneities in the
height of the compressed snow film. There is also a contribution from shearing the rubber-snow interface if slip occurs at this interface.
\end{itemize}
    
For rubber tires, tread design and sipes play a critical role. Compared to smooth surfaces, treads with sipes generate additional plowing at the 
edges of the bent rubber segments. In addition, compacted snow can enter the sipes, resulting in a snow-snow interface (of compacted snow) 
that is sheared during slip. These effects significantly enhance snow grip \cite{giessler2007,ripka2009,snow1}. 
Tread depth and void ratio also affect performance. Comparable observations have been reported for footwear on snow surfaces \cite{shoes}. 
Analogous to artificial tread structures, natural surface adaptations can also play a role in snow friction.

Recent experiments on bear paw pads have highlighted the importance of papillae morphology for frictional performance on snow \cite{orndorf2022}. 
Polar bears exhibit papillae that are significantly taller than those of brown and black bears, leading to a higher height-width ($H/D$) ratio. 
Friction tests with three-dimensional printed model surfaces demonstrated that shear stress on snow increases with increasing $H/D$, 
an effect comparable to the function of spikes in human footwear. 
As a result, polar bear paw pads achieve a $1.33$-$1.50$ times higher shear stress compared with other bear species. 
Notably, polar bear paw pads also exhibit smoother roughness at short wavelengths. 
This is attributed to the presence of a thin water film at the ice-snow interface, which prevents roughness at these scales from penetrating the film and contributing to friction. 
Therefore, during evolution, polar bears did not develop pronounced short-wavelength roughness.

These observations emphasize the general principle that surface microgeometry, whether engineered in rubber treads or evolved in biological systems, can critically influence snow friction through its effect on interfacial shear stress.

Finally, we note that measurements of rubber-ice friction on ice surfaces with a low concentration of snow crystals or with a thin snow film show drastically reduced friction coefficients \cite{Gnorich}. Pilots have also reported that ice-covered runway surfaces are much more slippery when loose snow is present \cite{sand,Paste}. 

It is not clear how these results can be understood, since ice sliding on ice generally gives relatively large friction coefficients, unless the temperature is close to the ice melting point or the sliding speed is so high that frictional heating becomes important \cite{IceIce,Pice}.

\vskip 0.3cm
{\bf 10 Summary and conclusions} 

This review has presented a comprehensive analysis of rubber friction on rigid, randomly rough surfaces, combining theoretical formulations and experimental comparisons.

Rubber friction has two primary contributions: viscoelastic energy dissipation due to time-dependent deformations induced by surface asperities, 
and adhesive contribution from the interfacial shear stress acting within the real area of contact. 
The analytical theory of rubber friction is established based on the Persson contact mechanics framework and provides 
quantitative predictions of frictional behavior over a wide range of conditions, incorporating the surface roughness power spectrum 
and the frequency-dependent viscoelastic modulus. When additional mechanisms such as strain softening and frictional heating 
are included, the theoretical predictions show good agreement with experimental results under stationary sliding conditions.

For non-stationary sliding (dynamic friction), other mechanisms become important and manifest as transient friction 
phenomena, such as breakloose friction. This can be explained by the interplay of two dynamic effects: the flash temperature, 
which influences the mechanical properties in the local contact area, and elastic effects, which enable pre-slip and alter the 
frictional response compared to uniform slip across the entire contact area.

The rubber friction theory is also applicable to different geometries (rolling or sliding) and conditions (with or without third-body layers). 
Rolling friction for cylinders and spheres can be interpreted as a limiting case where the contribution from shearing the contact area 
is absent and friction arises solely from viscoelastic dissipation. A theory for rolling friction \cite{theory}, which is in good agreement with experimental data, was reviewed. This approach can also describe the sliding of solid objects with arbitrary shapes on rubber if the contribution from the frictional shear stress in the area of real contact can be neglected, as may be the case for lubricated surfaces.

Open problems include the role of adhesion-induced enhancement of the real contact area and energy dissipation due to crack opening, both of which may be important for surfaces with small roughness. For rough surfaces, there is still no definitive understanding of the origin of the short-distance cut-off to be used when calculating the viscoelastic contribution to the friction force. Future research should aim to integrate these factors, as well as other possible effects (e.g., wear), into a unified model supported by systematic experimental validation across different materials, geometries, and environmental conditions.

\begin{figure}
\includegraphics[width=0.36\textwidth,angle=0.0]{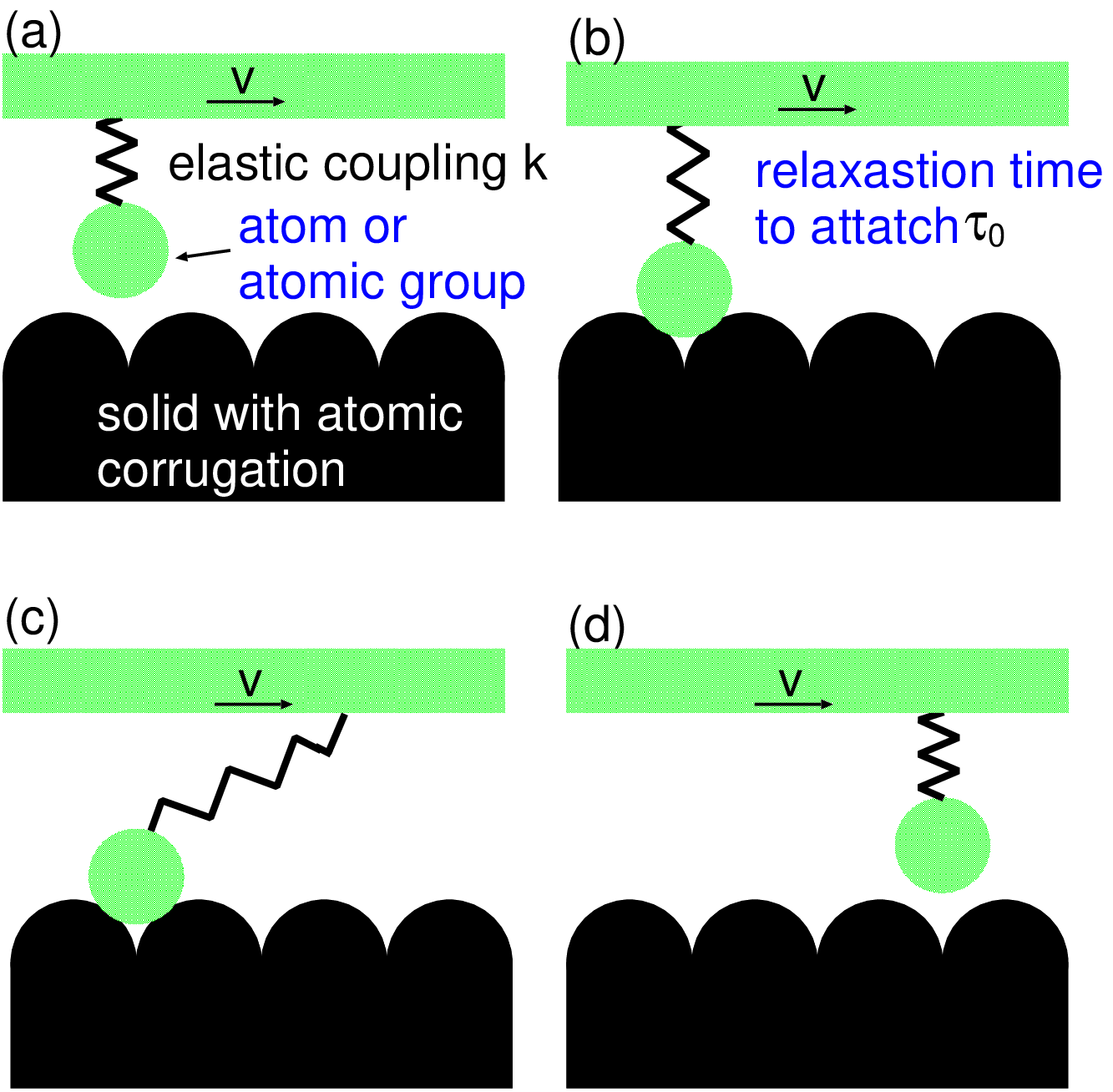}
\caption{\label{SchallamachPicNew1.eps}
A molecular segment of a small (nano-sized) region of the green solid is 
in a detached state (a), moves into contact due to thermal fluctuations (b),
remains attached while the solid locally deforms elastically (c), and finally detaches
when the force in the stretched ``bond'' becomes large enough (d).
Detachment is a thermally activated and stress-aided process. The elastic energy
stored in the stretched bond in (c) is dissipated as heat when the bond breaks.
}
\end{figure}

\vskip 0.3cm
{\bf Appendix A: Adhesive contribution to friction from binding-breaking-binding cycles}

Schallamach developed a fundamental theory for the adhesive contribution to sliding friction.
This theory was originally presented as a model for the adhesive contribution to sliding friction,
but is of more general validity. The basic idea is illustrated in Fig.~\ref{SchallamachPicNew1.eps}.

Consider a solid block (green in Fig.~\ref{SchallamachPicNew1.eps}) sliding on a substrate (velocity $v$).
A molecular segment of a small (nano-sized) region of the block, which we denote as an ``atom'' in what follows, is first in a detached state
(a). To bring it into adhesive contact with the substrate, a free energy barrier must be overcome,
which requires a sufficiently large thermal fluctuation. The average time the atom spends in the detached state is denoted by $\tau_0$.
After attachment, the atom remains attached for an average period $\tau_1$, while the solid locally deforms elastically (c).
When the force or stress in the stretched ``bond'' becomes large enough, the bond breaks.
This is a thermally activated and stress-aided process. When the bond breaks, the elastic energy
stored in the stretched bond is dissipated as heat. 

Consider sliding at the speed $v$. The fraction of time an atom spends in the attached state is $\tau_1/(\tau_0+\tau_1)$.
If $\langle f \rangle$ is the average (tangential) force of the stretched bond, then the
average tangential force acting on the sliding block from the atom is $F=\tau_1 \langle f \rangle /(\tau_0+\tau_1)$.
If there are $c$ atoms per unit surface
area that can participate in the attach–detach process, then the frictional shear stress is
$$\sigma_{\rm f} = {c \langle f\rangle \tau_1 \over \tau_0+\tau_1}.\eqno(A1)$$

Let $P$ be the probability for the ``atom'' to be in the attached state at time $t$, assuming $P(0)=1$.
We have
$${d P \over dt} = -w P\eqno(A2)$$
where $w$ is the probability per unit time 
for going from the attached state to the detached state.

Assume that the atom attaches to the substrate at time $t=0$. The probability that it is still in the same attached state at time
$t$ is $P(t)$. Hence, the average time in the attached state is
$$\tau_1 = \int_0^\infty dt \ P(t)\eqno(A3)$$
If $f(t)$ is the force in the spring at time $t$, the average force during the attachment is
$$\langle f \rangle = {1\over \tau_1} \int_0^\infty dt \ P(t) f(t). $$
If the binding potential keeping the atom attached to the substrate is short ranged, so that the displacement of the atom 
from the equilibrium position is small compared to the extension of the spring at the point of detachment, then
the spring force at time $t$ is $f(t)=kvt$, giving
$$\langle f \rangle = {kv\over \tau_1} \int_0^\infty dt \ P(t) t \eqno(A4)$$
which Schallamach approximated by $\langle f \rangle \approx k v \tau_1$. Using this in (A1) gives
$$\sigma_{\rm f} \approx {c k v \tau_1^2 \over \tau_0+\tau_1} = c k v \tau_0 {(\tau_1/\tau_0)^2\over 1+(\tau_1/\tau_0)}. \eqno (A5)$$
The fraction $N/N_0$ of the total number of atoms $N_0$ that are in the attached state is
$${N\over N_0} = {\tau_1\over \tau_0+\tau_1} = {(\tau_1/\tau_0) \over 1+(\tau_1/\tau_0)}.\eqno(A6)$$

The attachment [step (a) to (b)] may involve rearrangements of the atomic groups, 
which require overcoming a free energy barrier $\varepsilon_0$. The attachment rate is 
$$w_0 = \nu_0 e^{-\beta \varepsilon_0},\eqno(A7)$$
which we denote by $1/\tau_0$. In (A7) $\beta = 1 /k_{\rm B}T$ and $\nu_0$ is the ``attempt frequency,'' which we treat as a constant.
If $\varepsilon_1$ is the energy barrier to break the atom–substrate bond and $U(t)$ the elastic energy in the stretched bond, then
$$w = \nu_1 e^{-\beta [\varepsilon_1-U(t)]}.\eqno(A8)$$
If $u(t)$ is the elongation of the spring $k$, then $U(t) = k u^2(t)/2$. If an atom attaches at time $t=0$, then 
assuming a short-ranged interaction potential, $u(t) = vt$, giving $U(t) = k (vt)^2/2$. Schallamach instead
assumed a linear time dependency, $U(t) = \lambda k v t$, where $\lambda$ is a length of molecular dimension. 
This assumption simplifies the mathematical analysis but will not be used here. 

Introducing 
$$q(t) = \int_0^t dt' \nu_1 e^{-\beta [\epsilon_1-U(t')]}$$
we can write (A2) as
$${d \over dt} \left [P e^{q(t)} \right ] = 0,$$
or
$$P (t) = C e^{-q(t)}.$$
The integration constant $C$ is determined by the initial condition $P (0)=1$, which gives $C=1$, and
$$P (t) = e^{-q(t)}.\eqno(A9)$$
Thus,
$$\tau_1 = \int_0^\infty dt \ e^{-q(t)}.\eqno(A10)$$

If we write $\zeta = w_0 t =t/\tau_0$, then (A10) takes the form
$${\tau_1\over \tau_0} = \int_0^\infty d\zeta \ e^{-q(\zeta)},\eqno(A11)$$
where
$$q(\zeta) = Q \int_0^\zeta d\zeta' e^{R (\zeta')^2},\eqno(A12)$$
with
$$Q=\tau_0 \nu_1 e^{-\beta \varepsilon_1} = {\nu_1\over \nu_0} e^{-\beta (\varepsilon_1-\varepsilon_0)} 
={\nu_1\over \nu_0} e^{-\beta \Delta E},\eqno(A13)$$
and
$$R=\beta {1\over 2} k v^2 \tau_0^2. \eqno(A14)$$
Writing $x = \zeta \surd R$ and $y = \zeta' \surd R$, from (A11) we get
$${\tau_1\over \tau_0} = {1\over \surd R} \int_0^\infty dx \ e^{-q(x)},\eqno(A15)$$
$$q(x) = {Q\over \surd R} \int_0^x dy \ e^{y^2}. \eqno(A16)$$
We can write
$$ckv\tau_0 = \surd R \left ({2k c^2 \over \beta } \right )^{1/2}.\eqno(A17)$$
We define a reference velocity $v^*$ as
$$\surd R = v/v^*.\eqno(A18)$$
Using this we get 
$$v^* = \left ({ 2 k_{\rm B} T \over k \tau_0^2}\right )^{1/2}.\eqno(A19)$$
If we define the reference stress
$$\sigma^* = \left ({2k c^2 \over \beta } \right )^{1/2},\eqno(A20)$$
we can write (A5) as
$${\sigma_{\rm f} \over \sigma^* } \approx {v\over v^*} {(\tau_1/\tau_0)^2\over 1+(\tau_1/\tau_0)}. \eqno(A21)$$

Let us apply the results above to rubber sliding on a hard countersurface. If a small region (patch) 
of radius $R$ of the rubber adheres to the substrate, the effective spring constant $k \approx E R$, where
$E$ is the elastic modulus. Using $c=1/R^2$, this gives 
$$\sigma^* \approx \left ( {2 E k_{\rm B} T \over R^3} \right )^{1/2},\eqno(A22)$$
and
$$v^* = \left ({ 2 k_{\rm B} T \over E R \tau_0^2}\right )^{1/2}. \eqno(A23)$$ 

\begin{figure}
\includegraphics[width=0.47\textwidth,angle=0.0]{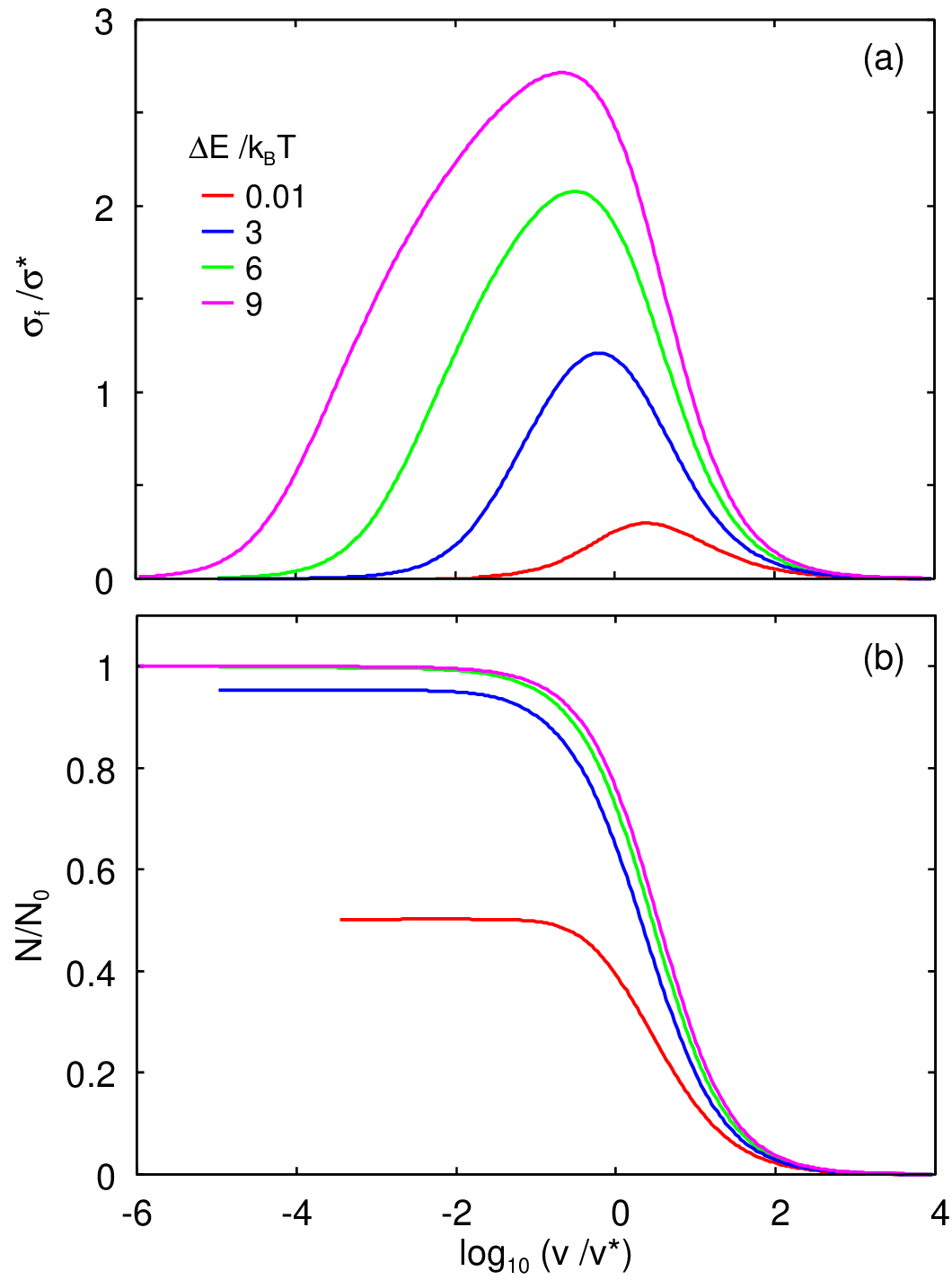}
\caption{\label{1logv.2sigma.3N.eps}
(a) The normalized frictional shear stress $\sigma_{\rm f}/\sigma^*$ and (b)
the ratio $N/N_0$ of the atoms in the attached state, as a function of the logarithm of the sliding speed.
Results are shown for several different binding energies $\Delta E =\varepsilon_1-\varepsilon_0$ and assuming $\nu_0 = \nu_1$.
}
\end{figure}

Fig.~\ref{1logv.2sigma.3N.eps} shows 
the normalized frictional shear stress $\sigma_{\rm f}/\sigma^*$, and 
the ratio $N/N_0$ of atoms in the attached state, as a function of the logarithm 
of the sliding speed for several different binding energies $\Delta E =\varepsilon_1-\varepsilon_0$.

Let us estimate $\sigma^*$ and $v^*$ for rubber at $T=300 \ {\rm K}$. 
We assume that the adhesion involves the rubber bead unit adhering to the substrate so that $R\approx 1 \ {\rm nm}$. Using (A22) with
$E\approx 10 \ {\rm MPa}$, we get $\sigma^* \approx 9 \ {\rm MPa}$. Assuming that $\tau_0$ is of the order of the
shortest Rouse relaxation time, which is the relaxation time characterizing the movement of the
bead units, typically $10^{-6} \ {\rm s}$, we get $v^* \approx 1 \ {\rm mm/s}$. 

\begin{figure}
\includegraphics[width=0.47\textwidth,angle=0.0]{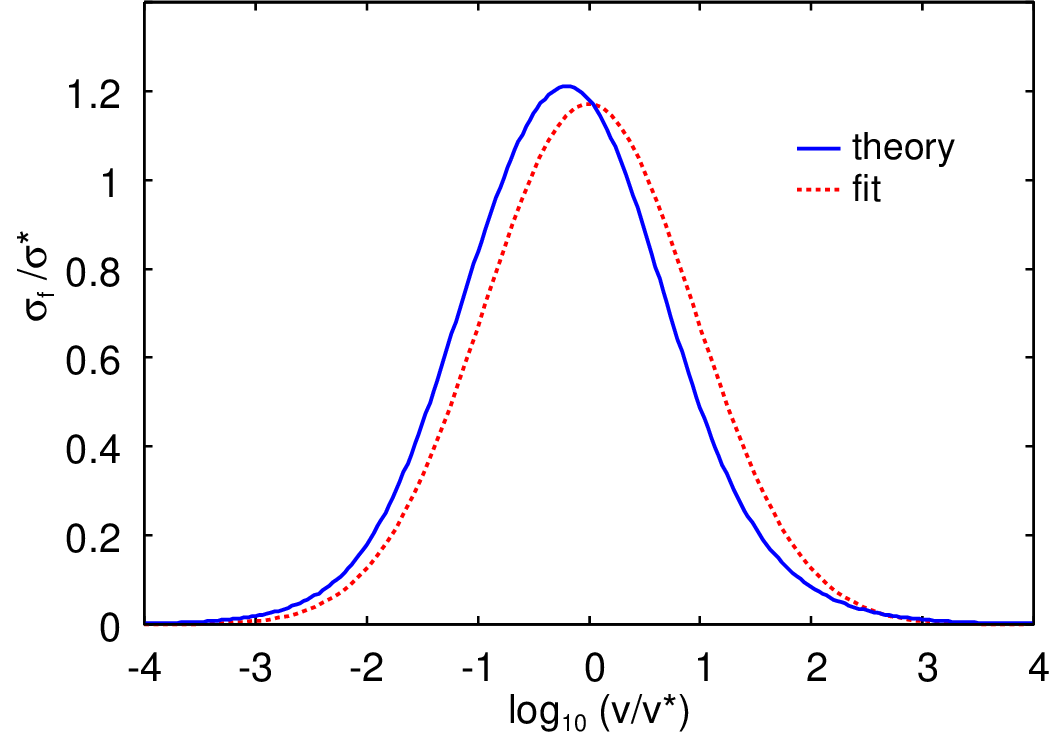}
\caption{\label{1log10v.2tau.Schallamach.theory.fit.eps}
The normalized frictional shear stress $\sigma_{\rm f}/\sigma^*$ as a function of the logarithm of the sliding speed for $\Delta E/k_{\rm B}T=3$.
The blue curve is the theory prediction (from Fig. \ref{1logv.2sigma.3N.eps}) and the dashed red curve a Gaussian fit function.
}
\end{figure}

Fig.~\ref{1log10v.2tau.Schallamach.theory.fit.eps} shows that for $\Delta E/k_{\rm B}T =3$ (a typical bead–substrate binding energy 
resulting from weak van der Waals interactions) $\sigma_{\rm f}/\sigma^*$ is well approximated by a Gaussian:
$${\sigma_{\rm f} \over \sigma^*} = a \, {\rm exp} \left (- b \, \left [ {\rm log}_{10} \left ( {v\over v^*}\right ) \right ]^2 \right )$$
with $a=1.172$ and $b=0.558$. The maximum of the frictional shear stress $\sigma_{\rm f}/\sigma^* \approx 1$, 
and the position of the maximum is roughly at $v/v^*=1$.
Thus, at room temperature we expect the adhesive friction coefficient to be a broad Gaussian as a function of
${\rm log}_{10} v$ with height $\approx \sigma^* \approx 10 \ {\rm MPa}$ and centered at $v\approx v^* \approx 1 \ {\rm mm/s}$. These results are consistent with 
experimental data, but the width (full width at half maximum) of the Gaussian is typically larger than predicted by the Schallamach theory
(typically, $\sim 4$ velocity decades rather than $\sim 2$ as in Fig.~\ref{1log10v.2tau.Schallamach.theory.fit.eps}). This can be explained
by a more complete theory that accounts for the viscoelastic nature of rubber~\cite{theory3}.

\end{document}